\documentclass[amsmath,twocolumn,pra,notitlepage,nofootinbib]{revtex4-1} %nofootinbib
\usepackage{amsthm,amsfonts,graphicx,verbatim, xcolor,bm}
\usepackage{hyperref}
\usepackage[utf8]{inputenc}
\usepackage{siunitx}

\let\oldmarginpar\marginpar
\renewcommand\marginpar[1]{\-\oldmarginpar[\raggedleft\footnotesize #1]%
{\raggedright\footnotesize #1}}
\renewcommand{\eqref}[1]{(\ref{#1})}
\newcommand{\circled}[1]{\textcircled{\raisebox{-0.9pt}{#1}}}

\renewcommand{\vec}[1]{{\bf #1}}

\newcommand{\sign}{\mathrm{sign}}

\newcommand{\ket}[1]{|#1\rangle}
\newcommand{\bra}[1]{\langle #1|}
\newcommand{\braket}[2]{\langle #1|#2\rangle}
\newcommand{\braOket}[3]{\langle #1|#2|#3\rangle}

\begin{document}
\title{Microscopic Study of the Coupled-Wire Construction and Plausible Realization in Spin-Dependent Optical Lattices}
\author{Valentin Cr\'epel$^{1,4}$, Benoit Estienne$^2$, Nicolas Regnault$^{3,4}$}
\affiliation{$^1$Center for Ultracold Atoms, Research Laboratory of Electronics, and Department of Physics \\ Massachusetts Institute of Technology, 77 Massachusetts Avenue, Cambridge, MA, USA
}
\affiliation{$^2$Sorbonne Universit\'e, CNRS, Laboratoire de Physique Th\'eorique et Hautes Energies, LPTHE, F-75005 Paris, France
}
\affiliation{$^3$Department  of  Physics,  Princeton  University,  NJ  08544,  USA}
\affiliation{$^4$Laboratoire  de  Physique  de  l’\'Ecole  normale  sup\'erieure, ENS, Universit\'e  PSL,  CNRS,  Sorbonne  Universit\'e, Universit\'e Paris-Diderot,  Sorbonne  Paris  Cit\'e,  Paris,  France
}

\begin{abstract}
Coupled-wire constructions offer particularly simple and powerful models to capture the essence of strongly correlated topological phases of matter. They often rely on effective theories valid in the low-energy and strong coupling limits, which impose severe constraints on the physical systems where they could be realized.
We investigate the microscopic relevance of a class of coupled-wire models and their possible experimental realization in cold-atom experiments. 
We connect with earlier results and prove the emergence of fractional quantum Hall states in the limit of strong inter-wire tunneling. Contrary to previous studies relying on renormalization group arguments, our microscopic approach exposes the connection between coupled-wire constructions and model wavefunctions in continuum Landau levels. 
Then, we use exact diagonalization methods to investigate the appearance of these fractional quantum Hall states in more realistic settings. We examine the parameter regimes where these strongly correlated phases arise, and provide a way to detect their appearance in cold-atom experiments through standard time-of-flight measurements.
Motivated by this experimental probe, we finally propose a realization of our model with cold-atom in spin-dependent optical lattices. Our estimates show that the previous fractional quantum Hall phases lie within experimentally accessible parameter regimes, giving a viable route towards their experimental study.
\end{abstract}

\maketitle

%\tableofcontents

\section{Introduction}

Our understanding of the fractional quantum Hall (FQH) effect has heavily relied on the use of model wavefunctions (WFs)~\cite{Laughlin_Nobel}. Their strength resides in their simultaneous predictive power and microscopic relevance. For instance, the Laughlin states~\cite{LaughlinWFandTheory,LaughlinBookGirvin} allow to successfully predict the existence of quasiparticle excitations with fractional charge~\cite{ExpFractionalChargeEtienne,ExpFractionalChargeHeiblum} and fractional statistics~\cite{FractionalStatQHE}. They also fully screen the short-range and largest pseudopotentials components of the Coulomb interaction projected in the lowest Landau level (LLL)~\cite{HaldaneHierarchiesPseudoPot,PseudoPotentialsTrugmanKivelson}. Such a feature ensures that it faithfully capture the low energy physics of a fractional quantum Hall system at filling factor $\nu=1/m$, while providing an excellent approximation of the system's ground state (GS).

Another powerful and elegant approach able to predict the universal behaviors of FQH phases is known as the coupled-wire (CW) construction~\cite{CoupledWireFirstKane,CoupledWireTeoKanePRB}. This approach has found applications in widely different models~\cite{WiresBiblio0,WiresBiblio1,WiresBiblio2,WiresBiblio3,WiresBiblio4}, sometimes far away from the conventional FQH phases~\cite{WiresFarFromFQH0,WiresFarFromFQH1,WiresFarFromFQH2}. In the CW construction, a higher-dimensional system is decomposed into a collection of one-dimensional subsystems, the quantum wires. Specific targeted topological phases arise due to suitable couplings between the wires and careful choices of interactions~\cite{CoupledWire_Review}. The effectiveness of the CW construction relies on powerful bosonization and renormalization techniques~\cite{GiamarchiQuantumOneD,Voit_Luttinger}. In the FQH case, this construction provides an intuitive understanding of the bulk-edge correspondence~\cite{ShortCourse_BulkEdge}: the edge effective theory is used for the wires to generate the bulk physics~\cite{CoupledWire_BulkEdge}. Often, the interacting CW models are not chosen for their experimental relevance, but rather to provide an intuitive picture for a given topological phase or to ease the renormalization group analysis.

CW models may however find a greater microscopic relevance and a wider range of application in ultracold quantum gases~\cite{Zoller_ColdAtomCoupledWires}. These physical systems have been envisioned for the realization of FQH phases for the last decade~\cite{Review_TopoMattercoldGas_Goldman,Review_TopocoldGases_Nascimbene}, due to the many experimental advances in the generation of artificial magnetic fields~\cite{LLLBEC_Cornell,LLLBEC_Delibard,LLLBEC_Fletcher,Spielman_FirstExperiment,Jaksch_GaugeFieldOpticalLatt,Review_ArtificialGauge_Dalibard}. In this prospect, the implementation of CW models with cold-atoms in optical lattices displays several advantages. First, cooling of one-dimensional wires below the Doppler limit~\cite{Vuletic_AllOpticalBEC}, and strong artificial magnetic fluxes~\cite{Bloch_StrongEffectiveMagneticFields} have already been demonstrated in such setups. Then, ultracold collisions between neutral particle in $s-$ or $p$-wave~\cite{Swave_CollisionDalibard,Pwave_intAtomicClock} naturally realize the idealized interactions of certain CW models~\cite{CoupledWireFirstKane}. Finally, sub-wavelengths spacing of the wire should provide the necessary tunneling coefficients to create flat Chern bands~\cite{Zoller_ColdAtomCoupledWires}. More recently, experimental breakthroughs in ultracold gases with a synthetic dimension~\cite{DalibardNascimbene_HallBulkSyntheticDimension,Mancini_ObservationChiralEdgestate} open another route towards experimental realizations of CW models, as theoretically envisioned and numerically evidenced in Refs.~\cite{Calvanese_Strinati_2017,Taddia_2017,CornfeldEran_ChiralCurrents,Schollwock_LadderChiralCurrent,LeHur_Ladders}. However, the long-range interactions in the synthetic dimension tend to stabilize crystal phases rather than FQH ones in these setups~\cite{Mazza_CrystalAndStripePhases,Takeshi_DevilStaircase}.

To support these prospective realization of FQH physics in optical lattices, a closer microscopic understanding of CW models is required. In fact, the low-energy and strong coupling limits, used in analytical treatments of CWs models, blurs crucial microscopic properties. The precise range of tunneling and interaction strengths required for the emergence of FQH phases remains unknown in most of these models. We still lack a microscopic characterization of CW ground states, for instance via model WFs, to connect with simple experimental probes such as density. The competing phases in these models and their distinctive features have yet to be identified. These problematics demand sustained efforts in the microscopic study of CW model, either analytically or numerically, but without recurring to effective low-energy theories.

This paper addresses some of these questions for a class of coupled-wire models. In particular, we show how and in which regimes the CW construction is microscopically related to the continuous description of the FQH effect. Moreover, we numerically study the phase diagram of a CW model in realistic experimental conditions.

The paper is organized as follows. In Sec.~\ref{sec:CylinderQHE}, we begin with a short review of continuum FQH system and of the pseudopotential approach for the Laughlin state. In Sec.~\ref{sec:IQHEFromWires}, we show the emergence of a Landau level structure for quantum wires strongly coupled by tunneling in an external magnetic field. Thanks to this correspondence, we adapt the pseudopotential approach to our coupled-wire system, and bridge the CW construction and model WFs (see Sec.~\ref{sec:AddingInteractions}). In Sec.~\ref{sec:ED_ForLaughlin}, we use exact diagonalization methods to sketch the phase diagram of an interacting CW model. We characterize the weakly-interacting phases and highlight an experimental probe to discriminate them from the Laughlin phase. Finally, we propose a plausible experimental realization of our model with cold-atoms in spin-dependent optical lattices (see Sec.~\ref{sec:ExperimentalRealization}). Our estimates indicate that the FQH-like phases could be observed in optical lattices for experimentally realistic parameters, provided temperature can be kept low enough.

\section{Quantum Hall Effect On The Cylinder Geometry} \label{sec:CylinderQHE}

In this section, we briefly review the quantized motion of charged particles on the cylinder geometry, \textit{i.e.} assuming periodic boundary condition along one direction. We introduce model interactions for which the Laughlin wavefunctions (WFs) are the exact densest ground state of the many-body problem~\cite{LaughlinWFandTheory}.

\subsection{Landau Levels} \label{ssec:CylinderLandauLevels}

We consider a two dimensional gas of $N$ particles of charge $e$ moving in the $(x,y)$ plane and subject to a perpendicular magnetic field $\mathbf{B}=\nabla \times \mathbf{A}$. We assume periodic boundary conditions along the $x$ direction and identify $x=0$ with $x=L$, thus mapping the problem to a cylinder. In the Landau gauge $\mathbf{A}= (\sign(e) B y,0)$, the momentum along the compact direction $k_x$ is a good quantum number and periodic boundary conditions impose
\begin{equation} \label{eq:DefinitionGamma}
k_x= \gamma k \, , \, \text{ with } \gamma = \frac{2\pi}{L} \, \text{ and } k \in \mathbb{Z} \, .
\end{equation}
The kinetic Hamiltonian 
\begin{equation}
H_{\rm kin} = \frac{(\vec{p}-e\vec{A})^2}{2m} \, , 
\end{equation}
splits the Hilbert space into Landau levels evenly spaced in energy by $\hbar\omega_c =\hbar |e| B / m$. Most of the QHE physics is already apparent in the Lowest Landau Level (LLL). Hence, ignoring spin degeneracies and assuming the temperature is low enough, we will focus on the LLL physics from now on. This subspace is spanned by momentum eigenfunctions of the form 
\begin{equation} \label{eq:LowestLandauLevelCylinder} 
\phi_k(x,y) = \frac{e^{i \gamma k x}}{\sqrt{\pi\sqrt{L}}} \exp\left(- \frac{(y-y_k)^2}{2 \ell_B^2} \right) \, ,
\end{equation} 
where 
\begin{equation}
	\ell_B = \sqrt{\frac{\hbar}{|e|B}} 
\end{equation}
denotes the magnetic length. In this geometry, the momentum label $k$ also determines the center of the Gaussian wave-packet in the $y$ direction, $y_k=\gamma k \ell_B^2$. We denote by $\tilde{c}_k^\dagger$ the particle creation operator in orbital $k$, thus $\phi_k(x,y) = \braOket{x,y}{\tilde{c}_k^\dagger}{0}$. For non-interacting fermions, the Integer Quantum Hall Effect (IQHE) occurs when a Landau level is completely filled. Here, we focus on the completely filled LLL with filling factor $\nu=1$. We can write the many-body WF as a Slater determinant involving orbitals of the form Eq.~\ref{eq:LowestLandauLevelCylinder}. Denoting by $z_j=x_j-i y_j$ the complex coordinate of the $j^{\rm th}$ particle, this many-body WF can be rewritten (up to a global normalization prefactor) as: 
\begin{equation} 
\Psi_{\nu=1}(z_1, \cdots , z_{N_e})= \prod_{i<j} \left(e^{i \gamma  z_i }-e^{i \gamma  z_j }\right) \prod_i e^{-\frac{y_i^2}{2\ell_B^2}}  \, .
\end{equation}

\subsection{Model Interactions In The LLL} \label{ssec:CylinderModelInteractiosn}

The study of the FQHE is much more difficult for standard perturbation methods are not available. Indeed, the Hamiltonian projected to the LLL consists solely of an interaction term projected to the flat and highly degenerate LLL. The theoretical understanding of the FQHE has thus heavily relied on trial WFs~\cite{MooreReadCFTCorrelator,CFTReviewForFQHE}. The most celebrated example is the Laughlin WF at filling $\nu=1/q$ with $q$ a positive integer. It reads~\cite{LaughlinWFandTheory} 
\begin{equation} \label{eq:LaughlinWFGeneralFilling}
\Psi_{1/q}(z_1, \cdots , z_{N_e})= \prod_{i<j} \left(e^{i \gamma  z_i }-e^{i \gamma  z_j }\right)^{q} \prod_i e^{-\frac{y_i^2}{2\ell_B^2}} \, ,
\end{equation} 
and describes fermionic (resp. bosonic) statistics when if $q$ is odd (resp. even). Although these trial WFs might not describe the exact ground state of a realistic microscopic Hamiltonian at filling factor $\nu=1/q$, they are believed to be adiabatically connected to the latter. Some of the physical content of these strongly interacting phases can then be revealed by the structure and analytical properties of the trial WFs. For instance, the plasma analogy enables predictions on Eq.~\ref{eq:LaughlinWFGeneralFilling} such as the existence of quasi-particles with fractional electric charge $e/q$, which were indeed observed experimentally~\cite{ExpFractionalChargeEtienne,ExpFractionalChargeHeiblum}. The physical relevance of the Laughlin WFs also stems in the ability to derive a microscopic interacting Hamiltonian whose densest ground state is exactly Eq.~\ref{eq:LaughlinWFGeneralFilling}~\cite{PseudoPotentialsTrugmanKivelson}. Those interactions Hamiltonian are called Haldane's pseudo-potentials~\cite{HaldaneHierarchiesPseudoPot}, which we now review.

We will focus on the model interactions whose densest ground state is either the bosonic Laughlin $\nu=1/2$ state or the fermionic Laughlin $\nu=1/3$ state, respectively denoted as $V^{(0)}$ and $V^{(1)}$, but the discussion can be extended to any Laughlin wavefunction~\cite{HaldaneBookQuantumHall,HaldaneHierarchiesPseudoPot,PseudoPotentialsTrugmanKivelson,PseudoPotentialsPapicThomale,PseudoPotentialsTwoBody}. Their action in real space is 
\begin{equation} \label{eq:PseudoPotInRealSpace} \begin{split}
V^{(0)}(z_1,z_2) & = \delta^{(2)} (z_1-z_2) \, , \\ V^{(1)}(z_1,z_2) & = -\nabla^2 \delta^{(2)} (z_1-z_2) \, ,
\end{split} \end{equation} 
properly symmetrized to account for the periodic boundary conditions in the $x$-direction. Their matrix elements in the LLL can be computed with Eq.~\ref{eq:LowestLandauLevelCylinder}~\cite{PseudoPotentialsPapicThomale,Pseudopot_Generalized_Simon,Pseudopot_Simon}. Up to an irrelevant prefactor, the parent Hamitonian for the bosonic Laughlin $\nu=1/2$ reads 
\begin{equation} \label{eq:ModelInnteractionLLLBosons} 
V^{(0)} = \!\! \sum_{u,k,l \in \mathbb{Z}} \left[ e^{-(\gamma\ell_B)^2 \left(k^2+l^2\right)} \right] \tilde{c}_{u+k}^\dagger \tilde{c}_{u-k}^\dagger \tilde{c}_{u+l} \tilde{c}_{u-l}
\end{equation} 
while the one for the fermionic Laughlin $\nu=1/3$ state is 
\begin{equation} \label{eq:ModelInnteractionLLLFermions} 
V^{(1)} = \!\! \sum_{u,k,l \in \mathbb{Z}} \left[ k l e^{-(\gamma\ell_B)^2 \left(k^2+l^2\right)} \right] \tilde{c}_{u+k}^\dagger \tilde{c}_{u-k}^\dagger \tilde{c}_{u+l} \tilde{c}_{u-l} \, .
\end{equation} 
Notice that the momentum conservation is readily satisfied and that the translational invariance along the cylinder axis is explicit as the matrix coefficients do not depend on the center of mass $u$. In the rest of the article, we will \textit{exactly} recover the LLL physics, including possible interactions and the Laughlin physics, starting from a tight binding model of one dimensional wires of free particles in a presence of a magnetic field.

\section{Landau Levels From Coupled Wires} \label{sec:IQHEFromWires}

\begin{figure}
	\centering
	\includegraphics[width=\columnwidth]{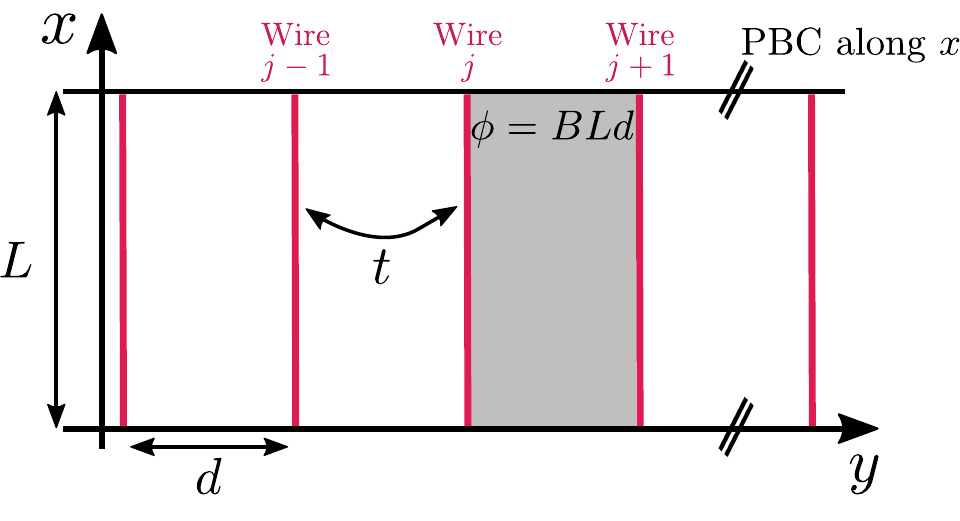}
	\caption{\emph{Periodic array of one-dimensional wires of length $L$ (in red). Periodic boundary conditions are assumed along $x$, the direction of the wires. They are equally spaced and centered at positions $y=jd$ with $j\in\mathbb{Z}$. A magnetic flux $\phi=BLd$ is threaded between each pair of consecutive wires (grey area).}}
	\label{fig:NotationsWires}
\end{figure}

In this section, we consider an array of one dimensional wires of free particles coupled by tunneling under a magnetic field. We prove that for large enough tunneling strengths, this system has all the features of the equivalent continuum Laudau problem studied in Sec.~\ref{sec:CylinderQHE}. At the single-particle level, eigenfunctions are labeled by their momentum along the direction of the wires which also determines their center of mass position as in Eq.~\ref{eq:LowestLandauLevelCylinder}. Altogether, these states form bands with non-trivial topological properties which are unveiled by adapting Laughlin's charge-pumping argument~\cite{Laughlin_QuantizedConductanceLaughlinArgument}. The results can be intuitively regarded as the consequence of taking the continuum limit in the $x$-direction of the Hofstadter model.

\subsection{Model} \label{ssec:ModelNonIntCoupledWires}

\begin{figure*}
	\centering
	\includegraphics[width=\textwidth]{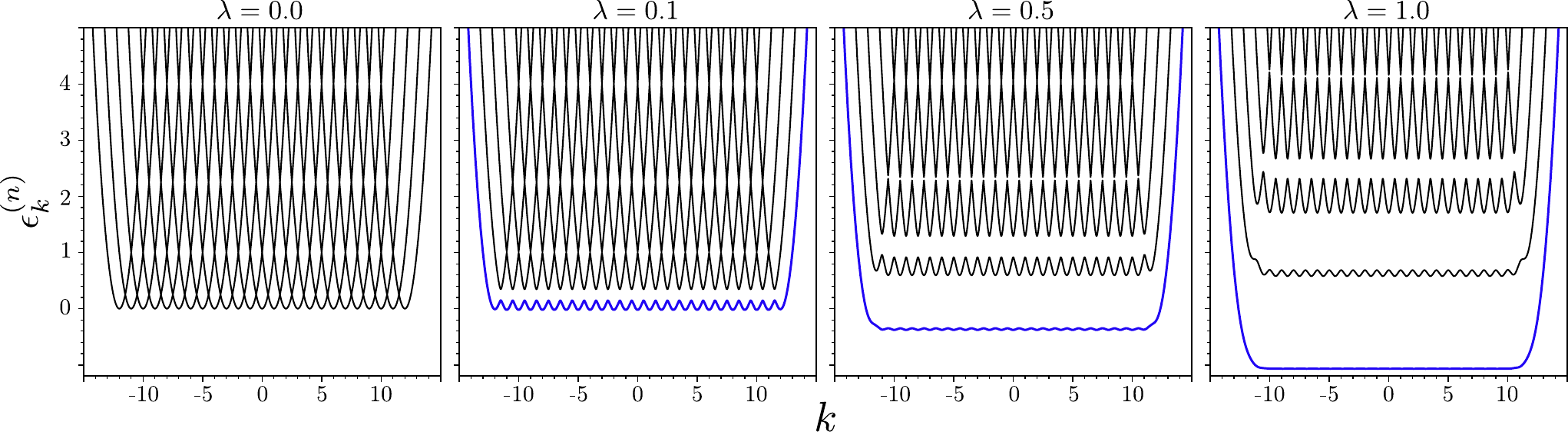}
	\caption{\emph{Spectrum of Eq.~\ref{eq:HamiltonianMomentumSector} for several tunneling strength $\lambda$ in a system of 25 wires with $\nu_w=1$ and open boundary conditions in $j$. The free parabolic branches obtained at $\lambda = 0$ hybridize near crossing points when tunneling between wire increases, until the lowest bands become flat. This happens to the lowest band (blue) for moderate tunneling strength $\lambda\sim 1$. Beyond this point, the system becomes exactly analogous to the Landau levels studied in Sec.~\ref{sec:CylinderQHE}.} 
	}
	\label{fig:TightBindingSpectrum}
\end{figure*}

We consider an array of equally spaced one-dimensional wires of free particles in the $(x,y)$ plane. As sketched in Fig.~\ref{fig:NotationsWires}, the wires are along the $x$ direction for which we assume periodic boundary conditions and identify $x=0$ to $x=L$. The distance between two consecutive wires is $d$ such that the $j$-th wire lies at $y=jd$. A magnetic field is applied perpendicular to the $(x,y)$ plane and we denote the non vanishing component of the vector potential as $A_x(y)=\sign(e)By$. The Hamiltonian consists of two terms, the kinetic energy from the unconfined $x$ direction and the tunneling between wires. We denote as $c_j^\dagger (x)$ the creation operator of a particle at position $x \in [0,L)$ on wire $j$. In second quantized form, the Hamiltonian reads 
\begin{widetext} \begin{equation} \label{eq:FullHamiltonianRealSpace} 
\mathcal{H} = \sum_{j\in \mathbb{Z}} \int {\rm d}x \, \Bigg( c_j^\dagger (x) \frac{(p_x-eA_x(jd))^2}{2m} c_j (x) - t \big( c_{j+1}^\dagger (x) c_j (x) + c_j^\dagger (x)c_{j+1} (x) \big) \Bigg) \, ,
\end{equation} \end{widetext} 
where $p_x = -i \hbar \partial_x$ denotes the momentum operator along $x$ and where we have introduced the tunneling strength $t$ between neighboring wires. As in Sec.~\ref{ssec:CylinderLandauLevels}, the momentum $p_x$ commutes with the kinetic Hamiltonian $\mathcal{H}$ and is quantized in unit of $h/L$ due to periodic boundary conditions. Introducing the Fourier components
\begin{equation} \label{eq:FourierTransformCreationAnnihilation}
c_{j,k}^\dagger = \frac{1}{\sqrt{L}} \int {\rm d}x \, e^{\frac{2i\pi}{L}kx} c_j^\dagger (x) \, ,  
\end{equation}
with $k\in \mathbb{Z}$ an integer, we can split the Hamiltonian of Eq.~\ref{eq:FullHamiltonianRealSpace} into different momentum sectors $\mathcal{H} = \sum_{k \in \mathbb{Z}} \mathcal{H}_k$ with
\begin{equation*} 
\mathcal{H}_k = \sum_{j\in \mathbb{Z}} \frac{(\hbar\gamma k - |e| B jd)^2}{2m} c_{j,k}^\dagger c_{j,k} -t \left( c_{j+1,k}^\dagger c_{j,k}+ h.c. \right) \, .
\end{equation*}
To simplify the notations, we introduce the natural kinetic energy scale 
\begin{equation} \label{eq:DefinitionenergyScale} 
E_0 = \frac{(eBd)^2}{2m} = \frac{\hbar^2}{2m}\Big(\frac{d}{\ell_B^2}\Big)^2 \, ,
\end{equation}
with $\ell_B$ the magnetic length, and we define the dimensionless tunneling parameter
\begin{equation}
\lambda = \frac{t}{E_0} \, .
\end{equation}
We also call $\nu_w$ the inverse number of flux quanta $\phi_0=h/e$ threading the $x-y$ plane between two consecutive wires
\begin{equation} \label{eq:DefinitionOfFlux}
\nu_w = \frac{\phi_0}{\phi} = \frac{2\pi \ell_B^2}{Ld}  \, ,
\end{equation}  
where the flux $\phi=BLd$ is depicted in Fig.~\ref{fig:NotationsWires}. For a finite size system with $N_w$ wires, it is equal to the total filling factor of wires in the system $\nu_w = N_w/N_\phi$ with $N_\phi$ the number of flux quanta threading the system. Using these notations, the Hamiltonian in the momentum sector $k$ becomes
\begin{equation} \label{eq:HamiltonianMomentumSector}
\frac{\mathcal{H}_k}{E_0} = \sum_{j \in \mathbb{Z}} ( j - \nu_w k )^2 c_{j,k}^\dagger c_{j,k} - \lambda (c_{j,k}^\dagger  c_{j+1,k}+c_{j+1,k}^\dagger c_{j,k}) \, .
\end{equation}
Its spectrum is depicted in Fig.~\ref{fig:TightBindingSpectrum} as a function of the momentum $k$. We observe that the free parabolic branches obtained for $\lambda = 0$ hybridize near crossing points when tunneling between wire increases. For $\lambda > 1/4$, the gap opened by the new avoided crossing of width become larger than the initial position of the crossing leading to a nearly flat low-energy spectrum. Higher in energy, the unbounded kinetic energy dominates and tunneling is negligible for the highly energetic bands. We can make these statements more precise by mapping the Schr\"odinger equation originating from $\mathcal{H}_k$ on Mathieu's differential equation, as detailed in App.~\ref{App:AllAboutMathieuEqn}. This enables to use the properties and asymptotics of the solutions of this differential equation~\cite[chap.~28]{NIST:DLMF}. 

For small tunneling strengths $\lambda \ll 1$, the perturbative picture of avoided crossing described above matches the exact solution and the eigenenergies of the system can be obtained as power series in $\lambda$ near the uncoupled wire point $\lambda=0$. The first order correction are simply those obtained within perturbation theory with a gap between the first two bands $\simeq \lambda$, while higher order term can be obtained iteratively~\cite{ContinuedFractionMathieuCharacteristic}. These power series however have a finite radius of convergence $\rho^{(n)}$ in each band $n$~\cite[chap.~28]{NIST:DLMF}, which set the transition between a perturbative regime $\lambda<\rho^{(n)}$ and the flat band regime $\lambda>\rho^{(n)}$ (as can be seen in Fig.~\ref{fig:TightBindingSpectrum}). These radii have been numerically estimated in Ref.~\cite[chap.~2.4]{MeixnerMathieuLectureNotes}. Their results $\rho^{(0)} \simeq 0.3672$ and $\rho^{(1)} \simeq 0.9425$ agree extremely well with both those presented in Fig.~\ref{fig:TightBindingSpectrum} and the study carried out in App.~\ref{App:AllAboutMathieuEqn}. When $\lambda>\rho^{(0)}$, we can no longer use the previous perturbative expansion and must rely on uniform semiclassical approximations~\cite{KurzPhDApproximationMathieuWF,GoldsteinCharacteristicFunction} to obtain estimates of, for instance, the spread of the lowest band
\begin{equation} \label{eq:ExponentiallySmallSpread}
\delta^{(0)} \underset{\lambda \gg 1}{\simeq} \sqrt{\frac{2}{\pi}} \left(8\sqrt{\lambda}\right)^{3/2} e^{-8\sqrt{\lambda}} \, .
\end{equation}
Due to the exponentially small spread of the energy bands, we shall refer to the limit $\lambda \gg 1$ as the flat-band regime. We show in App.~\ref{App:AllAboutMathieuEqn} that Eq.~\ref{eq:ExponentiallySmallSpread} agrees within a percent to numerical simulations even for moderate tunneling strengths $\lambda \sim 1$. Finally, we come back on the role of kinetic energy in highly excited bands $n \gg 1$. There, the flat band limit is extremely hard to achieve as $\rho^{(n)} \propto n^2$~\cite{RadiusConvergenceVolkmer} explaining why the parabolic profile still dominates for high energies in Fig.~\ref{fig:TightBindingSpectrum}.

\subsection{Strong Tunneling: Emergence Of Landau Levels} \label{ssec:NonIntLandauLevels}

Let us define as $d_k^{(n) \dagger}$ the operator creating a particle in band $n$ with momentum $k$, with energy $\epsilon_k^{(n)}$. The corresponding eigenfunction $\psi_k^{(n)}$ has the form 
\begin{equation} \label{eq:CoupledWireEigenfunction_GeneralMathieuImplicit}
\psi_k^{(n)}(x,j) = e^{i\gamma kx} g_k^{(n)}(j-\nu_w k) \, , 
\end{equation} 
where the $g_k^{(n)}$ is centered around zero and can be expressed in terms of Mathieu special functions (see Eq.~\ref{eq:AppMathieuExactExpressionForG}). We recover a structure analogous to the LLL on the cylinder studied in Sec.~\ref{sec:CylinderQHE} where the momentum label $k$ also determines their center $y_k$ along the cylinder:
\begin{equation} \label{eq:CenterWannierLikeOrbitals}
y_k = \nu_w dk = \gamma k \ell_B^2 \, .
\end{equation}

We now focus on the flat-band regime, and show that these eigenstates become analogous to the LLL orbitals of Sec.~\ref{sec:CylinderQHE}. This result can be rationalized if we interpret our model Eq.~\ref{eq:FullHamiltonianRealSpace} as a variant of the Hofstadter Hamiltonian in which the continuum limit has been taken along the wire direction $x$. In the strong tunneling limit, the system is asymptotically equivalent to an harmonic oscillator~\cite{SipsApproximationMathieuWF1,SipsApproximationMathieuWF2} of characteristic frequency $\hbar \omega_c = 2 \sqrt{\lambda} E_0$. The convergence is extremely fast for low-lying bands such that the functions $g_k^{(n)}$ for different momenta are exactly equal up to corrections exponentially small in $\sqrt{\lambda}$. In particular, they all have the same expression $g^{(n)}$ and the same Taylor expansion up to arbitrarily large powers of $1/\sqrt{\lambda}$ when $\lambda\gg 1$. Similarly all eigenenergies within a given band are equal up to exponentially small corrections. For instance, we have the asymptotic behavior~\cite{KurzPhDApproximationMathieuWF,SemiclassicalWKBConnorSmith}:
\begin{equation} \label{eq:LLLdispersionStrongTunneling}
\epsilon_k^{(0)} \simeq \frac{1}{2} \hbar \omega_c + \delta^{(0)} \sin (2\pi \nu_w k) \, , 
\end{equation}
with $\delta^{(0)}$ given in Eq.~\ref{eq:ExponentiallySmallSpread}. As a consequence, we will approximate the system in the flat-band regime by perfectly flat bands of energy $\epsilon^{(n)}$ and corresponding eigenfunctions $\psi_k^{(n)}(x,j) = e^{\gamma kx} g^{(n)}(j-\nu_w k)$. As in Sec.~\ref{ssec:CylinderLandauLevels}, the single body wavefunctions of a given Landau level are shifted copies of the same envelope along $y$, and plane-waves along the compact dimension. Let us stress again that this approximation is really well satisfied for the lowest band, where we only require $\lambda>\rho^{(0)} \simeq 0.3672$ (see lower right corner of Fig.~\ref{fig:TightBindingSpectrum} for which $\lambda=1$).

Not only do the eingenvalues converge towards those of an harmonic oscillator, but the envelopes $g^{(n)}$ themselves uniformly converge to hermite functions, as shown in Ref.~\cite{AunolaDiscreteHarmonicOscillator}. In the lowest band, this reduces to: 
\begin{equation} \label{eq:LowestBandisgaussian} 
g^{(0)}(u) \xrightarrow[\lambda \gg 1]{} g^{(0)}(u)= \frac{1}{(\pi \sqrt{\lambda})^{1/4}} \exp \Big(-\frac{u^2}{2 \sqrt{\lambda} }\Big) \, .
\end{equation}
For larger tunneling strength $\lambda$, a similar behavior is observed in the higher bands of the system. Although it is not necessary for the quantum Hall physics to arise~\cite{Haldane_GeometricDescription,GeometryLaughlin_PapicHaldane,BandMassAnisotropy_PapicHaldane}, we can recover the initial isotropy of the Landau problem on the cylinder (see Sec.~\ref{sec:CylinderQHE}) by matching the width of the obtained Gaussian with the magnetic length. This is achieved by a fine tuning of either the tunneling strength or the inter-wire distance in order to get
\begin{equation} \label{eq:IsotropyCondition} 
\sqrt{\lambda} d^2= \ell_B^2 \quad \Longleftrightarrow \quad t = \frac{\hbar^2}{2md^2}  \, .
\end{equation}

We now summarize the results obtained in the flat-band regime, which for low-lying bands only requires moderate tunneling strengths $\lambda \gtrsim 1$. First the spectrum of our model Eq.~\ref{eq:FullHamiltonianRealSpace} is, up to exponentially small corrections, made of highly degenerate flat bands centered on $n \hbar \omega_c + \epsilon^{(0)}$ with $\hbar \omega_c = 2\sqrt{\lambda}E_0$ and $\epsilon^{(0)} = -2\lambda+\sqrt{\lambda} -1/16$~\cite{NIST:DLMF}. The eigenfunctions with momentum $k$ are centered around $y_k = \gamma k \ell_B^2$ and their envelope $g^{(n)}$ does not depend on the momentum label $k$, recovering the Landau level structure of Sec.~\ref{sec:CylinderQHE}. When $\lambda \gg 1$, the analogy can be pushed further and eigenfunctions of the lowest band have a Gaussian shape Eq.~\ref{eq:LowestBandisgaussian}, which exactly matches that of the cylinder LLL orbitals when Eq.~\ref{eq:IsotropyCondition} is fulfilled.

\subsection{Charge Pumping And Quantized Hall Conductance} \label{ssec:PumpingArgument}

The momentum periodicity of the spectrum obtained in Eq.~\ref{eq:LLLdispersionStrongTunneling} reflects another translational symmetry in our model. Writing $\nu_w = \frac{p_w}{q_w}$ with $p_w$ and $q_w$ relatively prime, we observe that the Hamiltonian Eq.~\ref{eq:FullHamiltonianRealSpace} is invariant under the action of the magnetic translation operator sending both $j\to (j+p_w)$ and $k\to (k+q_w)$. This symmetry explains why the spectrum at $k$ and $k+q_w$ are equivalent. Similarly, it allows to relate the eigenfunctions of Eq.~\ref{eq:CoupledWireEigenfunction_GeneralMathieuImplicit}, derived in App.~\ref{App:AllAboutMathieuEqn}, by the pseudo-periodic relation
\begin{equation} \label{eq:PseudoPeriodic_FromMagneticTranslation}
\psi_{k+q_w}^{(n)} (x,j+p_w) = e^{i\gamma q_w x} \psi_k^{(n)}(x, j) \, .
\end{equation}
From now on, we focus on the lowest band of the system but the discussion applies equally well to more energetic ones. The first consequence of Eq.~\ref{eq:PseudoPeriodic_FromMagneticTranslation} is that there are only $q_w$ distinct functions $g_k^{(0)}$ and we rewrite
\begin{equation} 
\psi_k^{(0)}(x,j) = e^{i\gamma kx} g_{r(k)}(j-\nu_w k) \, , 
\end{equation} 
with $r(k)$ the remainder after division of $k$ by $q_w$.

With this results at hand, we would like to repeat Laughlin's charge-pumping argument~\cite{Laughlin_QuantizedConductanceLaughlinArgument}. In other words, we want to show the quantized Hall conductivity of our model at filling $\nu=1$ and thus unveil the non-trivial topology of our model. This will make the analogy with a continuum LLL complete. The particle filling of the lowest band is obtained as
$\nu = n_{\rm 1D} \nu_w$ with
\begin{equation}
n_{\rm 1D} = \frac{N}{N_w} 
\end{equation}
the number of particles per wire. Here, we implicitly consider a finite size system of $N_w$ wires with open boundary conditions along $j$. However, we still use Eq.~\ref{eq:PseudoPeriodic_FromMagneticTranslation} which remains extremely well satisfied in the bulk of the system where we can neglect the edge effects (see lower-right panel of Fig.~\ref{fig:TightBindingSpectrum} where $N_w = 25$ and $q_w = 1$).

Laughlin's charge-pumping argument starts by introducing a time-dependent flux $\Phi = \theta (t) \Phi_0$, with $\Phi_0 = h/|e|$ the quantum of flux, threading the surface enclosed by the wires in a system at $\nu=1$\cite{Laughlin_QuantizedConductanceLaughlinArgument}. This situation can be described by the modified gauge potential $A_x(y) = \sign(e) B y + \Phi/L$. The flux is increased from $\theta (0)=0$ to $\theta (t_f)=q_w$ adiabatically, \textit{i.e.}, $\omega_c (\partial_t \theta) \ll 1$, in order to adiabatically follow the eigenstates of the lowest band. We thus have $\psi_k^{(0)}(x,j; \theta) = e^{ikx} f_\theta(k,j)$ where the function $f_\theta$ takes simple forms at specific values of $\theta$
\begin{equation} \begin{split}
f_0(k,j) & = g_{r(k)} (j-\nu_w k) \, , \\
f_1(k,j) & = g_{r(k+1)} (j-\nu_w (k-1)) \, , \\
& \vdots  \\
f_{q_w}(k,j) & = g_{r(k)} (j + p_w-\nu_w k) = f_0(k,j+p_w) \, .
\end{split} \end{equation}
We emphasized a few well chosen intermediate states which can be exactly described for any value of the tunneling strength $\lambda >0$. In the flat band limit, the discussion simplifies since all $g_{r}$ are equivalent. After a full ramp-up of the flux $\Phi$, all orbitals recover their original expressions, with a shift of their center of mass by $\Delta y = \nu_w p_w d$ in the $y$-direction (see Eq.~\ref{eq:CenterWannierLikeOrbitals}). This displacement leads to a current whose response is determined by the transverse conductance~\cite{Polarization_ChernLandauLevel_Moore,Polarization_RestaModernTheory}
\begin{equation}
\sigma_{xy} = \frac{L\Delta P_y}{\theta(t_f)\Phi_0} \, ,
\end{equation}
with $\Delta P_y$ the polarization induced by threading operation~\cite{Polarization_MartinOrtiz,Polarization_Vanderbilt,Polarization_KingSmith}. It can be computed as the density of displaced charged $\Delta P_y = \frac{|e| p_w N_\phi}{LN_w}$, where we have used that each of the $N_\phi$ was filled and the previously computed displacement $\Delta y$. Combining the different pieces, this gives the quantized Hall conductance
\begin{equation}
\sigma_{xy} = \frac{e^2}{h} \nu = \frac{e^2}{h} \, ,
\end{equation}
and reveals the non-trivial topology of the lowest band of the system.

\section{Interactions In The Flat Lowest Band: Fractional Quantum Hall States} \label{sec:AddingInteractions}

In this section, we want to extend our discussion to include interactions between particles originally on the same wire. For simplicity, we consider the flat-band regime and require the temperature of the system to satisfy $\delta^{(0)} \ll k_B T \ll \hbar \omega_c$ in order to project the whole dynamics onto the lowest band of the system. Our discussion will closely follow that of Sec.~\ref{ssec:CylinderModelInteractiosn}, and we will borrow the exact results known in continuum FQH systems to show that similar physics arise in the coupled wire model of Eq.~\ref{eq:FullHamiltonianRealSpace}.

\subsection{Projection Onto The Lowest Band} \label{ssec:ProjectionFormFactor}

\begin{figure}
\centering
\includegraphics[width=\columnwidth]{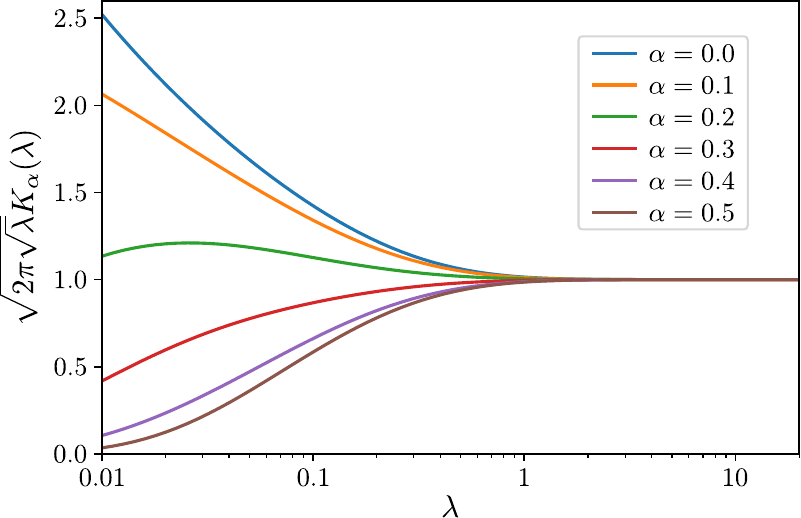}
\caption{\emph{Function $K_\alpha(\lambda)$ of Eq.~\ref{eq:NormalizationFormFactors} appearing in the expression of the form factors $\Gamma_{k,l}^u$. In the flat band regime of Sec.~\ref{ssec:NonIntLandauLevels} characterized by $\lambda > 1$, its is very well approximated by $(2\pi\sqrt{\lambda})^{-1/2}$. }}
\label{fig:ValidityApproxFormFactor}
\end{figure}

Consider on wire density-density interactions depending on the arbitrary potential $V$:
\begin{equation} \label{eq:NormalOrderedInteraction}
\mathcal{H}_{\rm int} = \sum_{j \in \mathbb{Z}} \int {\rm d}x \, {\rm d}x' \, V(x-x') : \rho_j(x) \rho_j(x') : \, , 
\end{equation}
where $\rho_j(x)= c_j^\dagger (x) c_j(x)$ is the density operator for the wire $j$. Using the Fourier components $V_q = \int dx V(x) \exp(-2i\pi qx/L)$, we can rewrite it as 
\begin{equation} 
\mathcal{H}_{\rm int} = \sum_{u,k,l \in \mathbb{Z}} V_{k-l} \sum_{j \in \mathbb{Z}} c_{j,u+k}^{\dagger} c_{j,u-k}^{\dagger}  c_{j,u+l} c_{j,u-l}  \, .
\end{equation}
Momentum conservation follows from translation invariance in the $x$-direction. Projecting this interaction onto the occupied band ($n=0$ in Eq.~\ref{eq:CoupledWireEigenfunction_GeneralMathieuImplicit}) yields
\begin{equation} \label{eq:ProjectedInteractionLowestBand}
\mathcal{H}_{\rm int} = \sum_{u,k,l \in \mathbb{Z}} V_{k-l} \Gamma_{k,l}^{u} \, d_{u+k}^{(0) \dagger}d_{u-k}^{(0) \dagger}d_{u+l}^{(0)} d_{u-l}^{(0)} \, .
\end{equation} 
The form factor $\Gamma_{k,l}^u$ involved has the form
\begin{align} \label{eq:InteractionFormFactor}
\Gamma_{k,l}^{u} = \sum_{j\in \mathbb{Z}} & g_{u+l}^{(0)} [j-\nu_w (u+l) ] g_{u-l}^{(0)} [j-\nu_w (u-l) ] \\ & \!\! \times \left[ g_{u+k}^{(0)} [j-\nu_w (u+k) ] g_{u-k}^{(0)} [j-\nu_w (u-k) ] \right]^* \, , \notag
\end{align}
which can be simplified in the flat-band limit, as detailed in Sec.~\ref{ssec:NonIntLandauLevels}. Using Eq.~\ref{eq:LowestBandisgaussian} within the flat band approximation, we can evaluate the form factor explicitly:
\begin{equation} \label{eq:GaussianInteractionFormFactor}
\Gamma_{k,l}^u = K_{\alpha(u)}(\lambda) e^{-(r \gamma \ell_B)^2 (k^2 + l^2)} \, , \,\, r = \left[\frac{\hbar^2}{2mt d^2}\right]^{\frac{1}{4}} \, ,
\end{equation}
where we have defined $\alpha(u) = \nu_w u$. This function only takes integer values when there is an equal number of wires and fluxes (as in Fig.~\ref{fig:TightBindingSpectrum}). It can take half-integer values when there are twice as many fluxes than wires, and so on. Here, the tunneling-dependent function $K$ satisfies 
\begin{equation} \label{eq:SymmetryKFunction}
K_\alpha (\lambda) = K_{-\alpha} (\lambda) = K_{\alpha +1} (\lambda) 
\end{equation}
and reads:
\begin{equation} \label{eq:NormalizationFormFactors}
K_\alpha(\lambda) = \frac{1}{\pi \sqrt{\lambda}} \sum_{j\in \mathbb{Z}} \exp\left(-\frac{2(j+\alpha)^2}{\sqrt{\lambda}}\right) \, . 
\end{equation}
We have plotted this function in Fig.~\ref{fig:ValidityApproxFormFactor} as a function of $\lambda$ for several values of $\alpha$ in the only relevant range $[0,1/2]$ (due to Eq.~\ref{eq:SymmetryKFunction}). We observe that it is well approximated by $(2\pi\sqrt{\lambda})^{-1/2}$ when $\lambda > 1$ which is assumed in the flat band regime considered here (see discussion in Sec.~\ref{ssec:NonIntLandauLevels}). As a consequence, we can write \begin{equation} \label{eq:MostGenericInteraction_GaussianFormFactor}
\mathcal{H}_{\rm int} = \!\! \sum_{u,k,l \in \mathbb{Z}} \left[ V_{k-l} e^{-(r\gamma\ell_B)^2(k^2+l^2)} \right] \; d_{u+k}^{(0) \dagger}d_{u-k}^{(0) \dagger}d_{u+l}^{(0)} d_{u-l}^{(0)} \, ,
\end{equation} 
where we have included the multiplicative factor $K_\alpha(\lambda>1)$ in the definition of the potential $V$. Due to the Gaussian form factors in Eq.~\ref{eq:MostGenericInteraction_GaussianFormFactor}, the correspondence with the model interactions of Eq.~\ref{eq:PseudoPotInRealSpace} becomes more precise and we now show that similar FQH physics can be stabilized.

\subsection{Laughlin States In The Continuum Limit} \label{ssec:ModelInteractionsLowestBand}

In the flat band limit, the orbitals of the lowest band Eq.~\ref{eq:CoupledWireEigenfunction_GeneralMathieuImplicit} can be recast in the more familiar form
\begin{equation} \label{eq:LowestBand_MatchContinuum}
\psi_k^{(0)}(x,j) = \frac{e^{i\gamma k x}}{(\pi\sqrt{\lambda})^{1/4}} \exp\left( - \frac{r^2 (y-y_k)^2}{2 \ell_B^2} \right) \, ,
\end{equation}
thanks to Eq.~\ref{eq:LowestBandisgaussian}. We have used the notation
\begin{equation}
r = \left[\frac{\hbar^2}{2mt d^2}\right]^{1/4} 
\end{equation}
and $y =jd$ for more straightforward comparison with Eq.~\ref{eq:LowestLandauLevelCylinder}. The anisotropy of the single particle state, measured by how much $r$ deviates from 1 (see Eq.~\ref{eq:IsotropyCondition}), can be incorporated into a rescaling of the coordinates $\tilde{x}=x/r$ and $\tilde{y} = ry$ and a re-definition of the complex coordinate $\tilde{z} = \frac{x}{r} - i ry$. This also impacts the natural inverse length $\tilde{\gamma} = r \gamma$. In this stretched coordinate system, the single-particle WFs $\psi_k^{(0)}$ exactly match their continuum counterparts of Eq.~\ref{eq:LowestLandauLevelCylinder}.

Furthermore, the projected interactions Eq.~\ref{eq:MostGenericInteraction_GaussianFormFactor} also become equivalent to Eq.~\ref{eq:PseudoPotInRealSpace}. Take for instance the one-dimensional version of the first pseudo-potential as on-wire interaction $V^{(0)}(x)= V_0 \delta(x)$. Eq.~\ref{eq:MostGenericInteraction_GaussianFormFactor} gives its expression after projection to the lowest flat band of the system
\begin{equation} 
\mathcal{H}_{\rm int}^{(0)} = V_0 \!\! \sum_{u,k,l \in \mathbb{Z}} e^{-(\tilde{\gamma}\ell_B)^2(k^2+l^2)}  d_{u+k}^{(0) \dagger}d_{u-k}^{(0) \dagger}d_{u+l}^{(0)} d_{u-l}^{(0)} \, ,
\end{equation}
which is exactly equivalent to Eq.~\ref{eq:ModelInnteractionLLLBosons} in a rescaled coordinate system. The pseudo-potential analysis~\cite{Haldane_HierarchiesPseudoPot} can be repeated to obtain the densest zero-energy state of $\mathcal{H}_{\rm int}^{(0)}$~\cite{HaldaneBookQuantumHall} for a particle filling factor $\nu = n_{\rm 1D} \nu_w = 1/2$. The later has the same coefficients in the many-body occupation number basis as the Laughlin state Eq.~\ref{eq:LaughlinWFGeneralFilling} at filling $\nu=1/2$. The only difference is the real-space representation of the many-body wavefunction $\Psi_2$ which depends on the rescaled coordinates
\begin{equation} \label{eq:IdealizedLaughlin2}
\Psi_2(\tilde{z}_1, \cdots , \tilde{z}_{N_e})= \prod_{i<j} \left(e^{i \tilde{\gamma}  \tilde{z}_i }-e^{i \tilde{\gamma} \tilde{z}_j }\right)^2 \prod_i e^{-\frac{\tilde{y}_i^2}{2\ell_B^2}} \, .
\end{equation}
With this formal mapping of our problem to the continuum, we also recover the exact $\nu=1/2$ bosonic Laughlin state when the isotropy condition Eq.~\ref{eq:IsotropyCondition} is satisfied, corresponding to the $r=1$ case. Following Haldane~\cite{Haldane_GeometricDescription}, we highlight that the coordinate rescaling does not wash out the topological properties of the Laughlin state. Indeed, it can be seen as a geometric deformation of the cyclotron orbits only, leaving the guiding centers' distribution and all the properties of the Laughlin state such as fractional excitations and braiding statistics intact~\cite{GeometryLaughlin_PapicHaldane}.

We can treat similarly the fermionic case with an on-wire interaction $V_{k-l}^{(1)} = - V_1 (k-l)^2$, corresponding to the one-dimensional version of the first Haldane pseudopotential Eq.~\ref{eq:PseudoPotInRealSpace}. Its projection onto the lowest flat band reads
\begin{equation} 
\mathcal{H}_{\rm int}^{(1)} = V_1 \sum_{u,k,l \in \mathbb{Z}} kl e^{-(\tilde{\gamma} \ell_B)^2(k^2+l^2)} d_{u+k}^{(0) \dagger}d_{u-k}^{(0) \dagger}d_{u+l}^{(0)} d_{u-l}^{(0)} \, , 
\end{equation}
which should this time be compared to Eq.~\ref{eq:ModelInnteractionLLLFermions}. We have used fermionic anti-commutation relations to cancel all contributions of monomials even under the transformations $k\to -k$ or $l \to -l$. Considering the $\nu=1/3$ case, the pseudo-potentials reasoning apply~\cite{PseudoPotentialsTrugmanKivelson} and the densest ground state of the interaction is 
\begin{equation} \label{eq:IdealizedLaughlin3}
\Psi_3(\tilde{z}_1, \cdots , \tilde{z}_{N_e})= \prod_{i<j} \left(e^{i \tilde{\gamma}  \tilde{z}_i }-e^{i \tilde{\gamma} \tilde{z}_j }\right)^3 \prod_i e^{-\frac{\tilde{y_i}^2}{2\ell_B^2}} \, .
\end{equation}
The discussion could be extended, as in Ref.~\cite{GeometryLaughlin_PapicHaldane}, to other Laughlin state at filling $\nu = 1/m$ with $m>3$.

By a formal mapping of the coupled-wire model in the flat-band limit to a continuum FQH system, we proved that it hosts strongly correlated FQH states for fractional filling of the lowest band. This results agrees with those presented in the pioneering works of Refs.~\cite{CoupledWireFirstKane,CoupledWireTeoKanePRB} and provides a new way to tackle the interacting coupled-wire problem without relying on effective low-energy theory or renormalization group arguments. Our results mainly come from the Gaussian form factors $\Gamma_{k,l}^u$ obtained thanks to the asymptotic behaviour Eq.~\ref{eq:LowestBandisgaussian}. This asymptotic regime is reached for $\lambda >1$ as discussed in Sec.~\ref{ssec:NonIntLandauLevels} (see also Fig.~\ref{fig:ValidityApproxFormFactor} for a more pictorial view).

\section{Exact Diagonalization Results} \label{sec:ED_ForLaughlin}

We have just proved that FQH-like phases arise in our model, when we consider the idealized limit $\lambda \gg 1$ with an infinite number of wires. Experimentally relevant situations are however likely to deviate from this asymptotic limit. The stabilization of the previous strongly correlated phases ultimately depends on the accessible range of tunneling $\lambda$ and interaction strength. In this section, we provide a full-fledged Exact-Diagonalization (ED) study of our interacting problem to precisely locate the transition towards the Laughlin state Eq.~\ref{eq:IdealizedLaughlin2}. Moreover, our calculations show how to experimentally discriminate the strongly correlated Laughlin phase from weakly interacting phases.

\subsection{Setup} \label{ssec:SetupED}

\begin{figure}
\centering
\includegraphics[width=\columnwidth]{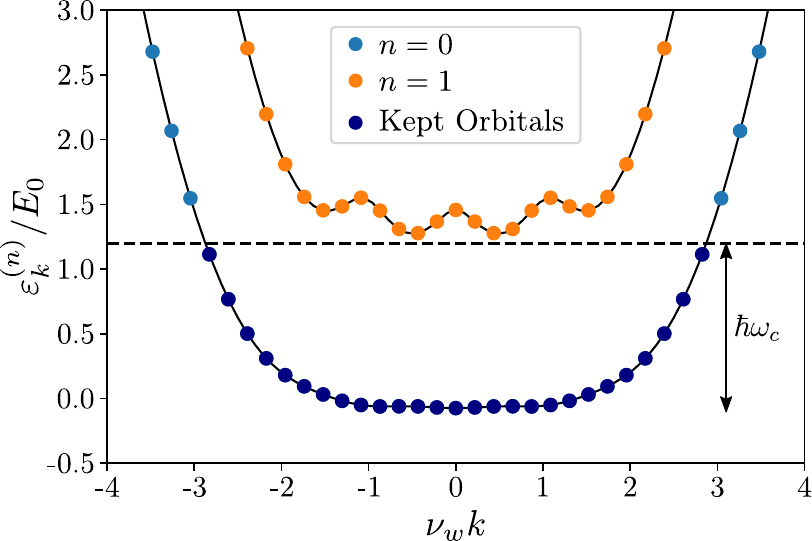}
\caption{\emph{Lowest two bands for a finite system of $N_w = 5$ wires with tunneling parameter $\lambda = 0.75$. The single particle state of the lowest (resp. first excited) band, labeled by their momentum $k$, are represented with blue (resp. orange) dots. To simulate a system near total filling factor $\nu = 1/2$, we tune the magnetic field strength in order to keep precisely $N_{\rm orb} = 2N+3$ orbitals below the single particle gap (darker blue). We only consider those orbitals in our ED calculations. Note that $N_{\rm orb}$ is larger than the natural number of orbitals for the Laughlin state.}}
\label{fig:SettingUpExactDiagonalization}
\end{figure}

Let us first fix the scope of our ED study. Anticipating the ultracold alkali vapors considered in Sec.~\ref{sec:ExperimentalRealization}, we focus on a bosonic system with on-wire contact interactions $V^{(0)}(x)= V_0 \delta(x)$. In all our finite size calculations, we work with $N \leq 12$ particles distributed in $N_w = 5$ wires\footnote{While we focus on the case $N_w = 5$ throughout Sec.~\ref{sec:ED_ForLaughlin}, we have also performed ED simulations for finite size systems with $3 \leq N_w \leq 7$. All these systems display similar physics.}. For simplicity, we assume that particles only occupy the $N_{\rm orb}$ orbitals centered around $k=0$ in the lowest energy band, as illustrated in Fig.~\ref{fig:SettingUpExactDiagonalization}. In other words, we keep all states of the lowest band with momentum $|k| \leq N_{\rm orb}/2$. The truncated many-body Hamiltonian splits into a dispersive and an interacting part $\mathcal{H}_{\rm tot} = \mathcal{H}_{\rm disp} + \mathcal{H}_{\rm int}$ with:
\begin{subequations} \label{eq:TotalEDHamiltonian} \begin{align}
\mathcal{H}_{\rm disp} & = \sum_{|k| \leq N_{\rm orb}/2} \varepsilon_k^{(0)}  d_k^{(0) \dagger}d_k^{(0)} \, , \\
\mathcal{H}_{\rm int}  & = V_0 \!\!\! \sum_{\substack{ p_1+p_2 = q_1+ q_2 \\ |p_i|, |q_i| \leq N_{\rm orb}/2}} \!\!\! \Gamma_{q_-, p_-}^{q_+} \, d_{q_1}^{(0) \dagger}d_{q_2}^{(0) \dagger}d_{p_1}^{(0)} d_{p_2}^{(0)} \, ,
\end{align} \end{subequations}
where we have used the short-hand notations $p_\pm = (p_1 \pm p_2)/2$ and $q_\pm = (q_1 \pm q_2)/2$. The $\Gamma$ coefficients have been defined in Eq.~\ref{eq:InteractionFormFactor}.

To simulate a system near filling factor $\nu = 1/2$, where we expect the bosonic Laughlin state Eq.~\ref{eq:IdealizedLaughlin2}, we tune the magnetic field strength in order to have precisely $N_{\rm orb} = 2N+3$ orbitals below the cyclotron energy $\hbar \omega_c$, as depicted in Fig.~\ref{fig:SettingUpExactDiagonalization}. While the densest Laughlin droplet only needs $2N-1$ orbitals to appear, the four extra orbitals avoid hard boundary conditions and let the system smoothly accommodate the dispersion relation at the edge.

\subsection{Laughlin States} \label{ssec:LaughlinFindingED}

\begin{figure*}
\centering
\includegraphics[width=\textwidth]{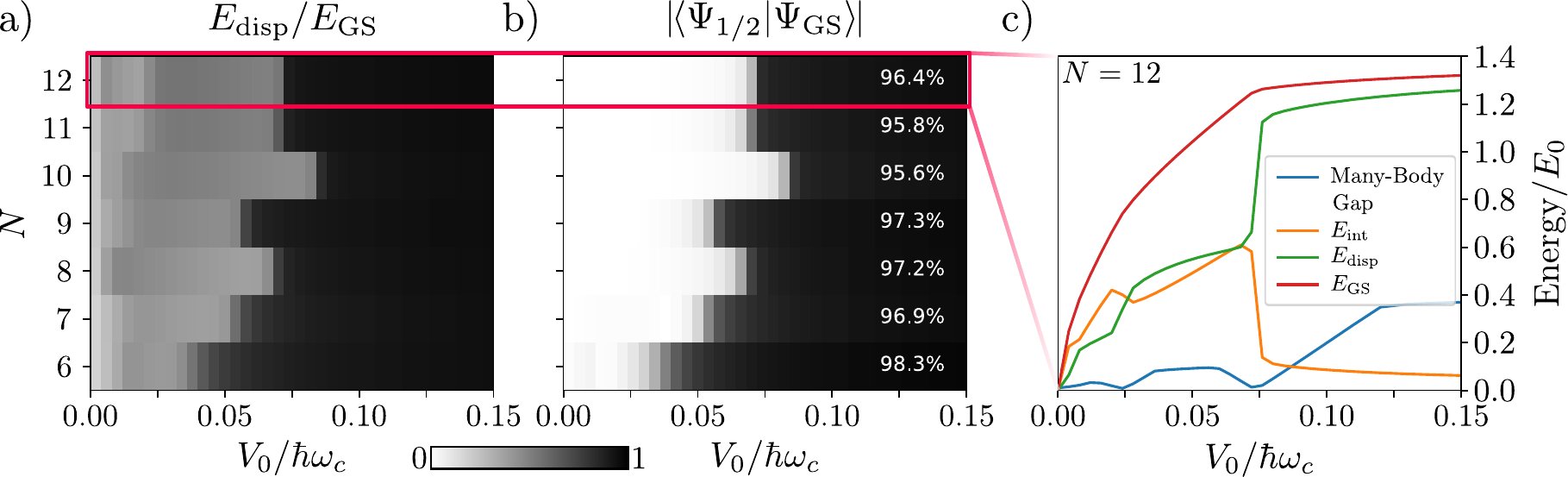}
\caption{\emph{a) Relative contribution of the dispersive energy $E_{\rm disp}$ to the total ground state energy $E_{\rm GS} = E_{\rm disp} + E_{\rm int}$ as a function of the interaction strength $V_0$, for different particle numbers $6 \leq N \leq 12$. Beyond $V_0 \simeq 0.075 \hbar \omega_c$, the system suddenly jumps to a phase where the interaction is almost entirely screened (black). This behavior is reminiscent of the Laughlin state $\ket{\Psi_{1/2}}$ described in Sec.~\ref{ssec:CylinderModelInteractiosn}. b) Overlap of the many-body ground state $\ket{\Psi_{GS}}$ with the Laughlin state $\ket{\Psi_{1/2}}$. The high overlaps, above 95\% for $V_0 = 0.15 \hbar \omega_c$ (highlighted in white for each $N$), allows us to identify the phase at large $V_0$ as the asymptotic $\nu=1/2$ Laughlin phase studied in Sec.~\ref{ssec:ModelInteractionsLowestBand}. c) Focus on ED results at $N=12$. The abrupt phase transition to the Laughlin state with $E_{\rm int} \ll E_{\rm disp}$ (respectively in orange and green) near $V_0 \simeq 0.075 \hbar \omega_c$ is accompanied with a closing of the many-body gap (bue). Other phases and phase transition can be observed, for instance near $V_0 \simeq 0.025 \hbar \omega_c$. Their nature is investigated in more detail in Sec.~\ref{ssec:IntermediatePhases}. }}
\label{fig:LaughlinTransitionLocation}
\end{figure*}

We first diagonalize the Hamiltonian~\ref{eq:TotalEDHamiltonian} for $\lambda = 0.75$ and a variable interaction strength $V_0$. We observe features reminiscent of the Laughlin phase for strong interactions. The many-body ground state $\ket{\Psi_{GS}}$ is characterized by the relative contribution of its dispersive and interacting energies, $E_{\rm disp} = \braOket{\Psi_{GS}}{\mathcal{H}_{\rm disp}}{\Psi_{GS}}$ and $E_{\rm int} = \braOket{\Psi_{GS}}{\mathcal{H}_{\rm int}}{\Psi_{GS}}$. We denote the total energy by $E_{\rm GS} = E_{\rm disp} + E_{\rm int}$. Our numerical results are depicted in Fig.~\ref{fig:LaughlinTransitionLocation}a-c, where we observe an abrupt change of behaviour near $V_0 \simeq 0.075 \hbar \omega_c$. Beyond this point, the ground state almost entirely screens the interaction and all its energy comes from the dispersion of the lowest band. This feature is reminiscent of the Laughlin state which is an exact zero energy state of $\mathcal{H}_{\rm int}$ (see Sec.~\ref{ssec:ModelInteractionsLowestBand}).

To confirm our intuition and to identify the sharp feature of Fig.~\ref{fig:LaughlinTransitionLocation}c as the transition towards the asymptotic Laughlin phase studied in Sec.~\ref{ssec:ModelInteractionsLowestBand}, we compute the overlap of $\ket{\Psi_{GS}}$ with the Laughlin state $\ket{\Psi_{1/2}}$ at filling factor $\nu = 1/2$ obtained on the cylinder geometry (see Eq.~\ref{eq:LaughlinWFGeneralFilling}). We show in App.~\ref{App:MatchAspectRatio} how the perimeter of the cylinder $L_{\rm cyl}$ should be chosen to match the aspect ratio of our anisotropic system. The overlaps presented in Fig.~\ref{fig:LaughlinTransitionLocation}b show a jump from zero to almost one at the transition, and remain above 95\% for $V_0 = 0.15 \hbar\omega_c$ for all the system size considered. These results provide clear evidence that the states at large interactions $V_0 > 0.075 \hbar\omega_c$ belong to the asymptotic Laughlin phase studied in Sec.~\ref{ssec:ModelInteractionsLowestBand}. For large enough interactions, the Laughlin physics arise in our finite size system.

\begin{figure}
\centering
\includegraphics[width=0.85\columnwidth]{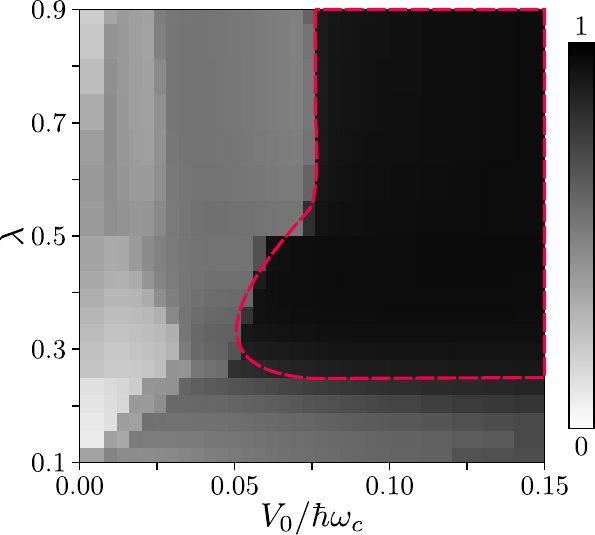}
\caption{\emph{
Phase diagram corresponding to Eq.~\ref{eq:TotalEDHamiltonian} obtained for $N=12$ particles. As in Fig.~\ref{fig:LaughlinTransitionLocation}a, we use $E_{\rm disp}/E_{\rm GS}$ to distinguish the weakly-interacting phases from the Laughlin one. The parameter range where the FQH physics can be realized is surrounded by a dashed dark blue line.
}}
\label{fig:PhaseDiagN12}
\end{figure}

Finally, we vary the tunneling parameter $\lambda$ in order to locate the boundaries of the Laughlin phase in the $(\lambda, V_0)$ parameter space. Our finite-size calculations suggest that the Laughlin physics can be stabilized near half-filling in our model provided $\lambda > 0.5$ and $V_0 > 0.075 \hbar \omega_c$. As shown in Fig.~\ref{fig:LaughlinTransitionLocation}a, we use $E_{\rm disp}/E_{\rm GS}$ as a probe to characterize the strongly correlated FQH phase. This quantity is extracted for several values of $(\lambda, V_0)$ and $N=12$ particles to obtain the phase diagram of Fig.~\ref{fig:PhaseDiagN12}. For $\lambda > 0.5$, the transition towards the Laughlin phase lies very close to the previously extracted value $ V_0 = 0.075 \hbar \omega_c$. This simply reflects the exponentially small dispersion of the lowest band in this tunneling regime (see Sec.~\ref{ssec:NonIntLandauLevels}). For small $\lambda$, the width of the lowest band increases forming well separated potential wells (see also Fig.~\ref{fig:MomentumDensityOccupation}a). This dispersion makes it harder for the system to build long-range correlation. There is however an intermediate region $0.225 \leq \lambda \leq 0.5$ where an FQH state can still be observed for large $V_0$. There, the density oscillations introduced by the dispersion of the lowest band are commensurate with those of the Laughlin state $\ket{\Psi_{1/2}}$ on a cylinder with a small perimeter~\cite{YellowBook}. Our finite-size numerics tend to suggest that this commensuration effect moves the transition towards lower interaction strengths. The wires eventually decouple, and we do not see any signatures of the Laughlin phase in our finite-size numerics for $\lambda < 0.225$.

\subsection{Experimental Signatures: Momentum-Space Density Distribution} \label{ssec:PeaksInMomentumSpace}

Having understood under which conditions the Laughlin physics arise in our model, we now highlight that the momentum-space density distribution provides clear signatures allowing to discriminate the FQH state from the weakly interacting phases of our model. This probe is accessible in cold-atom experiments with standard time-of-flight (TOF) measurements, which further motivates the experimental proposal of Sec.~\ref{sec:ExperimentalRealization}.

\begin{figure*}
\centering
\includegraphics[width=0.85\textwidth]{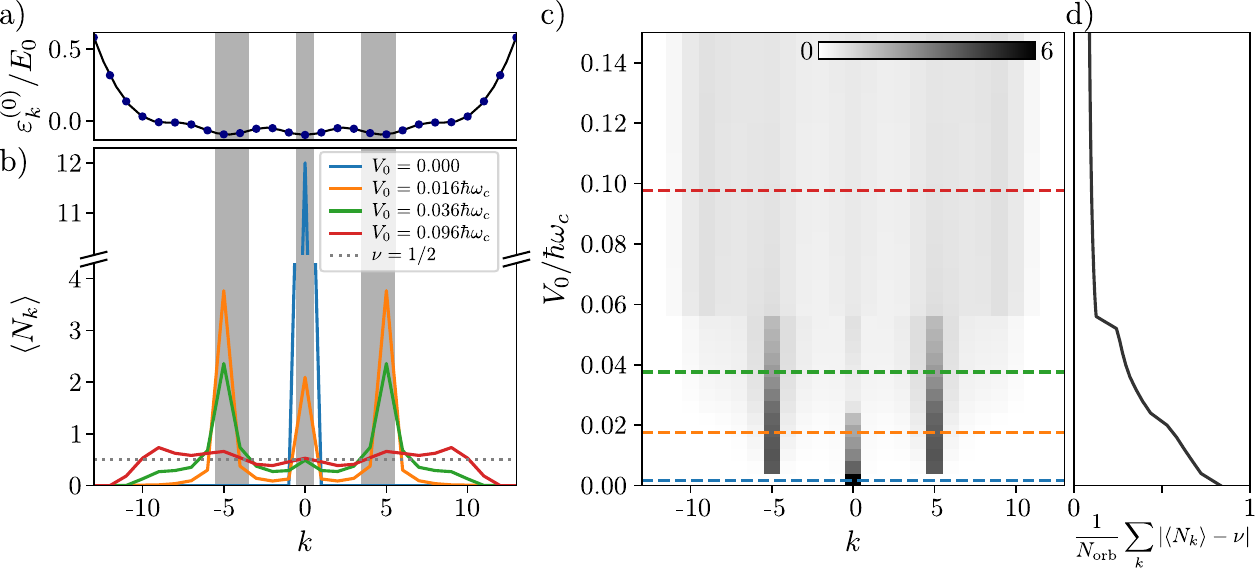}
\caption{\emph{a) Dispersion relation of the lowest band for $\lambda = 0.375$ and $N_w = 5$ wires. There and below, we highlight the position of the lowest energy orbitals with grey shades. b) Mean occupation number of momentum states $\langle N_k \rangle = \braOket{\Psi_{\rm GS}}{d_k^{(0) \dagger}d_k^{(0)}}{\Psi_{\rm GS}}$ as a function of the momentum $k$ for various values of the interaction parameter $V_0$. We show the numerical results for a system of $N=12$ particles. The weakly interacting phases, investigated further in Sec.~\ref{ssec:IntermediatePhases}, show strong peaks near the minima of the dispersion relation. On the contrary deep in the Laughlin phase (red line), the momentum-space density distribution is almost flat and equal to $\nu = 1/2$ except near the edges where we observe density fluctuations characteristics of this FQH state~\cite{LaughlinBookGirvin}. c) Same as b) over a finer grid in $V_0$, in order to highlight the transition between the weakly interacting and Laughlin phases near $V_0 = 0.055 \hbar \omega_c$. d) The mean deviation of $\langle N_k \rangle$ from $\nu = 1/2$ can be used to probe the transition towards FQH-like states. The non-zero contributions at large $V_0$ are mostly due to the empty orbitals near the edge of the system (here $|k| \geq 12$).} }
\label{fig:MomentumDensityOccupation}
\end{figure*}

We have computed the mean occupation of all momentum states $ \langle N_k \rangle = \braOket{\Psi_{\rm GS}}{d_k^{(0) \dagger}d_k^{(0)}}{\Psi_{\rm GS}}$, the quantity usually accessed through TOF measurements, for the ground state of Eq.~\ref{eq:TotalEDHamiltonian}. Our numerical results are exemplified in Fig.~\ref{fig:MomentumDensityOccupation}a-d for $N=12$ particle and $\lambda = 0.375$. We observe two strikingly different behaviors depending on the interaction strength. In the weakly interacting phases $V_0 < 0.055\hbar \omega_c$, particles are gathered around the minima of the dispersion relation, which are highlighted in gray in Fig.~\ref{fig:MomentumDensityOccupation}a and b. On the contrary, in the Laughlin phase for $V_0 > 0.055\hbar \omega_c$, the momentum-space density distribution is almost flat and all orbitals in the bulk approach $\langle N_k \rangle \simeq \nu = 1/2$. For large interaction strengths, the ground state also presents the typical density fluctuations of the Laughlin state near the edges of the system~\cite{LaughlinBookGirvin}.

Fig.~\ref{fig:MomentumDensityOccupation}c shows the same quantity $\langle N_k \rangle$ over a finer grid in the interaction parameter $V_0$. It highlights that the transition between the two previous behaviors is abrupt, hence showing that the momentum-space density distribution can be used as a probe of the transition towards the FQH-like states identified in Sec.~\ref{ssec:LaughlinFindingED}. As an illustrative example, we show in Fig.~\ref{fig:MomentumDensityOccupation}d that the mean deviation of $\langle N_k \rangle$ from $\nu=1/2$ is discontinuous at the transition, heralding the FQH-like state.

For all the considered parameters $(\lambda, V_0)$ and system sizes, we have witnessed the same signatures of the weakly interacting and Laughlin phases. This promotes the momentum-space density distribution as a simple probe to appraise the appearance of the FQH-like states predicted in Sec.~\ref{ssec:ModelInteractionsLowestBand} in our model. Let us insist on the experimental realization of Fig.~\ref{fig:MomentumDensityOccupation}d in cold-atom experiments where the momentum space density distribution is accessible with TOF measurements, and the interaction strength $V_0$ can be tuned by varying a static magnetic field near a Feschbach resonance.

\subsection{Nature Of The Intermediate Phase} \label{ssec:IntermediatePhases}

To complete our understanding of the phase diagram, we finally investigate the weakly-interacting phases of our model in more details. They are well captured within a mean-field approach that we briefly detail here, before comparing it to numerical results.

\subsubsection{Very-Weakly Interacting Phase: Bogoliubov Theory} \label{sssec:BogoTheoryText}

For $V_0 = 0$, the ground state is a Bose-Einstein Condensate (BEC) with all particles occupying the lowest energy orbital with momentum $k=0$. Formally, it can be written as
\begin{equation} \label{eq:BecAtNoInteraction}
\ket{\Psi_{\rm GS}(V_0 = 0)} = \frac{1}{\sqrt{N!}} \left(d_0^{(0) \dagger}\right)^N \ket{\emptyset} \, 
\end{equation}
with $\ket{\emptyset}$ the state with no bosons. This situation is depicted with blue lines in Fig.~\ref{fig:MomentumDensityOccupation}, where we indeed measure
\begin{equation}
\langle N_k (V_0 = 0) \rangle = \begin{cases} 
N & {\rm if :} \quad k=0 \\ 0 & {\rm if :} \quad k \neq 0 \\
\end{cases} \, .
\end{equation}

Switching on an infinitesimally small interaction strength $V_0$, the ground state can be obtained with a standard Bogoliubov analysis~\cite{Bogoliubov_TheorySuperconductivity}, as detailed in App.~\ref{App:BogoliubovTheory}. Here, we summarize the main ideas of this approach and adapt them to our finite-size systems.

Within the Bogoliubov approximation, the weak interactions with the BEC slightly admix the original orbitals with momentum $\pm k$, with $k>0$, together. Due to the weak interactions, excitations can still be described as quasi-particles of the mean-field Hamiltonian. To remain in the vacuum state for these new excitations, the BEC is depleted by the creation of particle pairs with non-zero momenta. The weakly-interacting ground-state is more easily described in the limit $\sqrt{N} \gg 1$ where we can neglect the particle number fluctuations in the BEC. In this regime, we have the simple form
\begin{equation} \label{eq:BogoGSTheoryThermoLimit}
\ket{\Psi_{\rm Bogo}(V_0)} \underset{\sqrt{N} \gg 1}{=} \exp\left( - \sum_{k>0} t_k \, d_k^{(0) \dagger} d_{-k}^{(0) \dagger} \right) \ket{\Psi_{\rm GS}(0)} \, ,
\end{equation}
derived in App.~\ref{App:BogoliubovTheory}, where we provide an explicit expression for the coefficients $t_k$. 

In order to compare the weakly-interacting theory to our ED results, we must adapt Eq.~\ref{eq:BogoGSTheoryThermoLimit} to finite size systems, where $\sqrt{N}$ is not much larger than one, and ensure particle number conservation. We thus consider the following ansatz
\begin{equation} \label{eq:BogoGSFiniteSize_OnePeak}
\ket{\Psi_{\rm Bogo}(V_0)} = e^{ - \sum_{k>0} \frac{t_k}{N} \, d_k^{(0) \dagger} d_{-k}^{(0) \dagger} d_0^{(0)} d_0^{(0)} } \ket{\Psi_{\rm GS}(0)} \, . 
\end{equation}
The coefficients $\{t_k\}_{k > 0}$ are variationally optimized around their $\sqrt{N} \gg 1$ theoretical value to correct for small finite-size effects\footnote{The number of variational parameters, $(N_{\rm orb}-1)/2 = 13$ for $N=12$ particles, remains much smaller than the many-body Hilbert space dimension, equal to 33 427 622 for the same parameters.}. The depletion of the BEC is explicitly accounted for by gluing the operator $d_0^{(0)} d_0^{(0)} / N$ to the pair creation operator in the exponential of Eq.~\ref{eq:BogoGSFiniteSize_OnePeak}. This formally implements the required particle number conservation.

\begin{figure}
\centering
\includegraphics[width=\columnwidth]{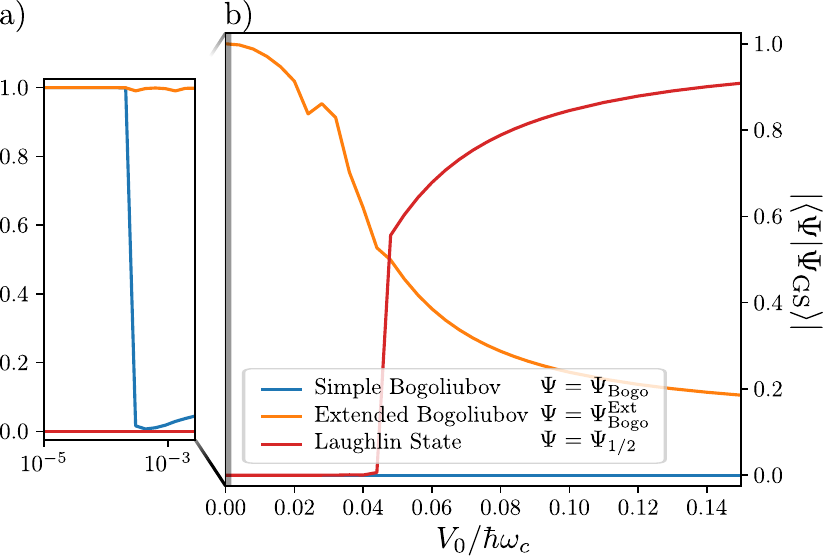}
\caption{\emph{ 
a) For very weak interactions, the standard Bogoliubov approach $\ket{\Psi_{\rm Bogo}(V_0)}$ well captures the physics of our model. As explained in Sec.~\ref{sssec:ExtensionBogoText}, it fails to describe the other weakly interacting phases which display multiple peaks in Fig.~\ref{fig:MomentumDensityOccupation}. For $V_0 > \SI{2.5e-4}{} \, \hbar \omega_c$, we have to rely on an extended ansatz $\ket{\Psi_{\rm Bogo}^{\rm Ext}(V_0)}$ adapted to the multiple peaks in the momentum-space density distribution of the weakly interacting phases depicted in Fig.~\ref{fig:MomentumDensityOccupation} (see App.~\ref{App:BogoliubovTheory}).
b) Overlap of the ED ground state $\ket{\Psi_{GS}(V_0)}$ with the extended Bogoliubov ansatz  $\ket{\Psi_{\rm Bogo}^{\rm Ext}(V_0)}$ (orange) and the bosonic Laughlin state $\ket{\Psi_{1/2}}$ at filling factor $\nu = 1/2$ (red). These two states accurately describe the physics of our model on each side of the transition. All the results presented in this figure were obtained for $N=12$ particles and $\lambda = 0.25$.
}}
\label{fig:OverlapsBogoLaughlin}
\end{figure}

The finite size ansatz Eq.~\ref{eq:BogoGSFiniteSize_OnePeak} almost perfectly captures the ED ground states for very-weak interaction strengths $V_0 \leq \SI{2.5e-4}{} \, \hbar \omega_c$, as shown by their overlaps in Fig.~\ref{fig:OverlapsBogoLaughlin}a. However, it quickly fails to capture the nature of the other intermediate phases, characterized by more than one peak in their momentum-space density distribution (see orange and green lines of Fig.~\ref{fig:MomentumDensityOccupation}).

\subsubsection{Extension To Other Mean-Field Phases} \label{sssec:ExtensionBogoText}

Because of the multiple local minima of the lowest band (see Fig.~\ref{fig:MomentumDensityOccupation}a) and the finite range of the interaction (see Eq.~\ref{eq:GaussianInteractionFormFactor}), even weak interactions tend to favor the creation of multiple condensates with very different momenta. Indeed, particles have a similar dispersion energy in all the potential wells of the lowest band $\varepsilon_k^{(0)} \simeq \varepsilon_0^{(0)}$. However, particles in distant minima of the dispersion relation hardly interact due to the finite range of the the form factors $\Gamma$ (see Eq.~\ref{eq:InteractionFormFactor}). A rough estimate shows that two BECs with each $N/2$ particles located in minima with large momentum difference has an interaction energy twice smaller than single BEC gathering $N$ particles at $k=0$. This simple argument is confirmed by Fig.~\ref{fig:OverlapsBogoLaughlin}a, where we observe that Eq.~\ref{eq:BogoGSFiniteSize_OnePeak} correctly captures our ED ground state for $V_0 \simeq \SI{2.5e-4}{} \, \hbar \omega_c$ which should be compared to the energy difference between the two local minima $N (  \varepsilon_k^{(0)} - \varepsilon_0^{(0)}  ) \simeq \SI{2.4e-4}{} \, \hbar \omega_c$ (the last factor of $N$ accounts for the different scalings of the dispersive and interacting Hamiltonian with respect to density). The same reasoning shows that the BEC at $k=0$ is destabilized towards intermediate phases with multiple macroscopically populated momentum states for an interaction strength $V_0 \propto 1/N$. Therefore, we only expect to see these latest in the thermodynamic limit (see orange and green lines in Fig.~\ref{fig:MomentumDensityOccupation}).

The previous Bogoliubov approach can be extended to the case of multiple smaller condensates, and allows to capture the other intermediate phases observed in Fig.~\ref{fig:LaughlinTransitionLocation}c and Fig.~\ref{fig:MomentumDensityOccupation}c. The idea is to conserve the variational parameters of Eq.~\ref{eq:BogoGSFiniteSize_OnePeak}, which measure the depletion of the condensates by pair-creation in order to accommodate the weak interactions of the system, while changing the state they act on to describe the presence of multiple smaller BECs. The explicit expression of this generalized ansatz $\ket{\Psi_{\rm Bogo}^{\rm Ext}(V_0)}$ can be found in App.~\ref{App:BogoliubovTheory}. As can be seen in Fig.~\ref{fig:OverlapsBogoLaughlin}a, this adjusted ansatz overcomes the limits of Eq.~\ref{eq:BogoGSFiniteSize_OnePeak} and perfectly captures the phases for weak interactions.

In Fig.~\ref{fig:OverlapsBogoLaughlin}b, we compare our ED results with this generalized Bogoliubov ansatz $\ket{\Psi_{\rm Bogo}^{\rm Ext}(V_0)}$ accross the transition from the non-interacting regime to the Laughlin state. We observe that all the mean-field phases are well captured by this mean-field approach, which eventually breaks down when the Laughlin phase arise.

\section{Possible Realization In Spin-Dependent Optical Lattice} \label{sec:ExperimentalRealization}

\begin{figure*}
\includegraphics[width=\textwidth]{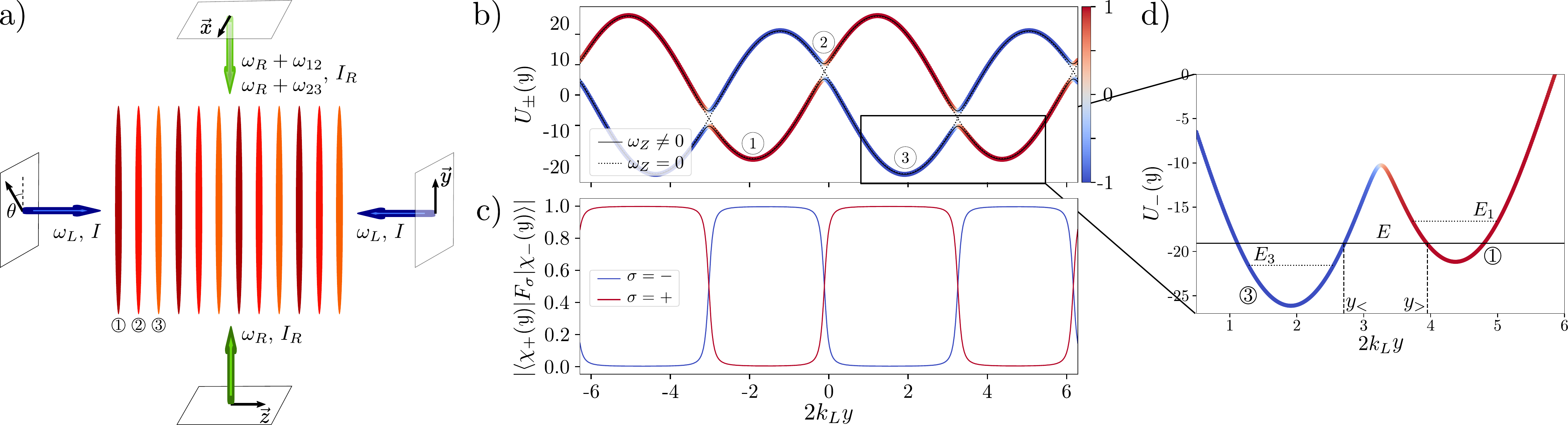}
\caption{\emph{
$a$) Sketch of the experimental setup. Two strong laser beams (blue arrows) counter-propagating along $z$ in a lin-$\theta$-lin configuration create a spin-dependent optical lattice. We identify three inequivalent Wannier centers within a unit cell of the lattice (see text and $b$), here schematically depicted with three different tones of orange. The Raman beams (green arrows) counter-propagating along $y$ generate complex tunneling coefficients between the wires \circled{2} and their neighbors, thus generating an artificial gauge field mimicking the Landau vector potential. $b$) Spin-dependent adiabatic potential obtained by diagonalization of $U_{\rm OL} + U_{\rm ZM}$ fr $\theta=80^\circ$, $U_0 = 45E_R$ and $\hbar\omega_Z = 5E_R$. In absence of transverse magnetic field (dotted lines), the spin eigenvectors $\ket{\chi_\pm(z)}$ are eigenstates of $F_z$. They are admixed close to the avoided crossing opened by the transversed magnetic field (solid lines), as can be seen from the background line color representing the spin polarization $\braOket{\chi_\pm(z)}{F_z}{\chi_\pm(z)}$. Within one unit cell of size $\pi/k_L$, we three potential wells denoted as \circled{1}-\circled{2}-\circled{3} give rise to localized Wannier states entering our tight-binding description of the system. $c$) While \circled{1} and \circled{3} are naturally coupled by tunneling matrix elements in the tight binding limit (see text), we need to engineer light-assisted hopping elements to couple them to \circled{2}. Due to the polarization of $\ket{\chi_\pm (z)}$, the transitions \circled{1} $\to$ \circled{2} and \circled{2} $\to$ \circled{3} can only de driven by a $F_-$ operator. $d$) Close-up of Fig.~\ref{fig:AdiabaticPotential}b around the potential wells $\circled{1}$ and $\circled{3}$. To estimate the tunneling matrix element between the Wannier states at energy $E_1$ and $E_2$, we use a semiclassical analysis and see this reduced problem as simpler double-well potential~\cite{SemicalssicalDoubleWellPotentialTunneling}. The results of this semiclassical analysis involves the mean energy $E = \frac{E_1+E_2}{2}$ and the turning points $z_<$ and $z_>$ which are schematically pictured.
}}
\label{fig:AdiabaticPotential}
\end{figure*}

In this section, we propose a plausible realization of the coupled-wire model Eq.~\ref{eq:FullHamiltonianRealSpace} where a one-dimensional spin-dependent optical lattice creates the initial wires~\cite{DeutschJessen_ApproximationAndDiabaticPotentials}. This configurations allows to cool the gas to sub-recoil temperatures, mitigating the effects of inter-band mixing. Moreover, the sub-wavelength spacing of potential wells in the spin-dependent potential allows to reach the strong tunneling regime of Sec.~\ref{ssec:NonIntLandauLevels}. The spin degree of freedom of the potential is also used to selectively drive Raman-assisted hopping between wires~\cite{SpinFlipTunneling40K}, which creates an artificial gauge field mimicking a uniform magnetic field~\cite{Spielman_FirstExperiment}. We now explain in more details how these different elements can be combined.

\subsection{Building The Wires: Spin-Dependent Trapping Potential} \label{ssec:ExpSpinDependentTrap}

The first ingredient of our experimental proposal is a strong optical lattice obtained by interfering a pair of red-detuned laser beams counter-propagating along $y$ with intensity $I$ and angle $\theta$ between their linear polarizations as depicted in Fig.~\ref{fig:AdiabaticPotential}a. For simplicity, we consider alkali atoms whose ground state manifold, characterized by the hyperfine spin $\textbf{F}$, is immune to rank-2 tensor light-shifts~\cite{CesiumExample_AllPolaribilities}. As a consequence, the spin-dependent potentials created by the previous laser configuration, dubbed lin-$\theta$-lin~\cite{Guo_BelowDopplerCooling,Zoller_BelowDopplerCooling}, only linearly couples to the spin $\vec{F}$. More precisely, it is diagonal in the spin basis $F_y \ket{m_y} = m_y \ket{m_y}$ and reads~\cite{CohenTanoudji_AtomPhoton,TensorPolarizability_FastJtoJprime,DeutschJessen_TensorCartesian}:
\begin{equation} \label{eq:LatticeDependentUncoupled}
U_{\rm OL} = U_0 \left[ \cos\theta\cos(2 k_L y) + u \sin\theta \sin(2 k_L y) F_y \right] \, ,
\end{equation}
with $k_L$ the lasers' wavevector, $U_0$ an overall multiplication factor proportional to $I$ and where the coefficient $u$ gathers all the relevant informations on the excited hyperfine structure~\cite{TensorPolarizability_AllDefinitionsAndTensorDecomposition}. As in Ref.~\cite{DeutschJessen_ApproximationAndDiabaticPotentials}, we focus on the case $F=1/2$ and $u=1$ which constitutes the simplest realization of the ideas put forward in this article. Our conclusions can be extended to other atomic species. For instance, we refer the interested reader to Refs.~\cite{Spielman_FirstExperiment,Spielman_RashbaDresselhaus,Spielmann_AdiabaticSuperlattice,Spielmann_RFRaman} for an extensive theoretical study and many experimental details on the $F=1$ case of $^{87}$Rb.

The spin eigenstates $\ket{m_y}$ diagonalizing Eq.~\ref{eq:LatticeDependentUncoupled} are further coupled by a magnetic field in the $y-z$ plane giving the additional energy:
\begin{equation}
U_{\rm ZM} = \hbar \omega_Z (F_y + F_z) \, .
\end{equation}
In absence of transverse magnetic field $\omega_Z = 0$, we can rewrite the spin-dependent potentials as 
\begin{equation}
U_{\rm LO} \ket{\pm 1/2} = \tilde{U}_0 \cos[2k_L (y\pm \delta y)]\ket{\pm 1/2} \, ,
\end{equation}
with
\begin{equation}
\delta y = \tan^{-1} \left(\frac{\tan\theta}{2}\right) \, , \quad \tilde{U}_0 = \frac{U_0}{2} \sqrt{3\cos^2\theta +1} \, .
\end{equation}
The angle $\theta$ thus modulates the depth of the optical potential and shifts the $\ket{m_y = \pm 1/2}$ potential wells as shown with dotted lines in Fig.~\ref{fig:AdiabaticPotential}b. Upon applying the mixing term $U_{\rm ZM}$, the degeneracies are lifted and we obtain two distinct branches 
\begin{equation}
(U_{\rm OL} + U_{\rm ZM}) \ket{\chi_\pm(z)} = U_\pm(z) \ket{\chi_\pm(z)}
\end{equation}
with $U_+(z)> U_-(z)$. With an appropriate choice of the angle between the laser polarizations and magnetic field strength, we can design two decoupled lattices of double-well potentials with an adjustable energy barrier between them. This situation is depicted in Fig.~\ref{fig:AdiabaticPotential}b, where the underlying color indicates the polarization $\braOket{\chi_\pm(z)}{F_y}{\chi_\pm(z)}$ of the spin eigenstates.

We identify three almost equally separated potential wells $\circled{1}$, $\circled{2}$ and $\circled{3}$ which are depicted in Fig.~\ref{fig:AdiabaticPotential}b. We would like to derive an effective tight binding model for the later. Within an harmonic approximation, the small oscillatory motion in the wells have natural frequencies determined by Taylor expansion around the local minima
\begin{equation} \begin{split}
\hbar\omega_1 & = \hbar\omega_3 = \sqrt{2\tilde{U}_0 E_R} \, , \\ \hbar\omega_2 & = 2 \tilde{U}_0 \sqrt{\frac{E_R \sin(2k_L\delta z)}{\hbar \omega_Z}} \, ,
\end{split}
\end{equation}
where we have introduced the recoil energy $E_R = (2\hbar k_L)^2/2m$. As stated earlier, the original unperturbed optical lattice with $\omega_Z=0$ enables the cooling of the atomic cloud to sub-doppler temperatures $k_B T < E_R$~\cite{EIT_Cooling,Vuletic_AllOpticalBEC}. In this temperature regime and assuming $U_0 \gg E_R$, the previous energy scales $\hbar \omega_i$ are much larger than the thermal energy of the atoms. As a consequence, we can approximate the system's dynamics by keeping only the lowest energy Wannier state of each potential wells $\circled{1}$, $\circled{2}$ and $\circled{3}$ within a tight-binding approximation~\cite{RevModPhys_ManyBodyUltracoldGases,Gerbier_OpticalLattices}.

\subsection{Tunneling And Artificial Gauge Field} \label{ssec:ExpRamanCoupling}

To obtain the effective tight-binding model of our system, we first consider a single unit-cell of length $\pi/k_L$ along $y$. We denote the Wannier state centered around the potential well \circled{$i$} with the first quantized notation $\ket{i}$ (see Fig.~\ref{fig:AdiabaticPotential}b), and its on-site energy as $E_i$ for $i=1,2,3$. While the orbitals $\circled{1}$ and $\circled{3}$ lie in the same band and can overlap, hopping from these states to $\circled{2}$ is prohibited by the orthogonality condition $\braket{\chi_+}{\chi_-}=0$~\cite{SpinFlipTunneling40K}. Light-assisted tunneling may however be used to couple these orbitals with two-photons Raman transitions. These two different tunneling mechanisms are considered separately thereafter.

\paragraph{Real Tunneling Amplitude:} To evaluate the tunneling matrix element $t$ between $\circled{1}$ and $\circled{3}$, we follow the semiclassical treatment of a double well potential of Ref.~\cite{SemicalssicalDoubleWellPotentialTunneling} (see Fig.~\ref{fig:AdiabaticPotential}d). This approach does not account for the periodic structure of the lattice, usually underestimating the tunneling strengths of the original model~\cite{CorrectTightBinding_CosinePotential}. The investigation made in Ref.~\cite{SemicalssicalDoubleWellPotentialTunneling} gives the estimate
\begin{equation}
t = \frac{\hbar \omega_1}{2 \pi} e^{-\theta} \, , \quad \theta = \int_{y_<}^{y_>} \frac{{\rm d} y'}{\hbar} \sqrt{2m [U_-(y')-E]} \, ,
\end{equation}
where the turning points $y_{<}$ and $y_{>}$ and the energy $E$ are schematically depicted in Fig.~\ref{fig:AdiabaticPotential}d. For the parameters of Fig.~\ref{fig:AdiabaticPotential}, the numerical evaluation of these quantities give $\theta \simeq 2.5$ and
\begin{equation} \label{eq:EstimateTypicalRealTunneling}
t \simeq 0.75 E_R \, .
\end{equation}
This large value compared to other cold-atom experiments with optical lattices~\cite{Ketterle_RealizeHarperModel} is explained by the subwavelength separation of the potential wells $\circled{1}$ and $\circled{3}$ in the spin-dependent optical lattice Eq.~\ref{eq:LatticeDependentUncoupled}~\cite{Zoller_ColdAtomCoupledWires}.

\paragraph{Raman Assisted Tunneling:} We finally turn to the engineering of light assisted hopping towards the Wannier state $\circled{2}$, which not only couples the latter to the effective tight-binding model but also generates an artificial gauge field for our system. These coupling can be achieved with two additional laser beams counter-propagating along the $x$-direction, very far detuned from any atomic line. The first one is polarized along $y$ and has frequency $\omega_R$ while the second is polarized along $z$ and possesses two different tones $\omega_R + \omega_{12}$ and $\omega_R + \omega_{23}$, with $\omega_{ij} = (E_j - E_i)/\hbar$ (see Fig.~\ref{fig:AdiabaticPotential}a). They coherently couple spin states with $|\Delta m_y|=1$ through two-photon Raman transitions and thus allow to drive the two transitions \circled{1} $\to$ \circled{2} and \circled{2} $\to$ \circled{3}.

Adiabatically eliminating the very off-resonantly coupled excited atomic states~\cite{SteveChu_LaserCooling} and within the Rotating Wave Approximation (RWA)~\cite{Cohen_Tannoudji_RWAApprox}, the Raman Hamiltonian reads~\cite{SteveChu_Vuletic,SyntheticGaugeLanthanidesAtoms,Spielman_ClockTransitions}:
\begin{equation} \begin{split}
H_{\rm Ram} = & - \left[ (t_R e^{2ik_Lx} + t_R' e^{-2ik_Lx}) \ket{2}\bra{1} + h.c. \right] \\  & - \left[ (\tilde{t}_R e^{2ik_Lx} + \tilde{t}_R' e^{-2ik_Lx}) \ket{3}\bra{2} + h.c. \right] \, .
\end{split} \end{equation}
All four Rabi frequencies $t_R$, $t_R'$, $\tilde{t}_R$ and $\tilde{t}_R'$ are proportional to the Raman lasers' intensity $I_R$. However, they crucially involve different spinorial and spatial overlaps between Wannier states~\cite{SpinFlipTunneling40K,Spielmann_AdiabaticSuperlattice}, for instance 
\begin{equation}
t_R \propto \int {\rm d}y \; \braOket{\chi_+(y)}{F_-}{\chi_-(y)} \braket{2}{y}\braket{y}{1} \, ,
\end{equation}
while 
\begin{equation}
t_R' \propto \int {\rm d}y \; \braOket{\chi_+(y)}{F_+}{\chi_-(y)} \braket{2}{y}\braket{y}{1} \, .
\end{equation}
As observed in Fig.~\ref{fig:AdiabaticPotential}c, the strong polarization of the spin states $\ket{\chi_\pm (y)}$ leads to $t_R \gg t_R'$. Similarly $\tilde{t}_R \gg \tilde{t}_R'$ and for simplicity we assume $\tilde{t}_R = t_R$ such that the light-atom interaction with the Raman fields is described by
\begin{equation} \label{eq:ComplexTunneling}
H_{\rm Ram} = -t_R \left[e^{2ik_Lx} (\ket{2}\bra{1} + \ket{3}\bra{2}) + h.c. \right] \, .
\end{equation}
This formula obtained within the RWA is valid as long as $t_R \ll \min(E_2-E_1, E_3-E_1)$. For the parameters used in Fig.~\ref{fig:AdiabaticPotential} and Eq.~\ref{eq:EstimateTypicalRealTunneling}, this corresponds to $t_R \ll 30 E_R$. Under realistic experimental conditions, technical limitations will set in before the saturation of this inequality which is another nice feature of our spin dependent lattice. The intensity of the Raman lasers can be increased or the detuning to the excited states decreased in order to achieve $t_R \simeq t$, which we will assume from now on.

Eq.~\ref{eq:ComplexTunneling} shows that a particle moving along increasing $y$ picks a phase proportional to $x$ because of the momentum difference between the two Raman beams, which is reminiscent to the effect of a gauge field. This analogy can be made exact~\cite{Spielman_RashbaDresselhaus,Spielmann_RFRaman,SyntheticGaugeLanthanidesAtoms} (also see below) and was used experimentally to create artificial gauge field for neutral atoms~\cite{Spielman_FirstExperiment,DalibardNascimbene_HallBulkSyntheticDimension,Mancini_ObservationChiralEdgestate}. The introduction of the Raman laser beams thus creates an effective gauge field for the atoms and connects the Wannier state $\circled{2}$ to our tight-binding model.

\subsection{Matching The Parameters Of The Two Models: Experimental Feasibility} \label{ssec:ExpCoupledWires_ExperimentalScalings}

We have seen that the added pair of Raman laser beams allows to create an artificial gauge field thanks to spin-selective two-photon transitions between the two bands of the optical lattice while the sub-wavelength spacing of Wannier states provides significant tunneling amplitudes $t \simeq 0.75 E_R$. To make connection with Sec.~\ref{sec:IQHEFromWires}, we call $L$ the system length along the weakly confined $x$ direction and introduce $c_j^\dagger (x)$ the creation operator at position $x$ in the $(r+1)$-th Wannier orbital of the $q$-th unit cell with $j=3q+r$. the effective tight-binding Hamiltonian following from Secs.~\ref{ssec:ExpSpinDependentTrap} and~\ref{ssec:ExpRamanCoupling} is 
\begin{equation} \begin{split}
H = \sum_{j \in \mathbb{Z}} \int_0^{L} & {\rm d}x \, c_j^\dagger (x) \frac{p_x^2}{2m} c_j (x) \\ & \quad - t \left[ e^{i \phi'(j) x} c_{j+1}^\dagger (x) c_j (x) + h.c. \right] \, ,
\end{split} \end{equation}
where the phases follow from
\begin{equation}
\phi'(j) = \begin{cases} 2 k_L & {\rm if } \, r =0 ,1 \\ 0 & {\rm if } \, r =2 
\end{cases} \, .
\end{equation}
The connection with Eq.~\ref{eq:FullHamiltonianRealSpace} is cleaner after a gauge transformation $c_j^\dagger (y) \to \exp(4i k_L y /3) c_j^\dagger (y)$ under which $H$ becomes
\begin{widetext} \begin{equation}
H = \sum_{j \in \mathbb{Z}} \int_0^{L_y} {\rm d}y \, c_j^\dagger (y) \frac{(p_y - \frac{4\hbar k_L}{3} j)^2}{2m} c_j (y) - t \left[ e^{i \tilde{\phi}'(j) y} c_{j+1}^\dagger (y) c_j (y) + h.c. \right] \, , 
\end{equation} \end{widetext}
where the tunneling phases now sum to zero
\begin{equation}
\tilde{\phi}'(j) = \begin{cases} \frac{2 k_L}{3} & {\rm if } \, r =0 ,1 \\ -\frac{4 k_L}{3} & {\rm if } \, r =2 
\end{cases} \, ,
\end{equation}
showing that all the artificial gauge field have been transfered to the kinetic part of the Hamiltonian. Pushing the analogy further, we can identify
\begin{equation}
E_0 = \frac{16}{9} \frac{\hbar^2 k_L^2}{2m} = \frac{4}{9} E_R \, ,
\end{equation}
such that the typical tunneling parameters obtained in Eq.~\ref{eq:EstimateTypicalRealTunneling} translate into 
\begin{equation} \label{eq:ExperimentalLambda_ReachStrongTunneling}
\lambda = \frac{t}{E_0} \simeq 1.7 > 1 \, .
\end{equation}
The ultracold atomic system reaches the flat-band limit studied in Sec.~\ref{ssec:NonIntLandauLevels} for the experimentally relevant parameters used in Fig.~\ref{fig:AdiabaticPotential}. This emphasizes the experimental feasibility of our proposal, potentially leading to the observation of quantum Hall physics in large ultracold atoms ensembles. Sub-Doppler temperatures are however necessary if one wants to only populate the lowest band of the system because of the rather small cyclotron frequency $\hbar \omega \simeq E_R / \pi$ in the cold atomic case.

Finally, we comment on the possibility to reach the strongly correlated states of Sec.~\ref{sec:AddingInteractions}. Since bosons at ultracold temperature only interact with $s$-wave scattering, corresponding to the zero-th Haldane pseudo-potential Eq.~\ref{eq:PseudoPotInRealSpace}, it is natural to investigate the experimental conditions required to achieve a total filling fraction $\nu = 1/2$. For on-wire initial density $n_{\rm 1D} = N/N_w \simeq 100$, we would need a wire filling
\begin{equation}
\nu_w = \frac{3\pi}{2 k_L L_y} \simeq \frac{1}{200} \, .
\end{equation}
This estimation leads to a length $L_y \simeq 150 \lambda_L$ along the weakly confined direction with $\lambda_L$ the Raman laser wavelength, which corresponds to $L_y\sim 100\mu$m under typical experimental condition. This can be realized with a very weak confinement along $y$. Most of the quantities computed here can be improved by numerical factors of order one by either introducing an angle between the two Raman beams~\cite{Goldman_ReviewArtificialGauge} or by choosing very different wavelength for the lattice and Raman beams. However, Eq.~\ref{eq:ExperimentalLambda_ReachStrongTunneling} and the possible sub-Doppler temperatures already give very favorable estimates with regard to the experimental realization of quantum Hall physics in such systems.

\section{Conclusion} \label{sec:Conclusion}

In this article, we have provided a microscopic characterization of a class of coupled-wire models. First, we have shown the emergence of Landau levels for strong inter-wire tunneling. This equivalence with a continuum system in the deep fractional quantum Hall regime allows to adapt the pseudopotential approach to the coupled-wire system. In particular, we could exhibit model on-wire interactions stabilizing both bosonic and fermionic Laughlin phases in the thermodynamic limit. Turning our attention towards potential experimental implementations of our model in cold-atom setups, we have used exact diagonalization to observe and characterize the different phases for realistic parameters. We provided evidence that time-of-flight measurements can distinguish between the Laughlin and the other weakly-interacting Bogoliubov phases. Finally, we have proposed an experimental realization of our model with cold-atoms in a spin-dependent optical lattice. The spin-dependent trapping potential leads to sub-wavelength spacing of the one-dimensional atomic wires, and allows to reach the strong tunneling regime. Our estimates indicate that the FQH-like phases could be observed in optical lattices for experimentally realistic parameters if spontaneous heating is mitigated.

\section{Acknowledgments}

We are indebted to Ady Stern for discussions which have initiated the present work. We particularly thank Christophe Mora for sharing his knowledge and ideas on related topics. V.C. also wants to thank Jane Dunphy for useful comments on the manuscript. 

V.C., B.E. and N.R. were supported by the grants ANR TNSTRONG No. ANR-16-CE30-0025 and ANR TopO No. ANR-17-CE30-0013-01. N.R. was also supported by the Department of Energy Grant No.  DE-SC0016239, the National Science Foundation EAGER Grant No.DMR 1643312, Simons Investigator Grants No. 404513, ONR No. N00014-14-1-0330, and NSF-MRSEC No. DMR-142051, the Packard Foundation, the Schmidt Fund for Innovative Research. 

\bibliography{Biblio_SpinLattice}

\begin{thebibliography}{104}%
\makeatletter
\providecommand \@ifxundefined [1]{%
 \@ifx{#1\undefined}
}%
\providecommand \@ifnum [1]{%
 \ifnum #1\expandafter \@firstoftwo
 \else \expandafter \@secondoftwo
 \fi
}%
\providecommand \@ifx [1]{%
 \ifx #1\expandafter \@firstoftwo
 \else \expandafter \@secondoftwo
 \fi
}%
\providecommand \natexlab [1]{#1}%
\providecommand \enquote  [1]{``#1''}%
\providecommand \bibnamefont  [1]{#1}%
\providecommand \bibfnamefont [1]{#1}%
\providecommand \citenamefont [1]{#1}%
\providecommand \href@noop [0]{\@secondoftwo}%
\providecommand \href [0]{\begingroup \@sanitize@url \@href}%
\providecommand \@href[1]{\@@startlink{#1}\@@href}%
\providecommand \@@href[1]{\endgroup#1\@@endlink}%
\providecommand \@sanitize@url [0]{\catcode `\\12\catcode `\$12\catcode
  `\&12\catcode `\#12\catcode `\^12\catcode `\_12\catcode `\%12\relax}%
\providecommand \@@startlink[1]{}%
\providecommand \@@endlink[0]{}%
\providecommand \url  [0]{\begingroup\@sanitize@url \@url }%
\providecommand \@url [1]{\endgroup\@href {#1}{\urlprefix }}%
\providecommand \urlprefix  [0]{URL }%
\providecommand \Eprint [0]{\href }%
\providecommand \doibase [0]{http://dx.doi.org/}%
\providecommand \selectlanguage [0]{\@gobble}%
\providecommand \bibinfo  [0]{\@secondoftwo}%
\providecommand \bibfield  [0]{\@secondoftwo}%
\providecommand \translation [1]{[#1]}%
\providecommand \BibitemOpen [0]{}%
\providecommand \bibitemStop [0]{}%
\providecommand \bibitemNoStop [0]{.\EOS\space}%
\providecommand \EOS [0]{\spacefactor3000\relax}%
\providecommand \BibitemShut  [1]{\csname bibitem#1\endcsname}%
\let\auto@bib@innerbib\@empty
%</preamble>
\bibitem [{\citenamefont {Laughlin}(1999)}]{Laughlin_Nobel}%
  \BibitemOpen
  \bibfield  {author} {\bibinfo {author} {\bibfnamefont {R.~B.}\ \bibnamefont
  {Laughlin}},\ }\href@noop {} {\bibfield  {journal} {\bibinfo  {journal}
  {Reviews of Modern Physics}\ }\textbf {\bibinfo {volume} {71}},\ \bibinfo
  {pages} {863} (\bibinfo {year} {1999})}\BibitemShut {NoStop}%
\bibitem [{\citenamefont {Laughlin}(1983)}]{LaughlinWFandTheory}%
  \BibitemOpen
  \bibfield  {author} {\bibinfo {author} {\bibfnamefont {R.~B.}\ \bibnamefont
  {Laughlin}},\ }\href {\doibase 10.1103/PhysRevLett.50.1395} {\bibfield
  {journal} {\bibinfo  {journal} {Phys. Rev. Lett.}\ }\textbf {\bibinfo
  {volume} {50}},\ \bibinfo {pages} {1395} (\bibinfo {year}
  {1983})}\BibitemShut {NoStop}%
\bibitem [{\citenamefont {Laughlin}(1990)}]{LaughlinBookGirvin}%
  \BibitemOpen
  \bibfield  {author} {\bibinfo {author} {\bibfnamefont {R.~B.}\ \bibnamefont
  {Laughlin}},\ }\enquote {\bibinfo {title} {Elementary theory: the
  incompressible quantum fluid},}\ in\ \href {\doibase
  10.1007/978-1-4612-3350-3_7} {\emph {\bibinfo {booktitle} {The Quantum Hall
  Effect}}},\ \bibinfo {editor} {edited by\ \bibinfo {editor} {\bibfnamefont
  {R.~E.}\ \bibnamefont {Prange}}\ and\ \bibinfo {editor} {\bibfnamefont
  {S.~M.}\ \bibnamefont {Girvin}}}\ (\bibinfo  {publisher} {Springer New
  York},\ \bibinfo {address} {New York, NY},\ \bibinfo {year} {1990})\ pp.\
  \bibinfo {pages} {233--301}\BibitemShut {NoStop}%
\bibitem [{\citenamefont {Saminadayar}\ \emph {et~al.}(1997)\citenamefont
  {Saminadayar}, \citenamefont {Glattli}, \citenamefont {Jin},\ and\
  \citenamefont {Etienne}}]{ExpFractionalChargeEtienne}%
  \BibitemOpen
  \bibfield  {author} {\bibinfo {author} {\bibfnamefont {L.}~\bibnamefont
  {Saminadayar}}, \bibinfo {author} {\bibfnamefont {D.~C.}\ \bibnamefont
  {Glattli}}, \bibinfo {author} {\bibfnamefont {Y.}~\bibnamefont {Jin}}, \ and\
  \bibinfo {author} {\bibfnamefont {B.}~\bibnamefont {Etienne}},\ }\href
  {\doibase 10.1103/PhysRevLett.79.2526} {\bibfield  {journal} {\bibinfo
  {journal} {Phys. Rev. Lett.}\ }\textbf {\bibinfo {volume} {79}},\ \bibinfo
  {pages} {2526} (\bibinfo {year} {1997})}\BibitemShut {NoStop}%
\bibitem [{\citenamefont {de~Picciotto}\ \emph {et~al.}(1997)\citenamefont
  {de~Picciotto}, \citenamefont {Reznikov}, \citenamefont {Heiblum},
  \citenamefont {Umansky}, \citenamefont {Bunin},\ and\ \citenamefont
  {Mahalu}}]{ExpFractionalChargeHeiblum}%
  \BibitemOpen
  \bibfield  {author} {\bibinfo {author} {\bibfnamefont {R.}~\bibnamefont
  {de~Picciotto}}, \bibinfo {author} {\bibfnamefont {M.}~\bibnamefont
  {Reznikov}}, \bibinfo {author} {\bibfnamefont {M.}~\bibnamefont {Heiblum}},
  \bibinfo {author} {\bibfnamefont {V.}~\bibnamefont {Umansky}}, \bibinfo
  {author} {\bibfnamefont {G.}~\bibnamefont {Bunin}}, \ and\ \bibinfo {author}
  {\bibfnamefont {D.}~\bibnamefont {Mahalu}},\ }\href
  {http://dx.doi.org/10.1038/38241} {\bibfield  {journal} {\bibinfo  {journal}
  {Nature}\ }\textbf {\bibinfo {volume} {389}},\ \bibinfo {pages} {162 EP }
  (\bibinfo {year} {1997})}\BibitemShut {NoStop}%
\bibitem [{\citenamefont {Arovas}\ \emph {et~al.}(1984)\citenamefont {Arovas},
  \citenamefont {Schrieffer},\ and\ \citenamefont
  {Wilczek}}]{FractionalStatQHE}%
  \BibitemOpen
  \bibfield  {author} {\bibinfo {author} {\bibfnamefont {D.}~\bibnamefont
  {Arovas}}, \bibinfo {author} {\bibfnamefont {J.~R.}\ \bibnamefont
  {Schrieffer}}, \ and\ \bibinfo {author} {\bibfnamefont {F.}~\bibnamefont
  {Wilczek}},\ }\href {\doibase 10.1103/PhysRevLett.53.722} {\bibfield
  {journal} {\bibinfo  {journal} {Phys. Rev. Lett.}\ }\textbf {\bibinfo
  {volume} {53}},\ \bibinfo {pages} {722} (\bibinfo {year} {1984})}\BibitemShut
  {NoStop}%
\bibitem [{\citenamefont
  {Haldane}(1983{\natexlab{a}})}]{HaldaneHierarchiesPseudoPot}%
  \BibitemOpen
  \bibfield  {author} {\bibinfo {author} {\bibfnamefont {F.~D.~M.}\
  \bibnamefont {Haldane}},\ }\href {\doibase 10.1103/PhysRevLett.51.605}
  {\bibfield  {journal} {\bibinfo  {journal} {Phys. Rev. Lett.}\ }\textbf
  {\bibinfo {volume} {51}},\ \bibinfo {pages} {605} (\bibinfo {year}
  {1983}{\natexlab{a}})}\BibitemShut {NoStop}%
\bibitem [{\citenamefont {Trugman}\ and\ \citenamefont
  {Kivelson}(1985)}]{PseudoPotentialsTrugmanKivelson}%
  \BibitemOpen
  \bibfield  {author} {\bibinfo {author} {\bibfnamefont {S.~A.}\ \bibnamefont
  {Trugman}}\ and\ \bibinfo {author} {\bibfnamefont {S.}~\bibnamefont
  {Kivelson}},\ }\href {\doibase 10.1103/PhysRevB.31.5280} {\bibfield
  {journal} {\bibinfo  {journal} {Phys. Rev. B}\ }\textbf {\bibinfo {volume}
  {31}},\ \bibinfo {pages} {5280} (\bibinfo {year} {1985})}\BibitemShut
  {NoStop}%
\bibitem [{\citenamefont {Kane}\ \emph {et~al.}(2002)\citenamefont {Kane},
  \citenamefont {Mukhopadhyay},\ and\ \citenamefont
  {Lubensky}}]{CoupledWireFirstKane}%
  \BibitemOpen
  \bibfield  {author} {\bibinfo {author} {\bibfnamefont {C.~L.}\ \bibnamefont
  {Kane}}, \bibinfo {author} {\bibfnamefont {R.}~\bibnamefont {Mukhopadhyay}},
  \ and\ \bibinfo {author} {\bibfnamefont {T.~C.}\ \bibnamefont {Lubensky}},\
  }\href {\doibase 10.1103/PhysRevLett.88.036401} {\bibfield  {journal}
  {\bibinfo  {journal} {Phys. Rev. Lett.}\ }\textbf {\bibinfo {volume} {88}},\
  \bibinfo {pages} {036401} (\bibinfo {year} {2002})}\BibitemShut {NoStop}%
\bibitem [{\citenamefont {Teo}\ and\ \citenamefont
  {Kane}(2014)}]{CoupledWireTeoKanePRB}%
  \BibitemOpen
  \bibfield  {author} {\bibinfo {author} {\bibfnamefont {J.~C.~Y.}\
  \bibnamefont {Teo}}\ and\ \bibinfo {author} {\bibfnamefont {C.~L.}\
  \bibnamefont {Kane}},\ }\href {\doibase 10.1103/PhysRevB.89.085101}
  {\bibfield  {journal} {\bibinfo  {journal} {Phys. Rev. B}\ }\textbf {\bibinfo
  {volume} {89}},\ \bibinfo {pages} {085101} (\bibinfo {year}
  {2014})}\BibitemShut {NoStop}%
\bibitem [{\citenamefont {Meng}\ and\ \citenamefont
  {Sela}(2014)}]{WiresBiblio0}%
  \BibitemOpen
  \bibfield  {author} {\bibinfo {author} {\bibfnamefont {T.}~\bibnamefont
  {Meng}}\ and\ \bibinfo {author} {\bibfnamefont {E.}~\bibnamefont {Sela}},\
  }\href {\doibase 10.1103/PhysRevB.90.235425} {\bibfield  {journal} {\bibinfo
  {journal} {Phys. Rev. B}\ }\textbf {\bibinfo {volume} {90}},\ \bibinfo
  {pages} {235425} (\bibinfo {year} {2014})}\BibitemShut {NoStop}%
\bibitem [{\citenamefont {Cano}\ \emph {et~al.}(2015)\citenamefont {Cano},
  \citenamefont {Hughes},\ and\ \citenamefont {Mulligan}}]{WiresBiblio1}%
  \BibitemOpen
  \bibfield  {author} {\bibinfo {author} {\bibfnamefont {J.}~\bibnamefont
  {Cano}}, \bibinfo {author} {\bibfnamefont {T.~L.}\ \bibnamefont {Hughes}}, \
  and\ \bibinfo {author} {\bibfnamefont {M.}~\bibnamefont {Mulligan}},\ }\href
  {\doibase 10.1103/PhysRevB.92.075104} {\bibfield  {journal} {\bibinfo
  {journal} {Phys. Rev. B}\ }\textbf {\bibinfo {volume} {92}},\ \bibinfo
  {pages} {075104} (\bibinfo {year} {2015})}\BibitemShut {NoStop}%
\bibitem [{\citenamefont {Sagi}\ \emph {et~al.}(2015)\citenamefont {Sagi},
  \citenamefont {Oreg}, \citenamefont {Stern},\ and\ \citenamefont
  {Halperin}}]{WiresBiblio2}%
  \BibitemOpen
  \bibfield  {author} {\bibinfo {author} {\bibfnamefont {E.}~\bibnamefont
  {Sagi}}, \bibinfo {author} {\bibfnamefont {Y.}~\bibnamefont {Oreg}}, \bibinfo
  {author} {\bibfnamefont {A.}~\bibnamefont {Stern}}, \ and\ \bibinfo {author}
  {\bibfnamefont {B.~I.}\ \bibnamefont {Halperin}},\ }\href {\doibase
  10.1103/PhysRevB.91.245144} {\bibfield  {journal} {\bibinfo  {journal} {Phys.
  Rev. B}\ }\textbf {\bibinfo {volume} {91}},\ \bibinfo {pages} {245144}
  (\bibinfo {year} {2015})}\BibitemShut {NoStop}%
\bibitem [{\citenamefont {Meng}\ \emph {et~al.}(2015)\citenamefont {Meng},
  \citenamefont {Neupert}, \citenamefont {Greiter},\ and\ \citenamefont
  {Thomale}}]{WiresBiblio3}%
  \BibitemOpen
  \bibfield  {author} {\bibinfo {author} {\bibfnamefont {T.}~\bibnamefont
  {Meng}}, \bibinfo {author} {\bibfnamefont {T.}~\bibnamefont {Neupert}},
  \bibinfo {author} {\bibfnamefont {M.}~\bibnamefont {Greiter}}, \ and\
  \bibinfo {author} {\bibfnamefont {R.}~\bibnamefont {Thomale}},\ }\href
  {\doibase 10.1103/PhysRevB.91.241106} {\bibfield  {journal} {\bibinfo
  {journal} {Phys. Rev. B}\ }\textbf {\bibinfo {volume} {91}},\ \bibinfo
  {pages} {241106} (\bibinfo {year} {2015})}\BibitemShut {NoStop}%
\bibitem [{\citenamefont {Huang}\ \emph {et~al.}(2016)\citenamefont {Huang},
  \citenamefont {Chen}, \citenamefont {Gomes}, \citenamefont {Neupert},
  \citenamefont {Chamon},\ and\ \citenamefont {Mudry}}]{WiresBiblio4}%
  \BibitemOpen
  \bibfield  {author} {\bibinfo {author} {\bibfnamefont {P.-H.}\ \bibnamefont
  {Huang}}, \bibinfo {author} {\bibfnamefont {J.-H.}\ \bibnamefont {Chen}},
  \bibinfo {author} {\bibfnamefont {P.~R.~S.}\ \bibnamefont {Gomes}}, \bibinfo
  {author} {\bibfnamefont {T.}~\bibnamefont {Neupert}}, \bibinfo {author}
  {\bibfnamefont {C.}~\bibnamefont {Chamon}}, \ and\ \bibinfo {author}
  {\bibfnamefont {C.}~\bibnamefont {Mudry}},\ }\href {\doibase
  10.1103/PhysRevB.93.205123} {\bibfield  {journal} {\bibinfo  {journal} {Phys.
  Rev. B}\ }\textbf {\bibinfo {volume} {93}},\ \bibinfo {pages} {205123}
  (\bibinfo {year} {2016})}\BibitemShut {NoStop}%
\bibitem [{\citenamefont {Lecheminant}\ and\ \citenamefont
  {Tsvelik}(2017)}]{WiresFarFromFQH0}%
  \BibitemOpen
  \bibfield  {author} {\bibinfo {author} {\bibfnamefont {P.}~\bibnamefont
  {Lecheminant}}\ and\ \bibinfo {author} {\bibfnamefont {A.~M.}\ \bibnamefont
  {Tsvelik}},\ }\href {\doibase 10.1103/PhysRevB.95.140406} {\bibfield
  {journal} {\bibinfo  {journal} {Phys. Rev. B}\ }\textbf {\bibinfo {volume}
  {95}},\ \bibinfo {pages} {140406} (\bibinfo {year} {2017})}\BibitemShut
  {NoStop}%
\bibitem [{\citenamefont {Mong}\ \emph {et~al.}(2014)\citenamefont {Mong},
  \citenamefont {Clarke}, \citenamefont {Alicea}, \citenamefont {Lindner},
  \citenamefont {Fendley}, \citenamefont {Nayak}, \citenamefont {Oreg},
  \citenamefont {Stern}, \citenamefont {Berg}, \citenamefont {Shtengel},\ and\
  \citenamefont {Fisher}}]{WiresFarFromFQH1}%
  \BibitemOpen
  \bibfield  {author} {\bibinfo {author} {\bibfnamefont {R.~S.~K.}\
  \bibnamefont {Mong}}, \bibinfo {author} {\bibfnamefont {D.~J.}\ \bibnamefont
  {Clarke}}, \bibinfo {author} {\bibfnamefont {J.}~\bibnamefont {Alicea}},
  \bibinfo {author} {\bibfnamefont {N.~H.}\ \bibnamefont {Lindner}}, \bibinfo
  {author} {\bibfnamefont {P.}~\bibnamefont {Fendley}}, \bibinfo {author}
  {\bibfnamefont {C.}~\bibnamefont {Nayak}}, \bibinfo {author} {\bibfnamefont
  {Y.}~\bibnamefont {Oreg}}, \bibinfo {author} {\bibfnamefont {A.}~\bibnamefont
  {Stern}}, \bibinfo {author} {\bibfnamefont {E.}~\bibnamefont {Berg}},
  \bibinfo {author} {\bibfnamefont {K.}~\bibnamefont {Shtengel}}, \ and\
  \bibinfo {author} {\bibfnamefont {M.~P.~A.}\ \bibnamefont {Fisher}},\ }\href
  {\doibase 10.1103/PhysRevX.4.011036} {\bibfield  {journal} {\bibinfo
  {journal} {Phys. Rev. X}\ }\textbf {\bibinfo {volume} {4}},\ \bibinfo {pages}
  {011036} (\bibinfo {year} {2014})}\BibitemShut {NoStop}%
\bibitem [{\citenamefont {Fuji}(2019)}]{WiresFarFromFQH2}%
  \BibitemOpen
  \bibfield  {author} {\bibinfo {author} {\bibfnamefont {Y.}~\bibnamefont
  {Fuji}},\ }\href {\doibase 10.1103/PhysRevB.100.235115} {\bibfield  {journal}
  {\bibinfo  {journal} {Phys. Rev. B}\ }\textbf {\bibinfo {volume} {100}},\
  \bibinfo {pages} {235115} (\bibinfo {year} {2019})}\BibitemShut {NoStop}%
\bibitem [{\citenamefont {Meng}(2020)}]{CoupledWire_Review}%
  \BibitemOpen
  \bibfield  {author} {\bibinfo {author} {\bibfnamefont {T.}~\bibnamefont
  {Meng}},\ }\href {\doibase 10.1140/epjst/e2019-900095-5} {\bibfield
  {journal} {\bibinfo  {journal} {The European Physical Journal Special
  Topics}\ }\textbf {\bibinfo {volume} {229}},\ \bibinfo {pages} {527–543}
  (\bibinfo {year} {2020})}\BibitemShut {NoStop}%
\bibitem [{\citenamefont {Giamarchi}(2004)}]{GiamarchiQuantumOneD}%
  \BibitemOpen
  \bibfield  {author} {\bibinfo {author} {\bibfnamefont {T.}~\bibnamefont
  {Giamarchi}},\ }\href {https://books.google.fr/books?id=1MwTDAAAQBAJ} {\emph
  {\bibinfo {title} {Quantum Physics in One Dimension}}},\ International Series
  of Monogr\ (\bibinfo  {publisher} {Clarendon Press},\ \bibinfo {year}
  {2004})\BibitemShut {NoStop}%
\bibitem [{\citenamefont {Voit}(1993)}]{Voit_Luttinger}%
  \BibitemOpen
  \bibfield  {author} {\bibinfo {author} {\bibfnamefont {J.}~\bibnamefont
  {Voit}},\ }\href {http://stacks.iop.org/0953-8984/5/i=44/a=020} {\bibfield
  {journal} {\bibinfo  {journal} {Journal of Physics: Condensed Matter}\
  }\textbf {\bibinfo {volume} {5}},\ \bibinfo {pages} {8305} (\bibinfo {year}
  {1993})}\BibitemShut {NoStop}%
\bibitem [{\citenamefont {Asb{\'o}th}\ \emph {et~al.}(2016)\citenamefont
  {Asb{\'o}th}, \citenamefont {Oroszl{\'a}ny},\ and\ \citenamefont
  {P{\'a}lyi}}]{ShortCourse_BulkEdge}%
  \BibitemOpen
  \bibfield  {author} {\bibinfo {author} {\bibfnamefont {J.~K.}\ \bibnamefont
  {Asb{\'o}th}}, \bibinfo {author} {\bibfnamefont {L.}~\bibnamefont
  {Oroszl{\'a}ny}}, \ and\ \bibinfo {author} {\bibfnamefont {A.}~\bibnamefont
  {P{\'a}lyi}},\ }\href@noop {} {\bibfield  {journal} {\bibinfo  {journal}
  {Lecture notes in physics}\ }\textbf {\bibinfo {volume} {919}},\ \bibinfo
  {pages} {87} (\bibinfo {year} {2016})}\BibitemShut {NoStop}%
\bibitem [{\citenamefont {Fontana}\ \emph {et~al.}(2019)\citenamefont
  {Fontana}, \citenamefont {Gomes},\ and\ \citenamefont
  {Hernaski}}]{CoupledWire_BulkEdge}%
  \BibitemOpen
  \bibfield  {author} {\bibinfo {author} {\bibfnamefont {W.~B.}\ \bibnamefont
  {Fontana}}, \bibinfo {author} {\bibfnamefont {P.~R.~S.}\ \bibnamefont
  {Gomes}}, \ and\ \bibinfo {author} {\bibfnamefont {C.~A.}\ \bibnamefont
  {Hernaski}},\ }\href {\doibase 10.1103/PhysRevB.99.201113} {\bibfield
  {journal} {\bibinfo  {journal} {Phys. Rev. B}\ }\textbf {\bibinfo {volume}
  {99}},\ \bibinfo {pages} {201113} (\bibinfo {year} {2019})}\BibitemShut
  {NoStop}%
\bibitem [{\citenamefont {Budich}\ \emph {et~al.}(2017)\citenamefont {Budich},
  \citenamefont {Elben}, \citenamefont {Lacki}, \citenamefont {Sterdyniak},
  \citenamefont {Baranov},\ and\ \citenamefont
  {Zoller}}]{Zoller_ColdAtomCoupledWires}%
  \BibitemOpen
  \bibfield  {author} {\bibinfo {author} {\bibfnamefont {J.~C.}\ \bibnamefont
  {Budich}}, \bibinfo {author} {\bibfnamefont {A.}~\bibnamefont {Elben}},
  \bibinfo {author} {\bibfnamefont {M.}~\bibnamefont {Lacki}}, \bibinfo
  {author} {\bibfnamefont {A.}~\bibnamefont {Sterdyniak}}, \bibinfo {author}
  {\bibfnamefont {M.~A.}\ \bibnamefont {Baranov}}, \ and\ \bibinfo {author}
  {\bibfnamefont {P.}~\bibnamefont {Zoller}},\ }\href {\doibase
  10.1103/PhysRevA.95.043632} {\bibfield  {journal} {\bibinfo  {journal} {Phys.
  Rev. A}\ }\textbf {\bibinfo {volume} {95}},\ \bibinfo {pages} {043632}
  (\bibinfo {year} {2017})}\BibitemShut {NoStop}%
\bibitem [{\citenamefont {Goldman}\ \emph {et~al.}(2016)\citenamefont
  {Goldman}, \citenamefont {Budich},\ and\ \citenamefont
  {Zoller}}]{Review_TopoMattercoldGas_Goldman}%
  \BibitemOpen
  \bibfield  {author} {\bibinfo {author} {\bibfnamefont {N.}~\bibnamefont
  {Goldman}}, \bibinfo {author} {\bibfnamefont {J.~C.}\ \bibnamefont {Budich}},
  \ and\ \bibinfo {author} {\bibfnamefont {P.}~\bibnamefont {Zoller}},\ }\href
  {\doibase 10.1038/nphys3803} {\bibfield  {journal} {\bibinfo  {journal}
  {Nature Physics}\ }\textbf {\bibinfo {volume} {12}},\ \bibinfo {pages} {639}
  (\bibinfo {year} {2016})}\BibitemShut {NoStop}%
\bibitem [{\citenamefont {Aidelsburger}\ \emph {et~al.}(2018)\citenamefont
  {Aidelsburger}, \citenamefont {Nascimbene},\ and\ \citenamefont
  {Goldman}}]{Review_TopocoldGases_Nascimbene}%
  \BibitemOpen
  \bibfield  {author} {\bibinfo {author} {\bibfnamefont {M.}~\bibnamefont
  {Aidelsburger}}, \bibinfo {author} {\bibfnamefont {S.}~\bibnamefont
  {Nascimbene}}, \ and\ \bibinfo {author} {\bibfnamefont {N.}~\bibnamefont
  {Goldman}},\ }\href {\doibase https://doi.org/10.1016/j.crhy.2018.03.002}
  {\bibfield  {journal} {\bibinfo  {journal} {Comptes Rendus Physique}\
  }\textbf {\bibinfo {volume} {19}},\ \bibinfo {pages} {394 } (\bibinfo {year}
  {2018})},\ \bibinfo {note} {quantum simulation / Simulation
  quantique}\BibitemShut {NoStop}%
\bibitem [{\citenamefont {Schweikhard}\ \emph {et~al.}(2004)\citenamefont
  {Schweikhard}, \citenamefont {Coddington}, \citenamefont {Engels},
  \citenamefont {Mogendorff},\ and\ \citenamefont {Cornell}}]{LLLBEC_Cornell}%
  \BibitemOpen
  \bibfield  {author} {\bibinfo {author} {\bibfnamefont {V.}~\bibnamefont
  {Schweikhard}}, \bibinfo {author} {\bibfnamefont {I.}~\bibnamefont
  {Coddington}}, \bibinfo {author} {\bibfnamefont {P.}~\bibnamefont {Engels}},
  \bibinfo {author} {\bibfnamefont {V.~P.}\ \bibnamefont {Mogendorff}}, \ and\
  \bibinfo {author} {\bibfnamefont {E.~A.}\ \bibnamefont {Cornell}},\ }\href
  {\doibase 10.1103/PhysRevLett.92.040404} {\bibfield  {journal} {\bibinfo
  {journal} {Phys. Rev. Lett.}\ }\textbf {\bibinfo {volume} {92}},\ \bibinfo
  {pages} {040404} (\bibinfo {year} {2004})}\BibitemShut {NoStop}%
\bibitem [{\citenamefont {Bretin}\ \emph {et~al.}(2004)\citenamefont {Bretin},
  \citenamefont {Stock}, \citenamefont {Seurin},\ and\ \citenamefont
  {Dalibard}}]{LLLBEC_Delibard}%
  \BibitemOpen
  \bibfield  {author} {\bibinfo {author} {\bibfnamefont {V.}~\bibnamefont
  {Bretin}}, \bibinfo {author} {\bibfnamefont {S.}~\bibnamefont {Stock}},
  \bibinfo {author} {\bibfnamefont {Y.}~\bibnamefont {Seurin}}, \ and\ \bibinfo
  {author} {\bibfnamefont {J.}~\bibnamefont {Dalibard}},\ }\href {\doibase
  10.1103/PhysRevLett.92.050403} {\bibfield  {journal} {\bibinfo  {journal}
  {Phys. Rev. Lett.}\ }\textbf {\bibinfo {volume} {92}},\ \bibinfo {pages}
  {050403} (\bibinfo {year} {2004})}\BibitemShut {NoStop}%
\bibitem [{\citenamefont {Fletcher}\ \emph {et~al.}(2019)\citenamefont
  {Fletcher}, \citenamefont {Shaffer}, \citenamefont {Wilson}, \citenamefont
  {Patel}, \citenamefont {Yan}, \citenamefont {Crépel}, \citenamefont
  {Mukherjee},\ and\ \citenamefont {Zwierlein}}]{LLLBEC_Fletcher}%
  \BibitemOpen
  \bibfield  {author} {\bibinfo {author} {\bibfnamefont {R.~J.}\ \bibnamefont
  {Fletcher}}, \bibinfo {author} {\bibfnamefont {A.}~\bibnamefont {Shaffer}},
  \bibinfo {author} {\bibfnamefont {C.~C.}\ \bibnamefont {Wilson}}, \bibinfo
  {author} {\bibfnamefont {P.~B.}\ \bibnamefont {Patel}}, \bibinfo {author}
  {\bibfnamefont {Z.}~\bibnamefont {Yan}}, \bibinfo {author} {\bibfnamefont
  {V.}~\bibnamefont {Crépel}}, \bibinfo {author} {\bibfnamefont
  {B.}~\bibnamefont {Mukherjee}}, \ and\ \bibinfo {author} {\bibfnamefont
  {M.~W.}\ \bibnamefont {Zwierlein}},\ }\href@noop {} {\enquote {\bibinfo
  {title} {Geometric squeezing into the lowest landau level},}\ } (\bibinfo
  {year} {2019}),\ \Eprint {http://arxiv.org/abs/1911.12347} {arXiv:1911.12347
  [cond-mat.quant-gas]} \BibitemShut {NoStop}%
\bibitem [{\citenamefont {Lin}\ \emph {et~al.}(2009)\citenamefont {Lin},
  \citenamefont {Compton}, \citenamefont {Jim{\'e}nez-Garc{\'i}a},
  \citenamefont {Porto},\ and\ \citenamefont
  {Spielman}}]{Spielman_FirstExperiment}%
  \BibitemOpen
  \bibfield  {author} {\bibinfo {author} {\bibfnamefont {Y.-J.}\ \bibnamefont
  {Lin}}, \bibinfo {author} {\bibfnamefont {R.~L.}\ \bibnamefont {Compton}},
  \bibinfo {author} {\bibfnamefont {K.}~\bibnamefont {Jim{\'e}nez-Garc{\'i}a}},
  \bibinfo {author} {\bibfnamefont {J.~V.}\ \bibnamefont {Porto}}, \ and\
  \bibinfo {author} {\bibfnamefont {I.~B.}\ \bibnamefont {Spielman}},\ }\href
  {\doibase 10.1038/nature08609} {\bibfield  {journal} {\bibinfo  {journal}
  {Nature}\ }\textbf {\bibinfo {volume} {462}},\ \bibinfo {pages} {628}
  (\bibinfo {year} {2009})}\BibitemShut {NoStop}%
\bibitem [{\citenamefont {Jaksch}\ and\ \citenamefont
  {Zoller}(2003)}]{Jaksch_GaugeFieldOpticalLatt}%
  \BibitemOpen
  \bibfield  {author} {\bibinfo {author} {\bibfnamefont {D.}~\bibnamefont
  {Jaksch}}\ and\ \bibinfo {author} {\bibfnamefont {P.}~\bibnamefont
  {Zoller}},\ }\href {\doibase 10.1088/1367-2630/5/1/356} {\bibfield  {journal}
  {\bibinfo  {journal} {New Journal of Physics}\ }\textbf {\bibinfo {volume}
  {5}},\ \bibinfo {pages} {56–56} (\bibinfo {year} {2003})}\BibitemShut
  {NoStop}%
\bibitem [{\citenamefont {Dalibard}\ \emph {et~al.}(2011)\citenamefont
  {Dalibard}, \citenamefont {Gerbier}, \citenamefont
  {Juzeli\ifmmode~\bar{u}\else \={u}\fi{}nas},\ and\ \citenamefont
  {\"Ohberg}}]{Review_ArtificialGauge_Dalibard}%
  \BibitemOpen
  \bibfield  {author} {\bibinfo {author} {\bibfnamefont {J.}~\bibnamefont
  {Dalibard}}, \bibinfo {author} {\bibfnamefont {F.}~\bibnamefont {Gerbier}},
  \bibinfo {author} {\bibfnamefont {G.}~\bibnamefont
  {Juzeli\ifmmode~\bar{u}\else \={u}\fi{}nas}}, \ and\ \bibinfo {author}
  {\bibfnamefont {P.}~\bibnamefont {\"Ohberg}},\ }\href {\doibase
  10.1103/RevModPhys.83.1523} {\bibfield  {journal} {\bibinfo  {journal} {Rev.
  Mod. Phys.}\ }\textbf {\bibinfo {volume} {83}},\ \bibinfo {pages} {1523}
  (\bibinfo {year} {2011})}\BibitemShut {NoStop}%
\bibitem [{\citenamefont {Hu}\ \emph {et~al.}(2017)\citenamefont {Hu},
  \citenamefont {Urvoy}, \citenamefont {Vendeiro}, \citenamefont {Cr{\'e}pel},
  \citenamefont {Chen},\ and\ \citenamefont
  {Vuleti{\'c}}}]{Vuletic_AllOpticalBEC}%
  \BibitemOpen
  \bibfield  {author} {\bibinfo {author} {\bibfnamefont {J.}~\bibnamefont
  {Hu}}, \bibinfo {author} {\bibfnamefont {A.}~\bibnamefont {Urvoy}}, \bibinfo
  {author} {\bibfnamefont {Z.}~\bibnamefont {Vendeiro}}, \bibinfo {author}
  {\bibfnamefont {V.}~\bibnamefont {Cr{\'e}pel}}, \bibinfo {author}
  {\bibfnamefont {W.}~\bibnamefont {Chen}}, \ and\ \bibinfo {author}
  {\bibfnamefont {V.}~\bibnamefont {Vuleti{\'c}}},\ }\href {\doibase
  10.1126/science.aan5614} {\bibfield  {journal} {\bibinfo  {journal}
  {Science}\ }\textbf {\bibinfo {volume} {358}},\ \bibinfo {pages} {1078}
  (\bibinfo {year} {2017})},\ \Eprint
  {http://arxiv.org/abs/https://science.sciencemag.org/content/358/6366/1078.full.pdf}
  {https://science.sciencemag.org/content/358/6366/1078.full.pdf} \BibitemShut
  {NoStop}%
\bibitem [{\citenamefont {Aidelsburger}\ \emph {et~al.}(2011)\citenamefont
  {Aidelsburger}, \citenamefont {Atala}, \citenamefont {Nascimb\`ene},
  \citenamefont {Trotzky}, \citenamefont {Chen},\ and\ \citenamefont
  {Bloch}}]{Bloch_StrongEffectiveMagneticFields}%
  \BibitemOpen
  \bibfield  {author} {\bibinfo {author} {\bibfnamefont {M.}~\bibnamefont
  {Aidelsburger}}, \bibinfo {author} {\bibfnamefont {M.}~\bibnamefont {Atala}},
  \bibinfo {author} {\bibfnamefont {S.}~\bibnamefont {Nascimb\`ene}}, \bibinfo
  {author} {\bibfnamefont {S.}~\bibnamefont {Trotzky}}, \bibinfo {author}
  {\bibfnamefont {Y.-A.}\ \bibnamefont {Chen}}, \ and\ \bibinfo {author}
  {\bibfnamefont {I.}~\bibnamefont {Bloch}},\ }\href {\doibase
  10.1103/PhysRevLett.107.255301} {\bibfield  {journal} {\bibinfo  {journal}
  {Phys. Rev. Lett.}\ }\textbf {\bibinfo {volume} {107}},\ \bibinfo {pages}
  {255301} (\bibinfo {year} {2011})}\BibitemShut {NoStop}%
\bibitem [{\citenamefont {Dalibard}(1999)}]{Swave_CollisionDalibard}%
  \BibitemOpen
  \bibfield  {author} {\bibinfo {author} {\bibfnamefont {J.}~\bibnamefont
  {Dalibard}},\ }in\ \href@noop {} {\emph {\bibinfo {booktitle} {Proceedings of
  the International School of Physics-Enrico Fermi}}},\ Vol.\ \bibinfo {volume}
  {321}\ (\bibinfo {year} {1999})\ p.~\bibinfo {pages} {14}\BibitemShut
  {NoStop}%
\bibitem [{\citenamefont {Lemke}\ \emph {et~al.}(2011)\citenamefont {Lemke},
  \citenamefont {von Stecher}, \citenamefont {Sherman}, \citenamefont {Rey},
  \citenamefont {Oates},\ and\ \citenamefont {Ludlow}}]{Pwave_intAtomicClock}%
  \BibitemOpen
  \bibfield  {author} {\bibinfo {author} {\bibfnamefont {N.~D.}\ \bibnamefont
  {Lemke}}, \bibinfo {author} {\bibfnamefont {J.}~\bibnamefont {von Stecher}},
  \bibinfo {author} {\bibfnamefont {J.~A.}\ \bibnamefont {Sherman}}, \bibinfo
  {author} {\bibfnamefont {A.~M.}\ \bibnamefont {Rey}}, \bibinfo {author}
  {\bibfnamefont {C.~W.}\ \bibnamefont {Oates}}, \ and\ \bibinfo {author}
  {\bibfnamefont {A.~D.}\ \bibnamefont {Ludlow}},\ }\href {\doibase
  10.1103/physrevlett.107.103902} {\bibfield  {journal} {\bibinfo  {journal}
  {Physical Review Letters}\ }\textbf {\bibinfo {volume} {107}} (\bibinfo
  {year} {2011}),\ 10.1103/physrevlett.107.103902}\BibitemShut {NoStop}%
\bibitem [{\citenamefont {{Chalopin}}\ \emph {et~al.}(2020)\citenamefont
  {{Chalopin}}, \citenamefont {{Satoor}}, \citenamefont {{Evrard}},
  \citenamefont {{Makhalov}}, \citenamefont {{Dalibard}}, \citenamefont
  {{Lopes}},\ and\ \citenamefont
  {{Nascimbene}}}]{DalibardNascimbene_HallBulkSyntheticDimension}%
  \BibitemOpen
  \bibfield  {author} {\bibinfo {author} {\bibfnamefont {T.}~\bibnamefont
  {{Chalopin}}}, \bibinfo {author} {\bibfnamefont {T.}~\bibnamefont
  {{Satoor}}}, \bibinfo {author} {\bibfnamefont {A.}~\bibnamefont {{Evrard}}},
  \bibinfo {author} {\bibfnamefont {V.}~\bibnamefont {{Makhalov}}}, \bibinfo
  {author} {\bibfnamefont {J.}~\bibnamefont {{Dalibard}}}, \bibinfo {author}
  {\bibfnamefont {R.}~\bibnamefont {{Lopes}}}, \ and\ \bibinfo {author}
  {\bibfnamefont {S.}~\bibnamefont {{Nascimbene}}},\ }\href@noop {} {\bibfield
  {journal} {\bibinfo  {journal} {arXiv e-prints}\ ,\ \bibinfo {eid}
  {arXiv:2001.01664}} (\bibinfo {year} {2020})},\ \Eprint
  {http://arxiv.org/abs/2001.01664} {arXiv:2001.01664 [cond-mat.quant-gas]}
  \BibitemShut {NoStop}%
\bibitem [{\citenamefont {Mancini}\ \emph {et~al.}(2015)\citenamefont
  {Mancini}, \citenamefont {Pagano}, \citenamefont {Cappellini}, \citenamefont
  {Livi}, \citenamefont {Rider}, \citenamefont {Catani}, \citenamefont {Sias},
  \citenamefont {Zoller}, \citenamefont {Inguscio}, \citenamefont {Dalmonte}
  \emph {et~al.}}]{Mancini_ObservationChiralEdgestate}%
  \BibitemOpen
  \bibfield  {author} {\bibinfo {author} {\bibfnamefont {M.}~\bibnamefont
  {Mancini}}, \bibinfo {author} {\bibfnamefont {G.}~\bibnamefont {Pagano}},
  \bibinfo {author} {\bibfnamefont {G.}~\bibnamefont {Cappellini}}, \bibinfo
  {author} {\bibfnamefont {L.}~\bibnamefont {Livi}}, \bibinfo {author}
  {\bibfnamefont {M.}~\bibnamefont {Rider}}, \bibinfo {author} {\bibfnamefont
  {J.}~\bibnamefont {Catani}}, \bibinfo {author} {\bibfnamefont
  {C.}~\bibnamefont {Sias}}, \bibinfo {author} {\bibfnamefont {P.}~\bibnamefont
  {Zoller}}, \bibinfo {author} {\bibfnamefont {M.}~\bibnamefont {Inguscio}},
  \bibinfo {author} {\bibfnamefont {M.}~\bibnamefont {Dalmonte}},  \emph
  {et~al.},\ }\href@noop {} {\bibfield  {journal} {\bibinfo  {journal}
  {Science}\ }\textbf {\bibinfo {volume} {349}},\ \bibinfo {pages} {1510}
  (\bibinfo {year} {2015})}\BibitemShut {NoStop}%
\bibitem [{\citenamefont {Calvanese~Strinati}\ \emph
  {et~al.}(2017)\citenamefont {Calvanese~Strinati}, \citenamefont {Cornfeld},
  \citenamefont {Rossini}, \citenamefont {Barbarino}, \citenamefont {Dalmonte},
  \citenamefont {Fazio}, \citenamefont {Sela},\ and\ \citenamefont
  {Mazza}}]{Calvanese_Strinati_2017}%
  \BibitemOpen
  \bibfield  {author} {\bibinfo {author} {\bibfnamefont {M.}~\bibnamefont
  {Calvanese~Strinati}}, \bibinfo {author} {\bibfnamefont {E.}~\bibnamefont
  {Cornfeld}}, \bibinfo {author} {\bibfnamefont {D.}~\bibnamefont {Rossini}},
  \bibinfo {author} {\bibfnamefont {S.}~\bibnamefont {Barbarino}}, \bibinfo
  {author} {\bibfnamefont {M.}~\bibnamefont {Dalmonte}}, \bibinfo {author}
  {\bibfnamefont {R.}~\bibnamefont {Fazio}}, \bibinfo {author} {\bibfnamefont
  {E.}~\bibnamefont {Sela}}, \ and\ \bibinfo {author} {\bibfnamefont
  {L.}~\bibnamefont {Mazza}},\ }\href {\doibase 10.1103/physrevx.7.021033}
  {\bibfield  {journal} {\bibinfo  {journal} {Physical Review X}\ }\textbf
  {\bibinfo {volume} {7}} (\bibinfo {year} {2017}),\
  10.1103/physrevx.7.021033}\BibitemShut {NoStop}%
\bibitem [{\citenamefont {Taddia}\ \emph {et~al.}(2017)\citenamefont {Taddia},
  \citenamefont {Cornfeld}, \citenamefont {Rossini}, \citenamefont {Mazza},
  \citenamefont {Sela},\ and\ \citenamefont {Fazio}}]{Taddia_2017}%
  \BibitemOpen
  \bibfield  {author} {\bibinfo {author} {\bibfnamefont {L.}~\bibnamefont
  {Taddia}}, \bibinfo {author} {\bibfnamefont {E.}~\bibnamefont {Cornfeld}},
  \bibinfo {author} {\bibfnamefont {D.}~\bibnamefont {Rossini}}, \bibinfo
  {author} {\bibfnamefont {L.}~\bibnamefont {Mazza}}, \bibinfo {author}
  {\bibfnamefont {E.}~\bibnamefont {Sela}}, \ and\ \bibinfo {author}
  {\bibfnamefont {R.}~\bibnamefont {Fazio}},\ }\href {\doibase
  10.1103/physrevlett.118.230402} {\bibfield  {journal} {\bibinfo  {journal}
  {Physical Review Letters}\ }\textbf {\bibinfo {volume} {118}} (\bibinfo
  {year} {2017}),\ 10.1103/physrevlett.118.230402}\BibitemShut {NoStop}%
\bibitem [{\citenamefont {Cornfeld}\ and\ \citenamefont
  {Sela}(2015)}]{CornfeldEran_ChiralCurrents}%
  \BibitemOpen
  \bibfield  {author} {\bibinfo {author} {\bibfnamefont {E.}~\bibnamefont
  {Cornfeld}}\ and\ \bibinfo {author} {\bibfnamefont {E.}~\bibnamefont
  {Sela}},\ }\href@noop {} {\bibfield  {journal} {\bibinfo  {journal} {Physical
  Review B}\ }\textbf {\bibinfo {volume} {92}},\ \bibinfo {pages} {115446}
  (\bibinfo {year} {2015})}\BibitemShut {NoStop}%
\bibitem [{\citenamefont {Greschner}\ \emph {et~al.}(2015)\citenamefont
  {Greschner}, \citenamefont {Piraud}, \citenamefont {Heidrich-Meisner},
  \citenamefont {McCulloch}, \citenamefont {Schollw{\"o}ck},\ and\
  \citenamefont {Vekua}}]{Schollwock_LadderChiralCurrent}%
  \BibitemOpen
  \bibfield  {author} {\bibinfo {author} {\bibfnamefont {S.}~\bibnamefont
  {Greschner}}, \bibinfo {author} {\bibfnamefont {M.}~\bibnamefont {Piraud}},
  \bibinfo {author} {\bibfnamefont {F.}~\bibnamefont {Heidrich-Meisner}},
  \bibinfo {author} {\bibfnamefont {I.}~\bibnamefont {McCulloch}}, \bibinfo
  {author} {\bibfnamefont {U.}~\bibnamefont {Schollw{\"o}ck}}, \ and\ \bibinfo
  {author} {\bibfnamefont {T.}~\bibnamefont {Vekua}},\ }\href@noop {}
  {\bibfield  {journal} {\bibinfo  {journal} {Physical review letters}\
  }\textbf {\bibinfo {volume} {115}},\ \bibinfo {pages} {190402} (\bibinfo
  {year} {2015})}\BibitemShut {NoStop}%
\bibitem [{\citenamefont {Petrescu}\ and\ \citenamefont
  {Le~Hur}(2015)}]{LeHur_Ladders}%
  \BibitemOpen
  \bibfield  {author} {\bibinfo {author} {\bibfnamefont {A.}~\bibnamefont
  {Petrescu}}\ and\ \bibinfo {author} {\bibfnamefont {K.}~\bibnamefont
  {Le~Hur}},\ }\href@noop {} {\bibfield  {journal} {\bibinfo  {journal}
  {Physical Review B}\ }\textbf {\bibinfo {volume} {91}},\ \bibinfo {pages}
  {054520} (\bibinfo {year} {2015})}\BibitemShut {NoStop}%
\bibitem [{\citenamefont {Barbarino}\ \emph {et~al.}(2015)\citenamefont
  {Barbarino}, \citenamefont {Taddia}, \citenamefont {Rossini}, \citenamefont
  {Mazza},\ and\ \citenamefont {Fazio}}]{Mazza_CrystalAndStripePhases}%
  \BibitemOpen
  \bibfield  {author} {\bibinfo {author} {\bibfnamefont {S.}~\bibnamefont
  {Barbarino}}, \bibinfo {author} {\bibfnamefont {L.}~\bibnamefont {Taddia}},
  \bibinfo {author} {\bibfnamefont {D.}~\bibnamefont {Rossini}}, \bibinfo
  {author} {\bibfnamefont {L.}~\bibnamefont {Mazza}}, \ and\ \bibinfo {author}
  {\bibfnamefont {R.}~\bibnamefont {Fazio}},\ }\href {\doibase
  10.1038/ncomms9134} {\bibfield  {journal} {\bibinfo  {journal} {Nature
  Communications}\ }\textbf {\bibinfo {volume} {6}},\ \bibinfo {pages} {8134}
  (\bibinfo {year} {2015})}\BibitemShut {NoStop}%
\bibitem [{\citenamefont {Saito}\ and\ \citenamefont
  {Furukawa}(2017)}]{Takeshi_DevilStaircase}%
  \BibitemOpen
  \bibfield  {author} {\bibinfo {author} {\bibfnamefont {T.~Y.}\ \bibnamefont
  {Saito}}\ and\ \bibinfo {author} {\bibfnamefont {S.}~\bibnamefont
  {Furukawa}},\ }\href {\doibase 10.1103/PhysRevA.95.043613} {\bibfield
  {journal} {\bibinfo  {journal} {Phys. Rev. A}\ }\textbf {\bibinfo {volume}
  {95}},\ \bibinfo {pages} {043613} (\bibinfo {year} {2017})}\BibitemShut
  {NoStop}%
\bibitem [{\citenamefont {Moore}\ and\ \citenamefont
  {Read}(1991)}]{MooreReadCFTCorrelator}%
  \BibitemOpen
  \bibfield  {author} {\bibinfo {author} {\bibfnamefont {G.}~\bibnamefont
  {Moore}}\ and\ \bibinfo {author} {\bibfnamefont {N.}~\bibnamefont {Read}},\
  }\href {\doibase https://doi.org/10.1016/0550-3213(91)90407-O} {\bibfield
  {journal} {\bibinfo  {journal} {Nucl. Phys. B}\ }\textbf {\bibinfo {volume}
  {360}},\ \bibinfo {pages} {362 } (\bibinfo {year} {1991})}\BibitemShut
  {NoStop}%
\bibitem [{\citenamefont {Hansson}\ \emph {et~al.}(2017)\citenamefont
  {Hansson}, \citenamefont {Hermanns}, \citenamefont {Simon},\ and\
  \citenamefont {Viefers}}]{CFTReviewForFQHE}%
  \BibitemOpen
  \bibfield  {author} {\bibinfo {author} {\bibfnamefont {T.~H.}\ \bibnamefont
  {Hansson}}, \bibinfo {author} {\bibfnamefont {M.}~\bibnamefont {Hermanns}},
  \bibinfo {author} {\bibfnamefont {S.~H.}\ \bibnamefont {Simon}}, \ and\
  \bibinfo {author} {\bibfnamefont {S.~F.}\ \bibnamefont {Viefers}},\ }\href
  {\doibase 10.1103/RevModPhys.89.025005} {\bibfield  {journal} {\bibinfo
  {journal} {Rev. Mod. Phys.}\ }\textbf {\bibinfo {volume} {89}},\ \bibinfo
  {pages} {025005} (\bibinfo {year} {2017})}\BibitemShut {NoStop}%
\bibitem [{\citenamefont {Haldane}(1990)}]{HaldaneBookQuantumHall}%
  \BibitemOpen
  \bibfield  {author} {\bibinfo {author} {\bibfnamefont {F.~D.~M.}\
  \bibnamefont {Haldane}},\ }\href {\doibase 10.1007/978-1-4612-3350-3} {\emph
  {\bibinfo {title} {The Quantum Hall Effect}}},\ \bibinfo {edition} {3rd}\
  ed.\ (\bibinfo  {publisher} {Springer-Verlag New York},\ \bibinfo {year}
  {1990})\ \bibinfo {note} {an optional note}\BibitemShut {NoStop}%
\bibitem [{\citenamefont {Lee}\ \emph {et~al.}(2015)\citenamefont {Lee},
  \citenamefont {Papi\ifmmode~\acute{c}\else \'{c}\fi{}},\ and\ \citenamefont
  {Thomale}}]{PseudoPotentialsPapicThomale}%
  \BibitemOpen
  \bibfield  {author} {\bibinfo {author} {\bibfnamefont {C.~H.}\ \bibnamefont
  {Lee}}, \bibinfo {author} {\bibfnamefont {Z.}~\bibnamefont
  {Papi\ifmmode~\acute{c}\else \'{c}\fi{}}}, \ and\ \bibinfo {author}
  {\bibfnamefont {R.}~\bibnamefont {Thomale}},\ }\href {\doibase
  10.1103/PhysRevX.5.041003} {\bibfield  {journal} {\bibinfo  {journal} {Phys.
  Rev. X}\ }\textbf {\bibinfo {volume} {5}},\ \bibinfo {pages} {041003}
  (\bibinfo {year} {2015})}\BibitemShut {NoStop}%
\bibitem [{\citenamefont {Ortiz}\ \emph {et~al.}(2013)\citenamefont {Ortiz},
  \citenamefont {Nussinov}, \citenamefont {Dukelsky},\ and\ \citenamefont
  {Seidel}}]{PseudoPotentialsTwoBody}%
  \BibitemOpen
  \bibfield  {author} {\bibinfo {author} {\bibfnamefont {G.}~\bibnamefont
  {Ortiz}}, \bibinfo {author} {\bibfnamefont {Z.}~\bibnamefont {Nussinov}},
  \bibinfo {author} {\bibfnamefont {J.}~\bibnamefont {Dukelsky}}, \ and\
  \bibinfo {author} {\bibfnamefont {A.}~\bibnamefont {Seidel}},\ }\href
  {\doibase 10.1103/PhysRevB.88.165303} {\bibfield  {journal} {\bibinfo
  {journal} {Phys. Rev. B}\ }\textbf {\bibinfo {volume} {88}},\ \bibinfo
  {pages} {165303} (\bibinfo {year} {2013})}\BibitemShut {NoStop}%
\bibitem [{\citenamefont {Simon}\ \emph
  {et~al.}(2007{\natexlab{a}})\citenamefont {Simon}, \citenamefont {Rezayi},\
  and\ \citenamefont {Cooper}}]{Pseudopot_Generalized_Simon}%
  \BibitemOpen
  \bibfield  {author} {\bibinfo {author} {\bibfnamefont {S.~H.}\ \bibnamefont
  {Simon}}, \bibinfo {author} {\bibfnamefont {E.~H.}\ \bibnamefont {Rezayi}}, \
  and\ \bibinfo {author} {\bibfnamefont {N.~R.}\ \bibnamefont {Cooper}},\
  }\href {\doibase 10.1103/PhysRevB.75.075318} {\bibfield  {journal} {\bibinfo
  {journal} {Phys. Rev. B}\ }\textbf {\bibinfo {volume} {75}},\ \bibinfo
  {pages} {075318} (\bibinfo {year} {2007}{\natexlab{a}})}\BibitemShut
  {NoStop}%
\bibitem [{\citenamefont {Simon}\ \emph
  {et~al.}(2007{\natexlab{b}})\citenamefont {Simon}, \citenamefont {Rezayi},\
  and\ \citenamefont {Cooper}}]{Pseudopot_Simon}%
  \BibitemOpen
  \bibfield  {author} {\bibinfo {author} {\bibfnamefont {S.~H.}\ \bibnamefont
  {Simon}}, \bibinfo {author} {\bibfnamefont {E.~H.}\ \bibnamefont {Rezayi}}, \
  and\ \bibinfo {author} {\bibfnamefont {N.~R.}\ \bibnamefont {Cooper}},\
  }\href {\doibase 10.1103/PhysRevB.75.195306} {\bibfield  {journal} {\bibinfo
  {journal} {Phys. Rev. B}\ }\textbf {\bibinfo {volume} {75}},\ \bibinfo
  {pages} {195306} (\bibinfo {year} {2007}{\natexlab{b}})}\BibitemShut
  {NoStop}%
\bibitem [{\citenamefont
  {Laughlin}(1981)}]{Laughlin_QuantizedConductanceLaughlinArgument}%
  \BibitemOpen
  \bibfield  {author} {\bibinfo {author} {\bibfnamefont {R.~B.}\ \bibnamefont
  {Laughlin}},\ }\href {\doibase 10.1103/PhysRevB.23.5632} {\bibfield
  {journal} {\bibinfo  {journal} {Phys. Rev. B}\ }\textbf {\bibinfo {volume}
  {23}},\ \bibinfo {pages} {5632} (\bibinfo {year} {1981})}\BibitemShut
  {NoStop}%
\bibitem [{{\relax DLMF}()}]{NIST:DLMF}%
  \BibitemOpen
  {\relax DLMF},\ \href {http://dlmf.nist.gov/28} {\enquote {\bibinfo {title}
  {{\it NIST Digital Library of Mathematical Functions}},}\ }\bibinfo
  {howpublished} {http://dlmf.nist.gov/, Release 1.0.18 of 2018-03-27},\
  \bibinfo {note} {f.~W.~J. Olver, A.~B. {Olde Daalhuis}, D.~W. Lozier, B.~I.
  Schneider, R.~F. Boisvert, C.~W. Clark, B.~R. Miller and B.~V. Saunders,
  eds.}\BibitemShut {Stop}%
\bibitem [{\citenamefont {Abramov}\ and\ \citenamefont
  {Kurochkin}(2007)}]{ContinuedFractionMathieuCharacteristic}%
  \BibitemOpen
  \bibfield  {author} {\bibinfo {author} {\bibfnamefont {A.~A.}\ \bibnamefont
  {Abramov}}\ and\ \bibinfo {author} {\bibfnamefont {S.~V.}\ \bibnamefont
  {Kurochkin}},\ }\href {\doibase 10.1134/S0965542507030050} {\bibfield
  {journal} {\bibinfo  {journal} {Computational Mathematics and Mathematical
  Physics}\ }\textbf {\bibinfo {volume} {47}},\ \bibinfo {pages} {397}
  (\bibinfo {year} {2007})}\BibitemShut {NoStop}%
\bibitem [{\citenamefont {Meixner}\ \emph {et~al.}(1980)\citenamefont
  {Meixner}, \citenamefont {Sch{\"a}fke},\ and\ \citenamefont
  {Wolf}}]{MeixnerMathieuLectureNotes}%
  \BibitemOpen
  \bibfield  {author} {\bibinfo {author} {\bibfnamefont {J.}~\bibnamefont
  {Meixner}}, \bibinfo {author} {\bibfnamefont {F.~W.}\ \bibnamefont
  {Sch{\"a}fke}}, \ and\ \bibinfo {author} {\bibfnamefont {G.}~\bibnamefont
  {Wolf}},\ }\href@noop {} {\emph {\bibinfo {title} {{M}athieu Functions and
  Spheroidal Functions and Their Mathematical Foundations: Further Studies}}},\
  \bibinfo {series} {Lecture Notes in Mathematics}, Vol.\ \bibinfo {volume}
  {837}\ (\bibinfo  {publisher} {Springer-Verlag},\ \bibinfo {address}
  {Berlin-New York},\ \bibinfo {year} {1980})\ pp.\ \bibinfo {pages}
  {vii+126}\BibitemShut {NoStop}%
\bibitem [{\citenamefont {Kurz}(1979)}]{KurzPhDApproximationMathieuWF}%
  \BibitemOpen
  \bibfield  {author} {\bibinfo {author} {\bibfnamefont {M.}~\bibnamefont
  {Kurz}},\ }\emph {\bibinfo {title} {{F}ehlerabsch{\"a}tzungen zu
  asymptotischen {E}ntwicklungen der {E}igenwerte und {E}igenl{\"o}sungen der
  {M}athieuschen {D}ifferentialgleichung}},\ \href@noop {} {Ph.D. thesis},\
  \bibinfo  {school} {Universit{\"a}t Duisburg-Essen}, \bibinfo {address}
  {Essen, D 45177} (\bibinfo {year} {1979})\BibitemShut {NoStop}%
\bibitem [{\citenamefont {Goldstein}(1930)}]{GoldsteinCharacteristicFunction}%
  \BibitemOpen
  \bibfield  {author} {\bibinfo {author} {\bibfnamefont {S.}~\bibnamefont
  {Goldstein}},\ }\href {\doibase 10.1017/S0370164600026407} {\bibfield
  {journal} {\bibinfo  {journal} {Proceedings of the Royal Society of
  Edinburgh}\ }\textbf {\bibinfo {volume} {49}},\ \bibinfo {pages} {210–223}
  (\bibinfo {year} {1930})}\BibitemShut {NoStop}%
\bibitem [{\citenamefont {Volkmer}(1996)}]{RadiusConvergenceVolkmer}%
  \BibitemOpen
  \bibfield  {author} {\bibinfo {author} {\bibfnamefont {H.}~\bibnamefont
  {Volkmer}},\ }\href {\doibase https://doi.org/10.1006/jdeq.1996.0098}
  {\bibfield  {journal} {\bibinfo  {journal} {Journal of Differential
  Equations}\ }\textbf {\bibinfo {volume} {128}},\ \bibinfo {pages} {327 }
  (\bibinfo {year} {1996})}\BibitemShut {NoStop}%
\bibitem [{\citenamefont {Sips}(1949)}]{SipsApproximationMathieuWF1}%
  \BibitemOpen
  \bibfield  {author} {\bibinfo {author} {\bibfnamefont {R.}~\bibnamefont
  {Sips}},\ }\href@noop {} {\bibfield  {journal} {\bibinfo  {journal} {Trans.
  Amer. Math. Soc.}\ }\textbf {\bibinfo {volume} {66}},\ \bibinfo {pages}
  {93–134} (\bibinfo {year} {1949})}\BibitemShut {NoStop}%
\bibitem [{\citenamefont {Sips}(1959)}]{SipsApproximationMathieuWF2}%
  \BibitemOpen
  \bibfield  {author} {\bibinfo {author} {\bibfnamefont {R.}~\bibnamefont
  {Sips}},\ }\href@noop {} {\bibfield  {journal} {\bibinfo  {journal} {Trans.
  Amer. Math. Soc.}\ }\textbf {\bibinfo {volume} {90}},\ \bibinfo {pages}
  {340–368} (\bibinfo {year} {1959})}\BibitemShut {NoStop}%
\bibitem [{\citenamefont {Connor}\ \emph {et~al.}(1984)\citenamefont {Connor},
  \citenamefont {Uzer}, \citenamefont {Marcus},\ and\ \citenamefont
  {Smith}}]{SemiclassicalWKBConnorSmith}%
  \BibitemOpen
  \bibfield  {author} {\bibinfo {author} {\bibfnamefont {J.~N.~L.}\
  \bibnamefont {Connor}}, \bibinfo {author} {\bibfnamefont {T.}~\bibnamefont
  {Uzer}}, \bibinfo {author} {\bibfnamefont {R.~A.}\ \bibnamefont {Marcus}}, \
  and\ \bibinfo {author} {\bibfnamefont {A.~D.}\ \bibnamefont {Smith}},\ }\href
  {\doibase 10.1063/1.446581} {\bibfield  {journal} {\bibinfo  {journal} {The
  Journal of Chemical Physics}\ }\textbf {\bibinfo {volume} {80}},\ \bibinfo
  {pages} {5095} (\bibinfo {year} {1984})}\BibitemShut {NoStop}%
\bibitem [{\citenamefont {Aunola}(2003)}]{AunolaDiscreteHarmonicOscillator}%
  \BibitemOpen
  \bibfield  {author} {\bibinfo {author} {\bibfnamefont {M.}~\bibnamefont
  {Aunola}},\ }\href {\doibase 10.1063/1.1561156} {\bibfield  {journal}
  {\bibinfo  {journal} {Journal of Mathematical Physics}\ }\textbf {\bibinfo
  {volume} {44}},\ \bibinfo {pages} {1913} (\bibinfo {year} {2003})},\ \Eprint
  {http://arxiv.org/abs/https://doi.org/10.1063/1.1561156}
  {https://doi.org/10.1063/1.1561156} \BibitemShut {NoStop}%
\bibitem [{\citenamefont {Haldane}(2011)}]{Haldane_GeometricDescription}%
  \BibitemOpen
  \bibfield  {author} {\bibinfo {author} {\bibfnamefont {F.~D.~M.}\
  \bibnamefont {Haldane}},\ }\href {\doibase 10.1103/PhysRevLett.107.116801}
  {\bibfield  {journal} {\bibinfo  {journal} {Phys. Rev. Lett.}\ }\textbf
  {\bibinfo {volume} {107}},\ \bibinfo {pages} {116801} (\bibinfo {year}
  {2011})}\BibitemShut {NoStop}%
\bibitem [{\citenamefont {Johri}\ \emph {et~al.}(2016)\citenamefont {Johri},
  \citenamefont {Papic}, \citenamefont {Schmitteckert}, \citenamefont {Bhatt},\
  and\ \citenamefont {Haldane}}]{GeometryLaughlin_PapicHaldane}%
  \BibitemOpen
  \bibfield  {author} {\bibinfo {author} {\bibfnamefont {S.}~\bibnamefont
  {Johri}}, \bibinfo {author} {\bibfnamefont {Z.}~\bibnamefont {Papic}},
  \bibinfo {author} {\bibfnamefont {P.}~\bibnamefont {Schmitteckert}}, \bibinfo
  {author} {\bibfnamefont {R.~N.}\ \bibnamefont {Bhatt}}, \ and\ \bibinfo
  {author} {\bibfnamefont {F.~D.~M.}\ \bibnamefont {Haldane}},\ }\href
  {\doibase 10.1088/1367-2630/18/2/025011} {\bibfield  {journal} {\bibinfo
  {journal} {New Journal of Physics}\ }\textbf {\bibinfo {volume} {18}},\
  \bibinfo {pages} {025011} (\bibinfo {year} {2016})}\BibitemShut {NoStop}%
\bibitem [{\citenamefont {Yang}\ \emph {et~al.}(2012)\citenamefont {Yang},
  \citenamefont {Papic}, \citenamefont {Rezayi}, \citenamefont {Bhatt},\ and\
  \citenamefont {Haldane}}]{BandMassAnisotropy_PapicHaldane}%
  \BibitemOpen
  \bibfield  {author} {\bibinfo {author} {\bibfnamefont {B.}~\bibnamefont
  {Yang}}, \bibinfo {author} {\bibfnamefont {Z.}~\bibnamefont {Papic}},
  \bibinfo {author} {\bibfnamefont {E.~H.}\ \bibnamefont {Rezayi}}, \bibinfo
  {author} {\bibfnamefont {R.~N.}\ \bibnamefont {Bhatt}}, \ and\ \bibinfo
  {author} {\bibfnamefont {F.~D.~M.}\ \bibnamefont {Haldane}},\ }\href
  {\doibase 10.1103/PhysRevB.85.165318} {\bibfield  {journal} {\bibinfo
  {journal} {Phys. Rev. B}\ }\textbf {\bibinfo {volume} {85}},\ \bibinfo
  {pages} {165318} (\bibinfo {year} {2012})}\BibitemShut {NoStop}%
\bibitem [{Pol(2017)}]{Polarization_ChernLandauLevel_Moore}%
  \BibitemOpen
  \href@noop {} {\emph {\bibinfo {title} {Topological Aspects of Condensed
  Matter Physics: Lecture Notes of the Les Houches Summer School: Volume 103,
  August 2014}}}\ (\bibinfo  {publisher} {Oxford University Press},\ \bibinfo
  {address} {Oxford},\ \bibinfo {year} {2017})\ p.\ \bibinfo {pages}
  {704}\BibitemShut {NoStop}%
\bibitem [{\citenamefont {Resta}(1994)}]{Polarization_RestaModernTheory}%
  \BibitemOpen
  \bibfield  {author} {\bibinfo {author} {\bibfnamefont {R.}~\bibnamefont
  {Resta}},\ }\href {\doibase 10.1080/00150199408244722} {\bibfield  {journal}
  {\bibinfo  {journal} {Ferroelectrics}\ }\textbf {\bibinfo {volume} {151}},\
  \bibinfo {pages} {49} (\bibinfo {year} {1994})},\ \Eprint
  {http://arxiv.org/abs/https://doi.org/10.1080/00150199408244722}
  {https://doi.org/10.1080/00150199408244722} \BibitemShut {NoStop}%
\bibitem [{\citenamefont {Martin}\ and\ \citenamefont
  {Ortiz}(1997)}]{Polarization_MartinOrtiz}%
  \BibitemOpen
  \bibfield  {author} {\bibinfo {author} {\bibfnamefont {R.~M.}\ \bibnamefont
  {Martin}}\ and\ \bibinfo {author} {\bibfnamefont {G.}~\bibnamefont {Ortiz}},\
  }\href {\doibase https://doi.org/10.1016/S0038-1098(96)00719-3} {\bibfield
  {journal} {\bibinfo  {journal} {Solid State Communications}\ }\textbf
  {\bibinfo {volume} {102}},\ \bibinfo {pages} {121 } (\bibinfo {year}
  {1997})},\ \bibinfo {note} {highlights in Condensed Matter Physics and
  Materials Science}\BibitemShut {NoStop}%
\bibitem [{\citenamefont {Vanderbilt}\ and\ \citenamefont
  {King-Smith}(1993)}]{Polarization_Vanderbilt}%
  \BibitemOpen
  \bibfield  {author} {\bibinfo {author} {\bibfnamefont {D.}~\bibnamefont
  {Vanderbilt}}\ and\ \bibinfo {author} {\bibfnamefont {R.~D.}\ \bibnamefont
  {King-Smith}},\ }\href {\doibase 10.1103/PhysRevB.48.4442} {\bibfield
  {journal} {\bibinfo  {journal} {Phys. Rev. B}\ }\textbf {\bibinfo {volume}
  {48}},\ \bibinfo {pages} {4442} (\bibinfo {year} {1993})}\BibitemShut
  {NoStop}%
\bibitem [{\citenamefont {King-Smith}\ and\ \citenamefont
  {Vanderbilt}(1993)}]{Polarization_KingSmith}%
  \BibitemOpen
  \bibfield  {author} {\bibinfo {author} {\bibfnamefont {R.~D.}\ \bibnamefont
  {King-Smith}}\ and\ \bibinfo {author} {\bibfnamefont {D.}~\bibnamefont
  {Vanderbilt}},\ }\href {\doibase 10.1103/PhysRevB.47.1651} {\bibfield
  {journal} {\bibinfo  {journal} {Phys. Rev. B}\ }\textbf {\bibinfo {volume}
  {47}},\ \bibinfo {pages} {1651} (\bibinfo {year} {1993})}\BibitemShut
  {NoStop}%
\bibitem [{\citenamefont
  {Haldane}(1983{\natexlab{b}})}]{Haldane_HierarchiesPseudoPot}%
  \BibitemOpen
  \bibfield  {author} {\bibinfo {author} {\bibfnamefont {F.~D.~M.}\
  \bibnamefont {Haldane}},\ }\href {\doibase 10.1103/PhysRevLett.51.605}
  {\bibfield  {journal} {\bibinfo  {journal} {Phys. Rev. Lett.}\ }\textbf
  {\bibinfo {volume} {51}},\ \bibinfo {pages} {605} (\bibinfo {year}
  {1983}{\natexlab{b}})}\BibitemShut {NoStop}%
\bibitem [{\citenamefont {Francesco}\ \emph {et~al.}(1997)\citenamefont
  {Francesco}, \citenamefont {Mathieu},\ and\ \citenamefont
  {S\'en\'echal}}]{YellowBook}%
  \BibitemOpen
  \bibfield  {author} {\bibinfo {author} {\bibfnamefont {P.~D.}\ \bibnamefont
  {Francesco}}, \bibinfo {author} {\bibfnamefont {P.}~\bibnamefont {Mathieu}},
  \ and\ \bibinfo {author} {\bibfnamefont {D.}~\bibnamefont {S\'en\'echal}},\
  }\href@noop {} {\emph {\bibinfo {title} {Conformal Field Theory}}}\ (\bibinfo
   {publisher} {Springer-Verlag New York},\ \bibinfo {year} {1997})\BibitemShut
  {NoStop}%
\bibitem [{\citenamefont
  {Bogolyubov}(1947)}]{Bogoliubov_TheorySuperconductivity}%
  \BibitemOpen
  \bibfield  {author} {\bibinfo {author} {\bibfnamefont {N.}~\bibnamefont
  {Bogolyubov}},\ }\href@noop {} {\bibfield  {journal} {\bibinfo  {journal}
  {J.\ Phys.\ (USSR)}\ }\textbf {\bibinfo {volume} {11}},\ \bibinfo {pages}
  {23} (\bibinfo {year} {1947})}\BibitemShut {NoStop}%
\bibitem [{\citenamefont
  {Miller}(1979)}]{SemicalssicalDoubleWellPotentialTunneling}%
  \BibitemOpen
  \bibfield  {author} {\bibinfo {author} {\bibfnamefont {W.~H.}\ \bibnamefont
  {Miller}},\ }\href {\doibase 10.1021/j100471a015} {\bibfield  {journal}
  {\bibinfo  {journal} {The Journal of Physical Chemistry}\ }\textbf {\bibinfo
  {volume} {83}},\ \bibinfo {pages} {960} (\bibinfo {year} {1979})}\BibitemShut
  {NoStop}%
\bibitem [{\citenamefont {Deutsch}\ and\ \citenamefont
  {Jessen}(1998)}]{DeutschJessen_ApproximationAndDiabaticPotentials}%
  \BibitemOpen
  \bibfield  {author} {\bibinfo {author} {\bibfnamefont {I.~H.}\ \bibnamefont
  {Deutsch}}\ and\ \bibinfo {author} {\bibfnamefont {P.~S.}\ \bibnamefont
  {Jessen}},\ }\href {\doibase 10.1103/PhysRevA.57.1972} {\bibfield  {journal}
  {\bibinfo  {journal} {Phys. Rev. A}\ }\textbf {\bibinfo {volume} {57}},\
  \bibinfo {pages} {1972} (\bibinfo {year} {1998})}\BibitemShut {NoStop}%
\bibitem [{\citenamefont {Xu}\ \emph {et~al.}(2018)\citenamefont {Xu},
  \citenamefont {Morong}, \citenamefont {Hui}, \citenamefont {Scarola},\ and\
  \citenamefont {DeMarco}}]{SpinFlipTunneling40K}%
  \BibitemOpen
  \bibfield  {author} {\bibinfo {author} {\bibfnamefont {W.}~\bibnamefont
  {Xu}}, \bibinfo {author} {\bibfnamefont {W.}~\bibnamefont {Morong}}, \bibinfo
  {author} {\bibfnamefont {H.-Y.}\ \bibnamefont {Hui}}, \bibinfo {author}
  {\bibfnamefont {V.~W.}\ \bibnamefont {Scarola}}, \ and\ \bibinfo {author}
  {\bibfnamefont {B.}~\bibnamefont {DeMarco}},\ }\href {\doibase
  10.1103/PhysRevA.98.023623} {\bibfield  {journal} {\bibinfo  {journal} {Phys.
  Rev. A}\ }\textbf {\bibinfo {volume} {98}},\ \bibinfo {pages} {023623}
  (\bibinfo {year} {2018})}\BibitemShut {NoStop}%
\bibitem [{\citenamefont {Le~Kien}\ \emph {et~al.}(2013)\citenamefont
  {Le~Kien}, \citenamefont {Schneeweiss},\ and\ \citenamefont
  {Rauschenbeutel}}]{CesiumExample_AllPolaribilities}%
  \BibitemOpen
  \bibfield  {author} {\bibinfo {author} {\bibfnamefont {F.}~\bibnamefont
  {Le~Kien}}, \bibinfo {author} {\bibfnamefont {P.}~\bibnamefont
  {Schneeweiss}}, \ and\ \bibinfo {author} {\bibfnamefont {A.}~\bibnamefont
  {Rauschenbeutel}},\ }\href {\doibase 10.1140/epjd/e2013-30729-x} {\bibfield
  {journal} {\bibinfo  {journal} {The European Physical Journal D}\ }\textbf
  {\bibinfo {volume} {67}},\ \bibinfo {pages} {92} (\bibinfo {year}
  {2013})}\BibitemShut {NoStop}%
\bibitem [{\citenamefont {Finkelstein}\ \emph {et~al.}(1992)\citenamefont
  {Finkelstein}, \citenamefont {Berman},\ and\ \citenamefont
  {Guo}}]{Guo_BelowDopplerCooling}%
  \BibitemOpen
  \bibfield  {author} {\bibinfo {author} {\bibfnamefont {V.}~\bibnamefont
  {Finkelstein}}, \bibinfo {author} {\bibfnamefont {P.~R.}\ \bibnamefont
  {Berman}}, \ and\ \bibinfo {author} {\bibfnamefont {J.}~\bibnamefont {Guo}},\
  }\href {\doibase 10.1103/PhysRevA.45.1829} {\bibfield  {journal} {\bibinfo
  {journal} {Phys. Rev. A}\ }\textbf {\bibinfo {volume} {45}},\ \bibinfo
  {pages} {1829} (\bibinfo {year} {1992})}\BibitemShut {NoStop}%
\bibitem [{\citenamefont {Ta\"{\i}eb}\ \emph {et~al.}(1993)\citenamefont
  {Ta\"{\i}eb}, \citenamefont {Marte}, \citenamefont {Dum},\ and\ \citenamefont
  {Zoller}}]{Zoller_BelowDopplerCooling}%
  \BibitemOpen
  \bibfield  {author} {\bibinfo {author} {\bibfnamefont {R.}~\bibnamefont
  {Ta\"{\i}eb}}, \bibinfo {author} {\bibfnamefont {P.}~\bibnamefont {Marte}},
  \bibinfo {author} {\bibfnamefont {R.}~\bibnamefont {Dum}}, \ and\ \bibinfo
  {author} {\bibfnamefont {P.}~\bibnamefont {Zoller}},\ }\href {\doibase
  10.1103/PhysRevA.47.4986} {\bibfield  {journal} {\bibinfo  {journal} {Phys.
  Rev. A}\ }\textbf {\bibinfo {volume} {47}},\ \bibinfo {pages} {4986}
  (\bibinfo {year} {1993})}\BibitemShut {NoStop}%
\bibitem [{\citenamefont {Cohen-Tannoudji}\ \emph {et~al.}(1992)\citenamefont
  {Cohen-Tannoudji}, \citenamefont {Dupont-Roc},\ and\ \citenamefont
  {Grynberg}}]{CohenTanoudji_AtomPhoton}%
  \BibitemOpen
  \bibfield  {author} {\bibinfo {author} {\bibfnamefont {C.}~\bibnamefont
  {Cohen-Tannoudji}}, \bibinfo {author} {\bibfnamefont {J.}~\bibnamefont
  {Dupont-Roc}}, \ and\ \bibinfo {author} {\bibfnamefont {G.}~\bibnamefont
  {Grynberg}},\ }\href {https://books.google.com/books?id=m7gPAQAAMAAJ} {\emph
  {\bibinfo {title} {Atom-photon interactions: basic processes and
  applications}}},\ Wiley-Interscience publication\ (\bibinfo  {publisher} {J.
  Wiley},\ \bibinfo {year} {1992})\BibitemShut {NoStop}%
\bibitem [{\citenamefont {Gillen-Christandl}\ and\ \citenamefont
  {Copsey}(2011)}]{TensorPolarizability_FastJtoJprime}%
  \BibitemOpen
  \bibfield  {author} {\bibinfo {author} {\bibfnamefont {K.}~\bibnamefont
  {Gillen-Christandl}}\ and\ \bibinfo {author} {\bibfnamefont {B.~D.}\
  \bibnamefont {Copsey}},\ }\href {\doibase 10.1103/PhysRevA.83.023408}
  {\bibfield  {journal} {\bibinfo  {journal} {Phys. Rev. A}\ }\textbf {\bibinfo
  {volume} {83}},\ \bibinfo {pages} {023408} (\bibinfo {year}
  {2011})}\BibitemShut {NoStop}%
\bibitem [{\citenamefont {Deutsch}\ and\ \citenamefont
  {Jessen}(2010)}]{DeutschJessen_TensorCartesian}%
  \BibitemOpen
  \bibfield  {author} {\bibinfo {author} {\bibfnamefont {I.~H.}\ \bibnamefont
  {Deutsch}}\ and\ \bibinfo {author} {\bibfnamefont {P.~S.}\ \bibnamefont
  {Jessen}},\ }\href {\doibase https://doi.org/10.1016/j.optcom.2009.10.059}
  {\bibfield  {journal} {\bibinfo  {journal} {Optics Communications}\ }\textbf
  {\bibinfo {volume} {283}},\ \bibinfo {pages} {681 } (\bibinfo {year}
  {2010})}\BibitemShut {NoStop}%
\bibitem [{\citenamefont {Geremia}\ \emph {et~al.}(2006)\citenamefont
  {Geremia}, \citenamefont {Stockton},\ and\ \citenamefont
  {Mabuchi}}]{TensorPolarizability_AllDefinitionsAndTensorDecomposition}%
  \BibitemOpen
  \bibfield  {author} {\bibinfo {author} {\bibfnamefont {J.~M.}\ \bibnamefont
  {Geremia}}, \bibinfo {author} {\bibfnamefont {J.~K.}\ \bibnamefont
  {Stockton}}, \ and\ \bibinfo {author} {\bibfnamefont {H.}~\bibnamefont
  {Mabuchi}},\ }\href {\doibase 10.1103/PhysRevA.73.042112} {\bibfield
  {journal} {\bibinfo  {journal} {Phys. Rev. A}\ }\textbf {\bibinfo {volume}
  {73}},\ \bibinfo {pages} {042112} (\bibinfo {year} {2006})}\BibitemShut
  {NoStop}%
\bibitem [{\citenamefont {Campbell}\ \emph {et~al.}(2011)\citenamefont
  {Campbell}, \citenamefont {Juzeli\ifmmode~\bar{u}\else \={u}\fi{}nas},\ and\
  \citenamefont {Spielman}}]{Spielman_RashbaDresselhaus}%
  \BibitemOpen
  \bibfield  {author} {\bibinfo {author} {\bibfnamefont {D.~L.}\ \bibnamefont
  {Campbell}}, \bibinfo {author} {\bibfnamefont {G.}~\bibnamefont
  {Juzeli\ifmmode~\bar{u}\else \={u}\fi{}nas}}, \ and\ \bibinfo {author}
  {\bibfnamefont {I.~B.}\ \bibnamefont {Spielman}},\ }\href {\doibase
  10.1103/PhysRevA.84.025602} {\bibfield  {journal} {\bibinfo  {journal} {Phys.
  Rev. A}\ }\textbf {\bibinfo {volume} {84}},\ \bibinfo {pages} {025602}
  (\bibinfo {year} {2011})}\BibitemShut {NoStop}%
\bibitem [{\citenamefont {Anderson}\ \emph {et~al.}(2020)\citenamefont
  {Anderson}, \citenamefont {Trypogeorgos}, \citenamefont {Vald\'es-Curiel},
  \citenamefont {Liang}, \citenamefont {Tao}, \citenamefont {Zhao},
  \citenamefont {Andrijauskas}, \citenamefont {Juzeli\ifmmode~\bar{u}\else
  \={u}\fi{}nas},\ and\ \citenamefont
  {Spielman}}]{Spielmann_AdiabaticSuperlattice}%
  \BibitemOpen
  \bibfield  {author} {\bibinfo {author} {\bibfnamefont {R.~P.}\ \bibnamefont
  {Anderson}}, \bibinfo {author} {\bibfnamefont {D.}~\bibnamefont
  {Trypogeorgos}}, \bibinfo {author} {\bibfnamefont {A.}~\bibnamefont
  {Vald\'es-Curiel}}, \bibinfo {author} {\bibfnamefont {Q.-Y.}\ \bibnamefont
  {Liang}}, \bibinfo {author} {\bibfnamefont {J.}~\bibnamefont {Tao}}, \bibinfo
  {author} {\bibfnamefont {M.}~\bibnamefont {Zhao}}, \bibinfo {author}
  {\bibfnamefont {T.}~\bibnamefont {Andrijauskas}}, \bibinfo {author}
  {\bibfnamefont {G.}~\bibnamefont {Juzeli\ifmmode~\bar{u}\else
  \={u}\fi{}nas}}, \ and\ \bibinfo {author} {\bibfnamefont {I.~B.}\
  \bibnamefont {Spielman}},\ }\href {\doibase 10.1103/PhysRevResearch.2.013149}
  {\bibfield  {journal} {\bibinfo  {journal} {Phys. Rev. Research}\ }\textbf
  {\bibinfo {volume} {2}},\ \bibinfo {pages} {013149} (\bibinfo {year}
  {2020})}\BibitemShut {NoStop}%
\bibitem [{\citenamefont {Campbell}\ and\ \citenamefont
  {Spielman}(2016)}]{Spielmann_RFRaman}%
  \BibitemOpen
  \bibfield  {author} {\bibinfo {author} {\bibfnamefont {D.~L.}\ \bibnamefont
  {Campbell}}\ and\ \bibinfo {author} {\bibfnamefont {I.~B.}\ \bibnamefont
  {Spielman}},\ }\href {\doibase 10.1088/1367-2630/18/3/033035} {\bibfield
  {journal} {\bibinfo  {journal} {New Journal of Physics}\ }\textbf {\bibinfo
  {volume} {18}},\ \bibinfo {pages} {033035} (\bibinfo {year}
  {2016})}\BibitemShut {NoStop}%
\bibitem [{\citenamefont {Morigi}\ \emph {et~al.}(2000)\citenamefont {Morigi},
  \citenamefont {Eschner},\ and\ \citenamefont {Keitel}}]{EIT_Cooling}%
  \BibitemOpen
  \bibfield  {author} {\bibinfo {author} {\bibfnamefont {G.}~\bibnamefont
  {Morigi}}, \bibinfo {author} {\bibfnamefont {J.}~\bibnamefont {Eschner}}, \
  and\ \bibinfo {author} {\bibfnamefont {C.~H.}\ \bibnamefont {Keitel}},\
  }\href {\doibase 10.1103/PhysRevLett.85.4458} {\bibfield  {journal} {\bibinfo
   {journal} {Phys. Rev. Lett.}\ }\textbf {\bibinfo {volume} {85}},\ \bibinfo
  {pages} {4458} (\bibinfo {year} {2000})}\BibitemShut {NoStop}%
\bibitem [{\citenamefont {Bloch}\ \emph {et~al.}(2008)\citenamefont {Bloch},
  \citenamefont {Dalibard},\ and\ \citenamefont
  {Zwerger}}]{RevModPhys_ManyBodyUltracoldGases}%
  \BibitemOpen
  \bibfield  {author} {\bibinfo {author} {\bibfnamefont {I.}~\bibnamefont
  {Bloch}}, \bibinfo {author} {\bibfnamefont {J.}~\bibnamefont {Dalibard}}, \
  and\ \bibinfo {author} {\bibfnamefont {W.}~\bibnamefont {Zwerger}},\ }\href
  {\doibase 10.1103/RevModPhys.80.885} {\bibfield  {journal} {\bibinfo
  {journal} {Rev. Mod. Phys.}\ }\textbf {\bibinfo {volume} {80}},\ \bibinfo
  {pages} {885} (\bibinfo {year} {2008})}\BibitemShut {NoStop}%
\bibitem [{\citenamefont {Gerbier}\ and\ \citenamefont
  {Dalibard}(2010)}]{Gerbier_OpticalLattices}%
  \BibitemOpen
  \bibfield  {author} {\bibinfo {author} {\bibfnamefont {F.}~\bibnamefont
  {Gerbier}}\ and\ \bibinfo {author} {\bibfnamefont {J.}~\bibnamefont
  {Dalibard}},\ }\href {\doibase 10.1088/1367-2630/12/3/033007} {\bibfield
  {journal} {\bibinfo  {journal} {New Journal of Physics}\ }\textbf {\bibinfo
  {volume} {12}},\ \bibinfo {pages} {033007} (\bibinfo {year}
  {2010})}\BibitemShut {NoStop}%
\bibitem [{\citenamefont {Arzamasovs}\ and\ \citenamefont
  {Liu}(2017)}]{CorrectTightBinding_CosinePotential}%
  \BibitemOpen
  \bibfield  {author} {\bibinfo {author} {\bibfnamefont {M.}~\bibnamefont
  {Arzamasovs}}\ and\ \bibinfo {author} {\bibfnamefont {B.}~\bibnamefont
  {Liu}},\ }\href {\doibase 10.1088/1361-6404/aa8d2c} {\bibfield  {journal}
  {\bibinfo  {journal} {European Journal of Physics}\ }\textbf {\bibinfo
  {volume} {38}},\ \bibinfo {pages} {065405} (\bibinfo {year}
  {2017})}\BibitemShut {NoStop}%
\bibitem [{\citenamefont {Miyake}\ \emph {et~al.}(2013)\citenamefont {Miyake},
  \citenamefont {Siviloglou}, \citenamefont {Kennedy}, \citenamefont {Burton},\
  and\ \citenamefont {Ketterle}}]{Ketterle_RealizeHarperModel}%
  \BibitemOpen
  \bibfield  {author} {\bibinfo {author} {\bibfnamefont {H.}~\bibnamefont
  {Miyake}}, \bibinfo {author} {\bibfnamefont {G.~A.}\ \bibnamefont
  {Siviloglou}}, \bibinfo {author} {\bibfnamefont {C.~J.}\ \bibnamefont
  {Kennedy}}, \bibinfo {author} {\bibfnamefont {W.~C.}\ \bibnamefont {Burton}},
  \ and\ \bibinfo {author} {\bibfnamefont {W.}~\bibnamefont {Ketterle}},\
  }\href {\doibase 10.1103/PhysRevLett.111.185302} {\bibfield  {journal}
  {\bibinfo  {journal} {Phys. Rev. Lett.}\ }\textbf {\bibinfo {volume} {111}},\
  \bibinfo {pages} {185302} (\bibinfo {year} {2013})}\BibitemShut {NoStop}%
\bibitem [{\citenamefont {Kasevich}\ and\ \citenamefont
  {Chu}(1992)}]{SteveChu_LaserCooling}%
  \BibitemOpen
  \bibfield  {author} {\bibinfo {author} {\bibfnamefont {M.}~\bibnamefont
  {Kasevich}}\ and\ \bibinfo {author} {\bibfnamefont {S.}~\bibnamefont {Chu}},\
  }\href {\doibase 10.1103/PhysRevLett.69.1741} {\bibfield  {journal} {\bibinfo
   {journal} {Phys. Rev. Lett.}\ }\textbf {\bibinfo {volume} {69}},\ \bibinfo
  {pages} {1741} (\bibinfo {year} {1992})}\BibitemShut {NoStop}%
\bibitem [{\citenamefont {Cohen-Tannoudji}\ \emph {et~al.}(1973)\citenamefont
  {Cohen-Tannoudji}, \citenamefont {Dupont-Roc},\ and\ \citenamefont
  {Fabre}}]{Cohen_Tannoudji_RWAApprox}%
  \BibitemOpen
  \bibfield  {author} {\bibinfo {author} {\bibfnamefont {C.}~\bibnamefont
  {Cohen-Tannoudji}}, \bibinfo {author} {\bibfnamefont {J.}~\bibnamefont
  {Dupont-Roc}}, \ and\ \bibinfo {author} {\bibfnamefont {C.}~\bibnamefont
  {Fabre}},\ }\href {\doibase 10.1088/0022-3700/6/8/007} {\bibfield  {journal}
  {\bibinfo  {journal} {Journal of Physics B: Atomic and Molecular Physics}\
  }\textbf {\bibinfo {volume} {6}},\ \bibinfo {pages} {L214} (\bibinfo {year}
  {1973})}\BibitemShut {NoStop}%
\bibitem [{\citenamefont {Kerman}\ \emph {et~al.}(2000)\citenamefont {Kerman},
  \citenamefont {Vuleti\ifmmode~\acute{c}\else \'{c}\fi{}}, \citenamefont
  {Chin},\ and\ \citenamefont {Chu}}]{SteveChu_Vuletic}%
  \BibitemOpen
  \bibfield  {author} {\bibinfo {author} {\bibfnamefont {A.~J.}\ \bibnamefont
  {Kerman}}, \bibinfo {author} {\bibfnamefont {V.}~\bibnamefont
  {Vuleti\ifmmode~\acute{c}\else \'{c}\fi{}}}, \bibinfo {author} {\bibfnamefont
  {C.}~\bibnamefont {Chin}}, \ and\ \bibinfo {author} {\bibfnamefont
  {S.}~\bibnamefont {Chu}},\ }\href {\doibase 10.1103/PhysRevLett.84.439}
  {\bibfield  {journal} {\bibinfo  {journal} {Phys. Rev. Lett.}\ }\textbf
  {\bibinfo {volume} {84}},\ \bibinfo {pages} {439} (\bibinfo {year}
  {2000})}\BibitemShut {NoStop}%
\bibitem [{\citenamefont {Cui}\ \emph {et~al.}(2013)\citenamefont {Cui},
  \citenamefont {Lian}, \citenamefont {Ho}, \citenamefont {Lev},\ and\
  \citenamefont {Zhai}}]{SyntheticGaugeLanthanidesAtoms}%
  \BibitemOpen
  \bibfield  {author} {\bibinfo {author} {\bibfnamefont {X.}~\bibnamefont
  {Cui}}, \bibinfo {author} {\bibfnamefont {B.}~\bibnamefont {Lian}}, \bibinfo
  {author} {\bibfnamefont {T.-L.}\ \bibnamefont {Ho}}, \bibinfo {author}
  {\bibfnamefont {B.~L.}\ \bibnamefont {Lev}}, \ and\ \bibinfo {author}
  {\bibfnamefont {H.}~\bibnamefont {Zhai}},\ }\href {\doibase
  10.1103/PhysRevA.88.011601} {\bibfield  {journal} {\bibinfo  {journal} {Phys.
  Rev. A}\ }\textbf {\bibinfo {volume} {88}},\ \bibinfo {pages} {011601}
  (\bibinfo {year} {2013})}\BibitemShut {NoStop}%
\bibitem [{\citenamefont {Trypogeorgos}\ \emph {et~al.}(2018)\citenamefont
  {Trypogeorgos}, \citenamefont {Vald\'es-Curiel}, \citenamefont {Lundblad},\
  and\ \citenamefont {Spielman}}]{Spielman_ClockTransitions}%
  \BibitemOpen
  \bibfield  {author} {\bibinfo {author} {\bibfnamefont {D.}~\bibnamefont
  {Trypogeorgos}}, \bibinfo {author} {\bibfnamefont {A.}~\bibnamefont
  {Vald\'es-Curiel}}, \bibinfo {author} {\bibfnamefont {N.}~\bibnamefont
  {Lundblad}}, \ and\ \bibinfo {author} {\bibfnamefont {I.~B.}\ \bibnamefont
  {Spielman}},\ }\href {\doibase 10.1103/PhysRevA.97.013407} {\bibfield
  {journal} {\bibinfo  {journal} {Phys. Rev. A}\ }\textbf {\bibinfo {volume}
  {97}},\ \bibinfo {pages} {013407} (\bibinfo {year} {2018})}\BibitemShut
  {NoStop}%
\bibitem [{\citenamefont {Goldman}\ \emph {et~al.}(2014)\citenamefont
  {Goldman}, \citenamefont {Juzeli{\={u}}nas}, \citenamefont {Öhberg},\ and\
  \citenamefont {Spielman}}]{Goldman_ReviewArtificialGauge}%
  \BibitemOpen
  \bibfield  {author} {\bibinfo {author} {\bibfnamefont {N.}~\bibnamefont
  {Goldman}}, \bibinfo {author} {\bibfnamefont {G.}~\bibnamefont
  {Juzeli{\={u}}nas}}, \bibinfo {author} {\bibfnamefont {P.}~\bibnamefont
  {Öhberg}}, \ and\ \bibinfo {author} {\bibfnamefont {I.~B.}\ \bibnamefont
  {Spielman}},\ }\href {\doibase 10.1088/0034-4885/77/12/126401} {\bibfield
  {journal} {\bibinfo  {journal} {Reports on Progress in Physics}\ }\textbf
  {\bibinfo {volume} {77}},\ \bibinfo {pages} {126401} (\bibinfo {year}
  {2014})}\BibitemShut {NoStop}%
\bibitem [{\citenamefont {Bouchiat}\ \emph {et~al.}(1998)\citenamefont
  {Bouchiat}, \citenamefont {Vion}, \citenamefont {Joyez}, \citenamefont
  {Esteve},\ and\ \citenamefont {Devoret}}]{CooperPairBoxQubitFirstIdea}%
  \BibitemOpen
  \bibfield  {author} {\bibinfo {author} {\bibfnamefont {V.}~\bibnamefont
  {Bouchiat}}, \bibinfo {author} {\bibfnamefont {D.}~\bibnamefont {Vion}},
  \bibinfo {author} {\bibfnamefont {P.}~\bibnamefont {Joyez}}, \bibinfo
  {author} {\bibfnamefont {D.}~\bibnamefont {Esteve}}, \ and\ \bibinfo {author}
  {\bibfnamefont {M.~H.}\ \bibnamefont {Devoret}},\ }\href
  {http://stacks.iop.org/1402-4896/1998/i=T76/a=024} {\bibfield  {journal}
  {\bibinfo  {journal} {Physica Scripta}\ }\textbf {\bibinfo {volume} {1998}},\
  \bibinfo {pages} {165} (\bibinfo {year} {1998})}\BibitemShut {NoStop}%
\bibitem [{\citenamefont {Koch}\ \emph {et~al.}(2007)\citenamefont {Koch},
  \citenamefont {Yu}, \citenamefont {Gambetta}, \citenamefont {Houck},
  \citenamefont {Schuster}, \citenamefont {Majer}, \citenamefont {Blais},
  \citenamefont {Devoret}, \citenamefont {Girvin},\ and\ \citenamefont
  {Schoelkopf}}]{TransmonQubitShuntedCooperPair}%
  \BibitemOpen
  \bibfield  {author} {\bibinfo {author} {\bibfnamefont {J.}~\bibnamefont
  {Koch}}, \bibinfo {author} {\bibfnamefont {T.~M.}\ \bibnamefont {Yu}},
  \bibinfo {author} {\bibfnamefont {J.}~\bibnamefont {Gambetta}}, \bibinfo
  {author} {\bibfnamefont {A.~A.}\ \bibnamefont {Houck}}, \bibinfo {author}
  {\bibfnamefont {D.~I.}\ \bibnamefont {Schuster}}, \bibinfo {author}
  {\bibfnamefont {J.}~\bibnamefont {Majer}}, \bibinfo {author} {\bibfnamefont
  {A.}~\bibnamefont {Blais}}, \bibinfo {author} {\bibfnamefont {M.~H.}\
  \bibnamefont {Devoret}}, \bibinfo {author} {\bibfnamefont {S.~M.}\
  \bibnamefont {Girvin}}, \ and\ \bibinfo {author} {\bibfnamefont {R.~J.}\
  \bibnamefont {Schoelkopf}},\ }\href {\doibase 10.1103/PhysRevA.76.042319}
  {\bibfield  {journal} {\bibinfo  {journal} {Phys. Rev. A}\ }\textbf {\bibinfo
  {volume} {76}},\ \bibinfo {pages} {042319} (\bibinfo {year}
  {2007})}\BibitemShut {NoStop}%
\bibitem [{\citenamefont {Zaletel}\ and\ \citenamefont
  {Mong}(2012)}]{ZaletelMongMPS}%
  \BibitemOpen
  \bibfield  {author} {\bibinfo {author} {\bibfnamefont {M.}~\bibnamefont
  {Zaletel}}\ and\ \bibinfo {author} {\bibfnamefont {R.}~\bibnamefont {Mong}},\
  }\href {\doibase 10.1103/PhysRevB.86.245305} {\bibfield  {journal} {\bibinfo
  {journal} {Phys. Rev. B}\ }\textbf {\bibinfo {volume} {86}},\ \bibinfo
  {pages} {245305} (\bibinfo {year} {2012})}\BibitemShut {NoStop}%
\bibitem [{\citenamefont {{Estienne}}\ \emph {et~al.}()\citenamefont
  {{Estienne}}, \citenamefont {{Regnault}},\ and\ \citenamefont
  {{Bernevig}}}]{RegnaultConstructionMPS}%
  \BibitemOpen
  \bibfield  {author} {\bibinfo {author} {\bibfnamefont {B.}~\bibnamefont
  {{Estienne}}}, \bibinfo {author} {\bibfnamefont {N.}~\bibnamefont
  {{Regnault}}}, \ and\ \bibinfo {author} {\bibfnamefont {B.~A.}\ \bibnamefont
  {{Bernevig}}},\ }\href@noop {} {\bibfield  {journal} {\bibinfo  {journal}
  {Preprint at}\ }}\Eprint {http://arxiv.org/abs/1311.2936} {arXiv:1311.2936
  [cond-mat.str-el]} \BibitemShut {NoStop}%
\bibitem [{\citenamefont {Cr\'epel}\ \emph {et~al.}(2018)\citenamefont
  {Cr\'epel}, \citenamefont {Estienne}, \citenamefont {Bernevig}, \citenamefont
  {Lecheminant},\ and\ \citenamefont {Regnault}}]{MyFirstArticle}%
  \BibitemOpen
  \bibfield  {author} {\bibinfo {author} {\bibfnamefont {V.}~\bibnamefont
  {Cr\'epel}}, \bibinfo {author} {\bibfnamefont {B.}~\bibnamefont {Estienne}},
  \bibinfo {author} {\bibfnamefont {B.~A.}\ \bibnamefont {Bernevig}}, \bibinfo
  {author} {\bibfnamefont {P.}~\bibnamefont {Lecheminant}}, \ and\ \bibinfo
  {author} {\bibfnamefont {N.}~\bibnamefont {Regnault}},\ }\href {\doibase
  10.1103/PhysRevB.97.165136} {\bibfield  {journal} {\bibinfo  {journal} {Phys.
  Rev. B}\ }\textbf {\bibinfo {volume} {97}},\ \bibinfo {pages} {165136}
  (\bibinfo {year} {2018})}\BibitemShut {NoStop}%
\bibitem [{\citenamefont {P{\'\i}tajevsk{\'\i}j}\ \emph
  {et~al.}(2003)\citenamefont {P{\'\i}tajevsk{\'\i}j}, \citenamefont
  {S.~Stringari}, \citenamefont {Pitaevskii}, \citenamefont {Stringari},
  \citenamefont {Stringari},\ and\ \citenamefont
  {Press}}]{PitaevskiiStringari_BECBook}%
  \BibitemOpen
  \bibfield  {author} {\bibinfo {author} {\bibfnamefont {L.}~\bibnamefont
  {P{\'\i}tajevsk{\'\i}j}}, \bibinfo {author} {\bibfnamefont {L.}~\bibnamefont
  {S.~Stringari}}, \bibinfo {author} {\bibfnamefont {L.}~\bibnamefont
  {Pitaevskii}}, \bibinfo {author} {\bibfnamefont {S.}~\bibnamefont
  {Stringari}}, \bibinfo {author} {\bibfnamefont {S.}~\bibnamefont
  {Stringari}}, \ and\ \bibinfo {author} {\bibfnamefont {O.~U.}\ \bibnamefont
  {Press}},\ }\href {https://books.google.fr/books?id=rIobbOxC4j4C} {\emph
  {\bibinfo {title} {Bose-Einstein Condensation}}},\ International Series of
  Monographs on Physics\ (\bibinfo  {publisher} {Clarendon Press},\ \bibinfo
  {year} {2003})\BibitemShut {NoStop}%
\end{thebibliography}%

\appendix

\section{Mathieu Differential Equation And Asymptotics} \label{App:AllAboutMathieuEqn}

In this appendix, we show how the stationary Sch\"odinger corresponding to Eq.~\ref{eq:HamiltonianMomentumSector} maps onto Mathieu's differential equation, we review its solutions and analyze further the weak and strong tunneling limits. 

\subsection{Relevance Of Mathieu's Equation} \label{sApp:MappingOnMathieu}

In this section, we fix the momentum sector $k$, and we will omit the explicit dependence in momentum when it is unnecessary. We use the first quantized notations $\ket{j} = c_{j,k}^\dagger \ket{0}$ for the state localized on wire $j$. The stationary Schr\"odinger equation for the eigenstate $\braket{j}{\psi}=\psi(j)$ of Eq.~\ref{eq:HamiltonianMomentumSector} reads 
\begin{equation} \label{eq:AppMathieuEigenfunctionEqn} 
(j-j_0)^2 \psi(j) - \lambda \left[ \psi(j+1)+\psi(j-1) \right] = \mu \psi(j) 
\end{equation} 
with $j_0=\nu_w k$ and where $\mu$ is the eigenenergy in unit of $E_0$. The coupled-wire system studied in Sec.~\ref{sec:IQHEFromWires} is thus equivalent to the well studied Cooper pair box Hamiltonian~\cite{CooperPairBoxQubitFirstIdea,TransmonQubitShuntedCooperPair}, which is usually solved introducing the conjugate variable of $j$. Thus, we introduce the phase $\varphi \in [-\pi,\pi[$, canonically conjugated to the wire position $j \in \mathbb{Z}$:
\begin{equation} \begin{split}
\ket{\varphi} & = \sum_{j \in \mathbb{Z}} e^{i(j-j_0)\varphi} \ket{j} \, , \\ \ket{j} & = \int_{-\pi}^\pi \frac{{\rm d}\varphi}{2\pi} e^{-i(j-j_0)\varphi} \ket{\varphi} \, . 
\end{split} \end{equation} 
It yields
\begin{equation}  \label{eq:AppMathieuEquationStandard} 
\frac{{\rm d}^2 \psi(\varphi)}{{\rm d} \,  \varphi^2} +\left[\mu+2 \lambda \cos \varphi \right] \psi(\varphi) = 0 \, ,   
\end{equation} 
where $\psi(\varphi) = \braket{\varphi}{\psi}$. Notice that the inclusion of $j_0$ in the definition of the conjugate variable makes the function $\psi(\varphi)$ pseudo-periodic $\psi(\varphi+2\pi)=e^{2i\pi j_0}\psi(\varphi)$. This can also be seen as a gauge transformation in the $\varphi$ representation. After the final change of variable $x=(\varphi+\pi)/2$, this equation takes Mathieu's equation standard form~\cite[§~28]{NIST:DLMF} 
\begin{equation} 
w''(x)+\left[\gamma-2q \cos 2x \right] w(x) =0 \, ,
\end{equation} 
with $\gamma = 4\mu$ and $q=4\lambda$.

\subsection{Formal Solutions} \label{sApp:SolutionMathieuEquation}

Eq.~\ref{eq:AppMathieuEquationStandard} together with the pseudo-periodicity of $\psi$ allows to identify 
\begin{equation}
\psi(\varphi)={\rm me}_{\nu_M}(4\lambda,(\varphi+\pi)/2) \, ,
\end{equation}
where ${\rm me}_{\nu_M}$ denotes is the Floquet solution of the Mathieu's equation with characteristic exponent 
\begin{equation}
\nu_M=2(j_0+ \eta) \, ,
\end{equation}
with $\eta \in \mathbb{Z}$ labeling the solutions of Eq.~\ref{eq:AppMathieuEquationStandard}. To interpret $\eta$ as a band index, we shall sort the corresponding eigenvalues $\mu = (1/4) a_{\nu_M}(4\lambda)$ in ascending order, where we used Mathieu's characteristic function $a_{\nu_M}(q)$~\cite[chap.~28]{NIST:DLMF}. For the $n$-th band, the sorting function $\eta(j_0,n)$ is a non-trivial function of $j_0$ and of the band index whose explicit form is known~\cite{TransmonQubitShuntedCooperPair}. Denoting as $d_k^{(n) \dagger}$ the creation operator of a particle of momentum $k$ in band $n$ (see Sec.~\ref{ssec:NonIntLandauLevels}), and by $\ket{n}=d_k^{(n) \dagger}\ket{0}$ the corresponding wavefunction, we finally arrive at the closed form expression of the function $g_k^{(n)}$ used in Eq.~\ref{eq:CoupledWireEigenfunction_GeneralMathieuImplicit}:
\begin{widetext}
\begin{equation} \label{eq:AppMathieuExactExpressionForG}
\braket{j}{n} = g_k^{(n)} (j-j_0) = \int_{-\pi}^{\pi} \frac{{\rm d}\varphi}{2\pi} e^{i(j-j_0)\varphi} {\rm me}_{\left[2(j_0+ \eta(j_0,n))\right]}(4\lambda,(\varphi+\pi)/2) \, .
\end{equation} 
\end{widetext}

Our main interest in the formal mapping onto Mathieu's equation is to know whether the flat band approximation invoked in the main text is justified (see for instance Eq.~\ref{eq:LowestBandisgaussian} or Eq.~\ref{eq:LLLdispersionStrongTunneling}). Let us define $\delta^{(n)}(\lambda)$ as the spread of the $n$-th band of the spectrum of Eq.~\ref{eq:HamiltonianMomentumSector}. Using uniform semiclassical approximations~\cite{KurzPhDApproximationMathieuWF,GoldsteinCharacteristicFunction}, it is possible to obtain an expansion of the spread for large tunneling strength $\lambda \gg 1$: \begin{equation} \label{eq:AppMathieuAsymptoticsSpread} \begin{split}
\delta^{(n)}(\lambda) \simeq & \frac{2^{4n+3}}{n!} \left(
\frac{2}{\pi} \right)^{1/2} \left( 2 \sqrt{\lambda}\right)^{n+3/2}e^{-8\sqrt{\lambda}} \\ & \times {\left(1-\frac{6n^{2}+14n+7}{64\sqrt{\lambda}}+O\left(\frac{1}{\lambda}\right)\right)} \, ,
\end{split} \end{equation} 
We have numerically diagonalized the Hamiltonian Eq.~\ref{eq:HamiltonianMomentumSector} for several tunneling and we compare in Tab.~\ref{tab:NumericalTunnelingStrength} the asymptotic behavior Eq.~\ref{eq:AppMathieuAsymptoticsSpread} to the numerically extracted values. In practice, the exponentially small spread makes the flat band approximation very accurate, even for moderate tunneling strength, as evidenced by the large spread to gap ratio gathered in Tab.~\ref{tab:NumericalTunnelingStrength}.

\begin{table}
\centering
\caption{\emph{We numerically diagonalized Hamiltonian Eq.~\ref{eq:HamiltonianMomentumSector} for over 1000 wires. We compute the spread of the lowest band $\delta^{(0)}$ and compare it to the asymptotic limit given of Eq.~\ref{eq:AppMathieuAsymptoticsSpread}. We find that the flat band limit is extremely well satisfied even for very moderate tunneling strengths $\lambda\sim 1$. We also extract the flatness of the lowest band, defined as the ratio of its spread to the gap separating it to the first excited band. We find that the gap is orders of magnitude greater for $\lambda\geq 1$, justifying our projection onto the lowest band in Sec.~\ref{sec:AddingInteractions}.}}
\label{tab:NumericalTunnelingStrength}
\begin{tabular}{|c|c|c|c|}
\hline \hline
$\lambda$ & $\delta^{(0)}$ & Asymptotic Eq.~\ref{eq:AppMathieuAsymptoticsSpread} & Flatness \\ \hline 
\, 0.10 \,  & \, 1.649e-1 \, &   1.673e-1   &  \, 8.266e-1 \, \\
0.33  &  6.051e-2  &   6.333e-2   &   9.327e-2  \\
0.66  &  1.618e-2  &   1.680e-2   &   1.366e-2  \\
1.00  &  5.334e-3  &   5.394e-3   &   3.243e-3  \\
3.33  &  1.894e-5  &   1.899e-5   &   5.605e-6  \\
6.66  &  7.663e-8  &   7.673e-8   &   1.564e-8  \\ 
10.0  &  1.009e-9  &   1.010e-9   &   1.665e-10 \\ 
33.3  &  1.42e-14  &   2.144e-18  &   1.259e-15 \\ \hline \hline
\end{tabular}
\end{table}

\section{Laughlin State With Similar Aspect Ratio} \label{App:MatchAspectRatio}

In the main text, we have compared our ED ground states $\ket{\Psi_{GS}}$ with Laughlin states on a cylinder $\ket{\Psi_{1/2}(L_{\rm cyl})}$ of perimeter $L_{\rm cyl}$. In this appendix, we provide more details on the choice of $L_{\rm cyl}$. To differentiate this cylinder from the wire system, we respectively denote as $B^{\rm cyl}$, $\ell_B^{\rm cyl}$, $\omega_c^{\rm cyl}$ the magnetic field threading the cylinder, the corresponding magnetic length and cyclotron energy.

For an infinite system $N_w \gg 1$, we have derived a rescaling of the coordinates $x$ and $y$ which allows to directly compare the wire ground state with the cylinder Laughlin state (see Sec.~\ref{ssec:ModelInteractionsLowestBand}). In particular, equating the perimeter of the cylinder and the inter-orbital distance leads to
\begin{equation} \label{eq:AspectRatioInfiniteSize}
\left.
\begin{array}{ll}
L_{\rm cyl} = L / r \\
\frac{2 \pi (\ell_B^{\rm cyl})^2}{L_{\rm cyl}} = rd\nu_w
\end{array}
\right\}
\quad \Longrightarrow \quad \frac{L_{\rm cyl}}{\ell_B^{\rm cyl}} \underset{N_w \gg 1}{=} \frac{2\pi}{\nu_w} \lambda^{1/4} \, .
\end{equation}
This formula can be understood as matching the aspect ratio of the two models. However, it relies on the bulk properties derived in Sec.~\ref{sec:AddingInteractions} assuming a very large number of wires and cannot be used directly for the finite-size systems that we are interested with in Sec.~\ref{sec:ED_ForLaughlin}. We have thus relied on the following numerical approach.

\begin{figure}
\centering
\includegraphics[width=\columnwidth]{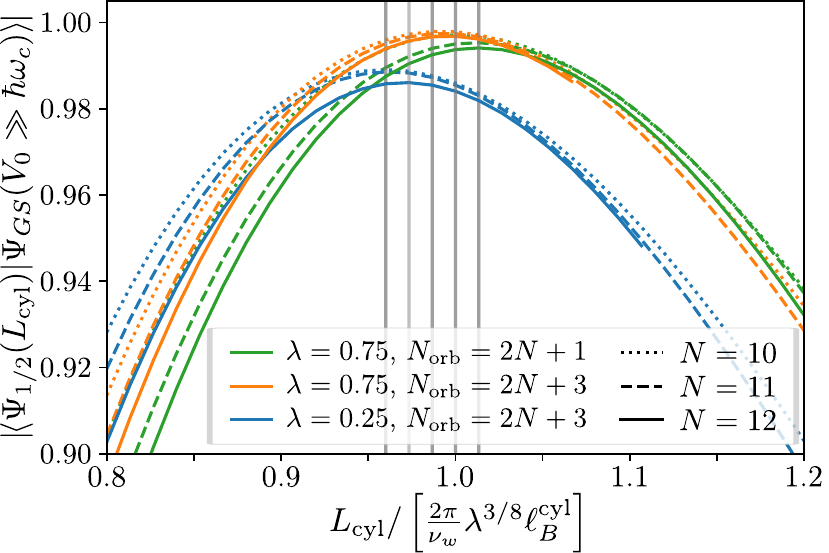}
\caption{\emph{
Overlap between the ED ground state obtained for very large interaction strengths and the Laughlin state on a perimeter $L_{\rm cyl}$. Different colors indicate different tunneling strength $\lambda$ or a different number of extra orbitals (see Sec.~\ref{ssec:SetupED}), while dotted, dashed and solid lines are respectively used for $N=$ 10, 11 and 12. The cylinder maximizing the overlap is highlighted with gray vertical bars. For all the considered parameters $(\lambda, N_\phi, N, N_{\rm orb})$, the optimal perimeter lies within ten percent of $\frac{2\pi}{\nu_w} \lambda^{3/8} \ell_B^{\rm cyl}$. This fact which is made apparent by the normalization of the $x$-axis. 
}}
\label{fig:AppLaughlinChoicePerim}
\end{figure}

Consider the ground state $\ket{\Psi_{GS}}$ obtained with the parameters $(\lambda, N_\phi, N, N_{\rm orb})$ and interaction strength $V_0$. To find $L_{\rm cyl}$, we first diagonalize the wire Hamiltonian Eq.~\ref{eq:TotalEDHamiltonian} for the same parameters $(\lambda, N_\phi, N, N_{\rm orb})$ but with $V_0 \gg 1$ (typically $V_0 = 5 \hbar \omega_c$). This leads to a new ground state $\ket{\Psi_{GS}(V_0 \gg \hbar\omega_c)}$. Then, we generate several Laughlin states $\ket{\Psi_{1/2}(L_{\rm cyl})}$ with different aspect ratios
\begin{equation} \label{eq:AspectRatioCylinder}
r_{\rm cyl} = 2 \pi N_{\rm orb} \left( \frac{\ell_B^{\rm cyl}}{L_{\rm cyl}} \right)^2 \, ,
\end{equation}
by varying te parameter $L_{\rm cyl}/\ell_B^{\rm cyl}$. Practically, we obtain the states $\ket{\Psi_{1/2}(L_{\rm cyl})}$ either by ED of the model interaction Eq.~\ref{eq:ModelInnteractionLLLBosons} or through exact matrix product states~\cite{ZaletelMongMPS,RegnaultConstructionMPS,MyFirstArticle}. We finally choose the perimeter maximizing the overlap $|\braket{\Psi_{1/2}(L_{\rm cyl})}{\Psi_{GS}(V_0 \gg \hbar\omega_c)}|$.

Our numerical results are presented in Fig.~\ref{fig:AppLaughlinChoicePerim}. The best overlaps obtained with this technique are never lower than 0.98, providing further evidence for the emergence of the Laughlin physics in our model (see Sec.~\ref{ssec:LaughlinFindingED}). They also increase with greater tunneling strengths $\lambda$, as expected from our analytical result of Sec.~\ref{ssec:ModelInteractionsLowestBand} in the limit $\lambda>1$. Moreover, we observe that the perimeters maximizing the overlaps lie close to
\begin{equation} \label{eq:EmpiricalLawL}
\frac{L_{\rm cyl}}{\ell_B^{\rm cyl}} \simeq 2 \pi \frac{\lambda^{3/8}}{\nu_w} \, .
\end{equation}
Among all the considered parameters $(\lambda, N_\phi, N, N_{\rm orb})$, we have only observed fluctuations of less than 10\% from this empirical formula. Therefore, we have used this empirical finding as an initial guess to limit our numerical burden.

While we are not able to justifies Eq.~\ref{eq:EmpiricalLawL} analytically, we can understand the scaling with $\lambda$ as follows. Equating $r_{\rm cyl}$ of Eq.~\ref{eq:AspectRatioCylinder} to the aspect ratio $r = N_w d/L$ of the finite wire system, we find:
\begin{equation}
\frac{L_{\rm cyl}}{\ell_B^{\rm cyl}} = \frac{2\pi}{\nu_w} \lambda^{1/4} \sqrt{\frac{N_{\rm orb} }{N_\phi}} \, .
\end{equation}
For an infinite system, it reduces to Eq.~\ref{eq:AspectRatioInfiniteSize} since $N_\phi \simeq N_{\rm orb}$. In our ED calculation however, the magnetic field $B$ of the wire system is tuned away from $B^{\rm cyl}$ such as to have precisely $N_{\rm orb}$ orbitals below the single particle gap. This introduce a difference between $N_\phi$ and $N_{\rm orb}$ in the calculations of Sec.~\ref{sec:ED_ForLaughlin} and explains the different scaling observed in Eq.~\ref{eq:EmpiricalLawL}. Considering the quadratic dispersion near the edge $\varepsilon_k^{(0)} \simeq (\nu_w k)^2 E_0$, and requiring that the orbital with momentum $k=N_{\rm orb}/2$ has energy $\hbar \omega_c \simeq 2 \sqrt{\lambda} E_0$ (see Eq.~\ref{ssec:SetupED}), we get
\begin{equation}
\frac{N_{\rm orb}}{N_\phi} \propto \lambda^{1/4} \quad \Rightarrow \quad \frac{L_{\rm cyl}}{\ell_B^{\rm cyl}} \propto \lambda^{3/8} \, .
\end{equation}

\section{Bogoliubov Excitations} \label{App:BogoliubovTheory}

In this appendix, we provide more details about the Bogoliubov approach sketched in Sec.~\ref{ssec:IntermediatePhases} used to describe the weakly-interacting phases in the model Eq.~\ref{eq:TotalEDHamiltonian}  (see, for instance, Ref.~\cite{PitaevskiiStringari_BECBook} for a more thorough and comprehensive review of the method). Due to both the finite-size of our system and to the presence of other local minima in the dispersion relation (see Fig.~\ref{fig:MomentumDensityOccupation}), the standard Bogoliubov analysis only holds for very small interaction strength. We thus provide a modified version to take into account the finite size and the presence of multiple disconnected minima in the dispersion relation.

\subsection{Very-Weakly Interacting Phase: Thermodynamic Limit} \label{ssec:StandardBogoliubov}

Let us first follow the standard derivation of the Bogoliubov quadratic Hamiltonian~\cite{PitaevskiiStringari_BECBook}. We assume that the macroscopically populated BEC of Eq.~\ref{eq:BecAtNoInteraction} is only slightly depleted, such that we can replace $N_0 = d_0^{(0) \dagger} d_0^{(0)} \simeq N$ and $d_0^{(0)} \simeq \sqrt{N_0}$ by their expectation value. The different terms of the Hamiltonian Eq.~\ref{eq:TotalEDHamiltonian} can then be sorted in decreasing order of importance, according to their scaling with $N$. Keeping only terms at least proportional to the particle number, we find the following Bogoliubov approximation of the many-body Hamiltonian
\begin{align} \label{eq:BogoHamiltonian}
	& \mathcal{H}_{\rm Bogo} = \varepsilon_0^{(0)}  N + V_0 N^2 v(0) \notag \\
	& + \sum_{k \neq 0} \underbrace{\left[ \left( \varepsilon_k^{(0)} - \varepsilon_0^{(0)} \right) + 2 V_0 N \left( 2 v(k) - v(0) \right) \right]}_{D(k)} d_k^{(0) \dagger} d_k^{(0)} \notag \\
	& + \sum_{k \neq 0}  V_0 N v(k) \left(d_k^{(0) \dagger} d_{-k}^{(0) \dagger} + d_k^{(0)}  d_{-k}^{(0)}  \right) \, ,
\end{align}
where $v(k) = \Gamma_{k/2,k/2}^{k/2} = \Gamma_{0,k}^0 = \Gamma_{k,0}^0$ is real. Note that the inversion symmetry of the problem implies $v(k) = v(-k)$ and $\varepsilon_k^{(0)} = \varepsilon_{-k}^{(0)}$. The first line of Eq.~\ref{eq:BogoHamiltonian} corresponds to the energy of the weakly interacting BEC $E_{\rm BEC}(N, V_0) = \varepsilon_0^{(0)}  N + V_0 N^2 v(0)$. The last two lines show the quadratic Hamiltonian which account for the entire many-body problem in the Bogoliubov approach.

This quadratic Hamiltonian can be diagonalized as
\begin{equation}
\mathcal{H}_{\rm Bogo} = E_{\rm BEC}(N, V_0) + \sum_{k \neq 0} \epsilon_k^B B_k^\dagger B_k \, , 
\end{equation}
where the operators $B_k$ are obtained by the squeezing (or Bogoliubov) transformation
\begin{equation}\label{eq:AppBogoTransfo}
d_k^{(0)} = u_k B_k - v_k B_{-k}^\dagger \, , \quad
d_{-k}^{(0) \dagger} = u_k B_{-k}^\dagger - v_k B_k \, .
\end{equation}
The real coefficients $u_k$ and $v_k$ must satisfy $u_k^2 - v_k^2 = 1$ for the new operators to obey bosonic commutation relations $[B_k , B_k^\dagger ] = 1$. They can thus be parametrized by a hyperbolic angle $u_k = \cosh \theta_k$ and $v_k = \sinh\theta_k$. Plugging Eq.~\ref{eq:AppBogoTransfo} into Eq.~\ref{eq:BogoHamiltonian}, this angle is chosen to make the  the $B_k B_{-k}$ and $B_k^\dagger B_{-k}^\dagger$ term vanish:
\begin{equation}
\tanh (2\theta_k) = \frac{2 V_0 N v(k)}{D(k)} \, .
\end{equation}
This allows to derive the following expressions for the quasiparticles eigen-energies and weights:
\begin{align}
\epsilon_k^B & = \sqrt{D(k)^2 - 4 [V_0 N v(k)]^2} \, , \\
u_k & = \sqrt{\frac{D(k)}{2\epsilon_k^B} + \frac{1}{2}} \, , \quad v_k = \sqrt{\frac{D(k)}{2\epsilon_k^B} - \frac{1}{2}} \, . \label{eq:UVfromBogoThemodynamic}
\end{align}

The ground state of the system is now defined as the vacuum state for the quasi-particle operators, \textit{i.e.} $B_k \ket{\Psi_{\rm Bogo}(V_0)} = 0$. It can be expressed as:
\begin{equation} \label{eq:BogoGSTheory}
\ket{\Psi_{\rm Bogo}(V_0)} = \exp\left( - \sum_{k>0} t_k \, d_k^{(0) \dagger} d_{-k}^{(0) \dagger} \right) \ket{\Psi_{\rm GS}(0)} \, ,
\end{equation}
with $t_k = v_k/u_k$. This equation makes clear that the system accommodates the weak interaction by the creation of particle pairs with non-zero momenta $\pm k$, as stated in the main text.

\subsection{Very-Weakly Interacting Phase: Finite Size Systems} \label{ssec:AppFiniteSizeProblems}

In order to compare the weakly-interacting theory to our ED results, we must adapt Eq.~\ref{eq:BogoGSTheory} to finite size systems, where $\sqrt{N}$ is not much larger than one, and ensure particle number conservation. We thus consider the following ansatz, reproduced from Eq.~\ref{eq:BogoGSFiniteSize_OnePeak}:
\begin{equation} \label{eqapp:BogoGSFiniteSize_OnePeak}
\ket{\tilde{\Psi}_{\rm Bogo}(V_0)} = e^{ - \sum_{k>0} \frac{t_k}{N} \, d_k^{(0) \dagger} d_{-k}^{(0) \dagger} d_0^{(0)} d_0^{(0)} } \ket{\Psi_{\rm GS}(0)} \, . 
\end{equation}
The coefficients $\{t_k\}_{k > 0}$ are variationally optimized around their $\sqrt{N} \gg 1$ theoretical value (Eq.~\ref{eq:UVfromBogoThemodynamic}) to correct for small finite-size effects. The depletion of the BEC is explicitly accounted for by gluing the operator $d_0^{(0)} d_0^{(0)} / N$ to the pair creation operator in the exponential of Eq.~\ref{eqapp:BogoGSFiniteSize_OnePeak}. This formally implement the required particle number conservation. 

As explained in the main text and numerically demonstrated in Fig.~\ref{fig:OverlapsBogoLaughlin}, Eq.~\ref{eqapp:BogoGSFiniteSize_OnePeak} almost perfectly captures the ED ground states for very-weak interactions $V_0 N \ll 1$. However, it fails to capture the nature of the other intermediate phases, characterized by more than one peak in their momentum-space density distribution (see Fig.~\ref{fig:MomentumDensityOccupation}).

\subsection{Other Weakly Interacting Phase} \label{ssec:AppOtherWeaklyInteractingPhases}

As explained in Sec.~\ref{ssec:IntermediatePhases}, the multiple local minima of the lowest band (see Fig.~\ref{fig:MomentumDensityOccupation}a) and the finite range of the interaction (see Eq.~\ref{eq:GaussianInteractionFormFactor}) tend to favor the creation of multiple condensates with very different momenta. The previous Bogoliubov approach can be adapted to this case as well. The idea is to conserve the variational parameters of Eq.~\ref{eq:BogoGSFiniteSize_OnePeak}, which measure the depletion of the condensates by pair-creation in order to accommodate the weak interactions of the system, while changing the initial state they act on. This initial state is composed of $P$ multiple BECs located near the minima of the dispersion relation with momenta $k_1, \cdots , k_P$, and hosting $N_1, \cdots N_P$ particles. They are can be written as
\begin{equation}
\ket{\left( k_i, N_i \right)_{i=1, \cdots , P} } = \prod_{i=1}^P \frac{1}{\sqrt{N_i}} \left(d_{k_i}^{(0) \dagger}\right)^{N_i} \ket{\emptyset} \, . 
\end{equation}
The original state $\ket{\Psi_{\rm GS}(0)}$, corresponding to the blue lines in Fig.~\ref{fig:MomentumDensityOccupation}, is recovered by choosing $P=1$, $k_1 = 0$ and $N_1 = N$. For all the other interaction strengths $V_0$ in the weakly-interacting phases, we fix these new parameters such that $\ket{\left( k_i, N_i \right)_{i=1, \cdots , P} }$ corresponds to the highest-weight number state in the many-body decomposition of the ED ground state $\ket{\Psi_{\rm GS}(V_0)}$. For instance, the case depicted in orange in Fig.~\ref{fig:MomentumDensityOccupation} has $P=3$ peaks located at $k_1 = -5$, $k_2=0$ and $k_3=5$, and hosting $N_1 = N_2 = N_3 = 4$ particle each (before depletion by the pair-creation operators).

Once the initial $\ket{\left( k_i, N_i \right)_{i=1, \cdots , P} }$ has been determined as detailed above, the Bogoliubov ansatz is obtained by appending the exponentiated particle-pair creation operator as in Eq.~\ref{eq:BogoGSFiniteSize_OnePeak}. This results in the following ansatz:
\begin{equation} \label{eq:BogoGSFiniteSize_ManyPeaks}
\ket{\Psi_{\rm Bogo}(V_0)} = \exp\left( - \sum_{ k>0} \frac{t_k}{N} \, d_k^{(0) \dagger} d_{-k}^{(0) \dagger} d_{k_i}^{(0)} d_{k_j}^{(0)} \right) \ket{\Psi_{\rm GS}(0)} \, .
\end{equation}
The overlaps between this ansatz and our numerical ground states are shown in Fig.~\ref{fig:OverlapsBogoLaughlin}. They nicely capture the underlying physics of our model before the transition towards the FQH-like state.

\end{document}